\documentclass[nofootinbib,aps]{revtex4}
\pdfoutput=1
\usepackage{graphicx}
\usepackage{hyperref}
\def\mpl{m_{\rm Pl}}
\begin{document}

%--%\chapter{TASI Lectures on Inflation}
\title{TASI Lectures on Inflation}

\author{William H. Kinney}

%--%\address{Department of Physics, University at Buffalo, SUNY,\\
%--%Buffalo, NY 14260-1500, USA\\
%--%$^*$E-mail: whkinney@buffalo.edu}
\affiliation{Department of Physics, University at Buffalo, SUNY,\\
Buffalo, NY 14260-1500, USA}
\email{whkinney@buffalo.edu}

\begin{abstract}
This series of lectures gives a pedagogical review of the subject of cosmological inflation. I discuss Friedmann-Robertson-Walker cosmology and the horizon and flatness problems of the standard hot Big Bang, and introduce inflation as a solution to those problems, focusing on the simple scenario of inflation from a single scalar field. I discuss quantum modes in inflation and the generation of primordial tensor and scalar fluctuations. Finally, I provide comparison of inflationary models to the WMAP satellite measurement of the Cosmic Microwave Background, and briefly discuss future directions for inflationary physics. The majority of the lectures should be accessible to advanced undergraduates or beginning graduate students with only a background in Special Relativity, although familiarity with General Relativity and quantum field theory will be helpful for the more technical sections. 
\end{abstract}

%--%\body
\maketitle

\section{Introduction}\label{sec:Intro}

Cosmology today is a vibrant scientific enterprise. New precision measurements are revealing a universe with surprising and unexpected properties, in particular the Dark Matter and Dark Energy which are now believed to be the dominant components of the cosmos. Galaxy surveys such as the Sloan Digital Sky Survey are making the first large-scale maps of the universe, and satellites such as WMAP are making exquisitely precise measurements of the Cosmic Microwave Background (CMB), the haze of relic photons left over from the Big Bang. In turn, these measurements are giving us clues which are helping to unravel one of the oldest and most profound questions people have ever asked: Where did the universe come from? In these lectures, I discuss what is currently the best motivated and most completely developed physical model for the first moments of the universe: cosmological inflation \cite{Guth:1980zm,Linde:1981mu,Albrecht:1982wi}.\footnote{Inflation in its current form was introduced by Guth, but similar ideas had been discussed before \cite{Gliner:1966,Gliner:1975}. A short history of the early development of inflation can be found in Ref. \cite{Linde:2007fr}.} Inflation naturally explains how the universe came to be so large, so old, and so flat, and provides a compellingly elegant and predictive mechanism for generating the primordial perturbations which gave rise to the rich structure we see in the universe today \cite{Starobinsky:1979ty,Mukhanov:1981xt,Hawking:1982cz,Hawking:1982my,Starobinsky:1982ee,Guth:1982ec,Bardeen:1983qw}. Inflation provides a link between the Outer Space of astrophysics and the Inner Space of particle physics, and gives us a window to physics at energy scales far beyond the reach of particle accelerators. Furthermore, inflation makes {\it testable} predictions, which have so far proven to be an excellent match to the data. 

The lectures are organized as follows:
\begin{itemize}
\item{Section \ref{sec:Intro} provides an introduction and a brief overview of General Relativity.}
\item{Section \ref{sec:FRWSpacetime} discusses the Friedmann-Robertson-Walker spacetime and the standard hot Big Bang picture of cosmology, including the Cosmic Microwave Background.}
\item{Section \ref{sec:Flatness} explains unresolved issues in the standard cosmology, in particular the horizon and flatness problems.}
\item{Section \ref{sec:InflationFromScalarFields} introduces inflation in scalar field theories.}
\item{Section \ref{sec:PerturbationsInInflation} discusses quantum fluctuations in inflation and the generation of cosmological perturbations.}
\item{Section \ref{sec:ObservationalConstraints} discusses the observational predictions of inflation, and current constraints from Cosmic Microwave Background measurements.}
\item{Section \ref{sec:Conclusions} discusses conclusions and the future outlook for inflationary physics.}
\item{Appendix \ref{sec:Appendix} describes in detail the generation of density perturbations during inflation.}
\end{itemize}
The lectures are at an advanced undergraduate or beginning graduate student level. Most of the the lectures should be accessible with only a background in Special Relativity. A working knowledge of General Relativity and quantum field theory are helpful for Sections \ref{sec:InflationFromScalarFields} and \ref{sec:PerturbationsInInflation} and for Appendix \ref{sec:Appendix}.  Where possible, I reference review articles for further reading on related topics. For other reviews on inflation, see Refs. \cite{Lyth:1998xn,Watson:2000hb,Riotto:2002yw,Lineweaver:2003ie,Kinney:2003xf,Trodden:2004st,Linde:2007fr,Baumann:2008bn}.

\subsection{The Metric}

\begin{figure}
\begin{center}
%--%\psfig{file=MinkowskiLightcone.eps,width=4.5in}
\includegraphics[width=4.5in]{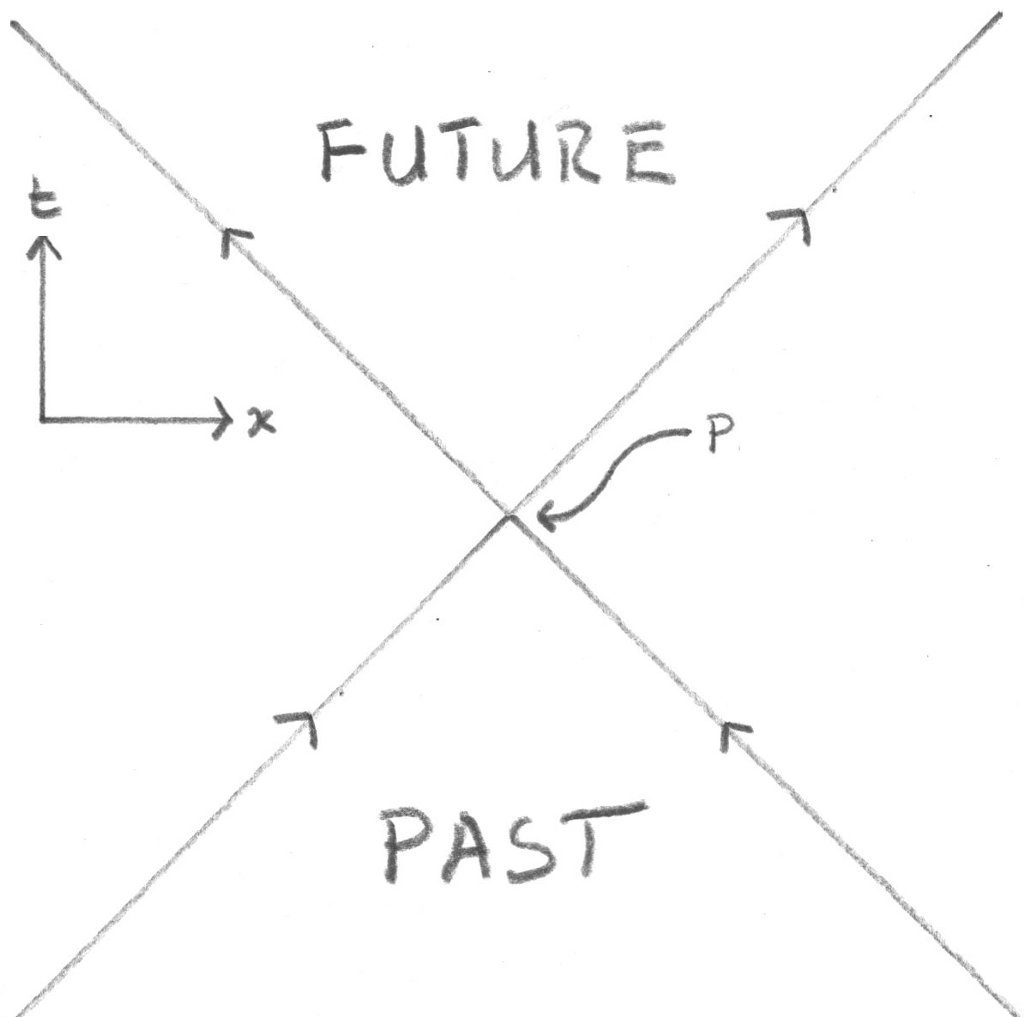}
\end{center}
\caption{Light cones in Minkowski Space. The past light cone defines the causal past of the event $P$, and the future light cone defines the causal future of $P$.}
\label{fig:MinkowskiLightcones}
\end{figure}

The fundamental object in General Relativity is the {\it metric}, which encodes the shape of the spacetime. A metric is a symmetric, bilinear form which defines distances on a manifold. For example, we can express Pythagoras' theorem in a Euclidean three-dimensional space,
\begin{equation}
\ell^2 = x^2 + y^2 + z^2,
\end{equation}
as a matrix product over the identity matrix $\delta_{ij} = {\rm diag}\left(1,1,1\right)$,
\begin{equation}
\ell^2 = \sum_{i,j = 1,3} \delta_{ij} x^i x^j.
\end{equation}
Therefore the identity matrix $\delta_{ij}$ can be identified as the metric for the Euclidean space: if we wish to describe a non-Euclidean manifold, we replace $\delta_{i j}$ with a more complicated matrix $g_{i j}$, which in general can depend on the coordinates $x^i$. For an arbitrary path through the space, we express distances on the manifold in differential form,
\begin{equation}
d\ell^2 = \sum_{i, j}{g_{ij} dx^i dx^j}.
\end{equation}
The distance along any path in the spacetime, or {\it world line}, is then given by integrating $d\ell$ along that path. A familiar example of a non-Euclidean space frequently used in physics is the Minkowski Space describing spacetime in Special Relativity. Distances along a world line in Minkowski Space are measured by the {\it proper time}, which is the time as measured by an observer traveling on that world line. The proper time  $s$ along a world line is given by the relation
\begin{eqnarray}
ds^2 &&= dt^2 - d{\bf x}^2\cr
&&= \sum_{\mu,\nu = 0, 3} \eta_{\mu \nu} dx^\mu dx^\nu,
\end{eqnarray}
where we take the speed of light $c = 1$. We express four-vectors as ${\tilde x} = (t, x, y, z) = (x^0, x^1, x^2, x^3)$, and $d{\bf x}^2 = dx^2 + dy^2 + dz^2$ is the Euclidean distance along a spatial interval. The metric $\eta_{\mu \nu}$ for Minkowski Space is given by
\begin{equation}
\eta_{\mu\nu} = \left(\begin{array}{cccc}
1& & & \\
 &-1& & \\
 & &-1& \\
 & & &-1
\end{array}\right).
\end{equation}
Anything traveling the speed of light has velocity $d\left\vert{\bf x}\right\vert / dt = 1$. Photons therefore always travel along world lines of zero proper time, $ds^2 = dt^2 - d{\bf x}^2 = 0$, called {\it null geodesics}. Massive particles travel along world lines with real proper time, $ds^2 > 0$, called {\it timelike geodesics}. Causally disconnected regions of spacetime are separated by {\it spacelike} intervals, with $ds^2 < 0$. The set of all null geodesics passing through a given point (or {\it event}) in spacetime is called the {\it light cone} (Fig. \ref{fig:MinkowskiLightcones}) The interior of the light cone, consisting of all null and timelike geodesics, defines the region of spacetime causally related to that event. 
%%%%%%%%%%%%%%%%%%%%%%%%%%%%%%%%%%%%%%%%%%%%%%%%%%%%%%%%%%%%%%%%%%%%%%%%%%%%%%%%%%%%%%%%%%%%%%%%%
%%%%%%%%%%%%%%%%%%%%%%%%%%%%%%%%%%%%%%%%%%%%%%%%%%%%%%%%%%%%%%%%%%%%%%%%%%%%%%%%%%%%%%%%%%%%%%%%%

\subsection{General Relativity and the Einstein Field Equation}

The Minkowski metric $\eta_{\mu \nu}$ of Special Relativity describes a Euclidean spacetime which is static, empty, and infinite in space and time. The addition of gravity to the picture requires General Relativity, which describes gravitational fields as curvature in the spacetime. The fundamental object in General Relativity is the metric $g_{\mu \nu}\left({\tilde x}\right)$, which describes the shape of the spacetime and in general depends on the spacetime coordinate ${\tilde x}$. As in Minkowski Space, lengths in curved spacetime are measured by the proper time $s$, with the proper time along a world line determined by the metric
\begin{equation}
ds^2 = \sum_{\mu,\nu = 0, 3} g_{\mu \nu}\left({\tilde x}\right) dx^\mu dx^\nu.
\end{equation}
As in Special Relativity, photons travel along null geodesics, with $ds^2 = 0$, and massive particles travel along timelike geodesics, with $ds^2 > 0$. However, unlike Special Relativity, null geodesics need not always be $45^\circ$ lines defining light cones, but can be curved by gravity. 

In General Relativity, the distribution of mass/energy in the spacetime determines the shape of the metric, and the metric in turn determines the evolution of the mass/energy. Electromagnetism provides a convenient analogy: in electromagnetism, the distribution of charges and currents determines the electromagnetic field, and the electromagnetic field in turn determines the evolution of the charges and currents. Given a current four-vector $J^{\mu}$, Maxwell's Equations are a set of linear, first-order partial differential equations that allow us to calculate the resulting electromagnetic field
\begin{equation}
\partial_\nu F^{\mu \nu} \equiv \sum_{\nu = 0, 3} \partial_\nu F^{\mu \nu} = \frac{4 \pi}{c} J^{\mu}.
\end{equation}
Here we have explicitly included the speed of light $c$ to highlight its role as an electromagnetic coupling constant. We also adopt the typical summation convention for relativity: repeated indices are implicitly summed over. In General Relativity, we describe the distribution of mass/energy in a covariant way by specifying a symmetric rank-2 {\it stress-energy} tensor $T_{\mu \nu}$, which acts as a source for the gravitational field similar to the way the current four-vector $J^{\mu}$ sources electromagnetism. The analog of Maxwell's Equations is the Einstein Field Equation, which can be written in the deceptively simple form
\begin{equation}
\label{eq:EFE}
G_{\mu \nu} = 8 \pi G T_{\mu \nu},
\end{equation}
where the coupling constant is Newton's gravitational constant $G$. The tensor $G_{\mu\nu}$, called the Einstein Tensor, is a symmetric $4 \times 4$ tensor consisting of the metric $g_{\mu\nu}$ and its first and second derivatives. The Einstein Field Equation therefore represents a set of ten coupled, nonlinear, second-order partial differential equations of ten free functions, which are the elements of the metric tensor $g_{\mu \nu}$. However, only six of these equations are actually indpendent, leaving four degrees of freedom. The physics of gravity is independent of coordinate system, and the additional degrees of freedom correspond to a choice of a coordinate system, or {\it gauge} on the four-dimensional space. Gravity is {\it much} more complicated than electromagnetism! As with any intractably complicated problem, we simplify the job by introducing a symmetry. In General Relativity there are a number of symmetries which allow either exact or perturbative solution to the Einstein Field Equations:
\begin{itemize}
\item{Vacuum: $T_{\mu \nu} = 0$. If we evaluate the Einstein Field Equations for small perturbations about an empty Minkowski Space, we find that they reduce at lowest order to a wave equation, and therefore General Relativity predicts the existence of gravity waves.}
\item{Spherical Symmetry. If we assume a spherically symmetric spacetime (also empty of matter, $T_{\mu \nu} = 0$) the Einstein Field Equation can be solved exactly, resulting in the Schwarzschild solution for black holes.}
\item{Homogeneity and Isotropy. If we assume that the stress-energy is distributed in a fashion which is homogeneous and isotropic, this is called a {\it Friedmann-Robertson-Walker} (FRW) space, and is the case of interest for cosmology. Since the homogeneity and isotropy remove all spatial dependence, the Einstein Field Equations reduce from a set of partial differential equations to a set of nonlinear ordinary differential equations in time. For particular types of homogeneous, isotropic matter, these equations can be solved exactly, and perturbations about those exact solutions can be handled self-consistently.}
\end{itemize}
Continuing the analogy with electromagnetism, the equivalent of charge conservation,
\begin{equation}
\partial_\mu J^\mu = \frac{\partial \rho}{\partial t} + \nabla \cdot {\bf j} = 0,
\end{equation}
in General Relativity is stress-energy conservation
\begin{equation}
\label{eq:SEconservation}
D_\nu T^{\mu \nu} = 0,
\end{equation}
where $D_\mu$ represents a covariant derivative, which is a generalization of the partial derivative to a curved manifold. We will also denote covariant derivatives with a semicolon, for example $T^{\mu\nu}{}_{\!;\nu} = 0$. Likewise, simple partial derivatives are denoted with a comma, $\partial f / \partial x^{\mu} \equiv \partial_\mu f \equiv f_{,\mu}$. As in the case of electromagnetism, where the charge conservation equation is not independent, but is instead a consequence of Maxwell's Equations, stress-energy conservation in General Relativity is a consequence of the Einstein Field Equations and does not independently constrain the solutions. In the next section, we discuss FRW spaces and their application to cosmology in more detail. 

\section{Friedmann-Robertson-Walker Spacetimes}
\label{sec:FRWSpacetime}

\subsection{The Friedmann Equation}
A {\it homogeneous} space is one which is translationally invariant, or the same at every point. An {\it isotropic} space is one which is rotationally invariant, or the same in every direction. The two are not the same: a space which is everywhere isotropic is necessarily homogeneous, but a space which is homogeneous is not necessarily isotropic. (Consider, for example a space with a uniform electric field: it is translationally invariant but not rotationally invariant.) It is possible to show \cite{Weinberg:1972Ch13} that the most general metric consistent with homogeneity and isotropy is obtained by multiplying a static spatial geometry with a time-dependent {\it scale factor} $a(t)$:
\begin{eqnarray}
\label{eq:generalFRWmetric}
ds^2 &=& dt^2 - a^2\left(t\right) d {\bf x}^2\cr
&=& dt^2 - a^2\left(t\right) \left[ \frac{dr^2}{1 - k r^2} + r^2 d\Omega^2\right],
\end{eqnarray}
where we have expressed the spatial line element in terms of spherical coordinates $r$,$\theta$,$\phi$, and the solid angle is given by the usual $d\Omega^2 = \sin{\theta} d\theta d\phi$. The constant $k$ defines the curvature of the spacetime, with $k = 0$ corresponding to flat (Euclidean) spatial sections, and $k = \pm 1$ corresponding to positive and negative curvatures, respectively. A spacetime of this general form is called a {\it Friedmann-Robertson-Walker} (FRW) spacetime. Likewise, the most general homogeneous, isotropic stress-energy is diagonal, with all of its spatial components identical,
\begin{equation}
\label{eq:fluidstressenergy}
T^{\mu}{}_{\!\nu} = \left(\begin{array}{cccc}
\rho\left(t\right)& & & \\
 &-p\left(t\right)& & \\
 & &-p\left(t\right)& \\
 & & &-p\left(t\right)
\end{array}\right),
\end{equation}
where we identify the energy density $\rho$ and the pressure $p$ from the continuity equation arising from stress-energy conservation,
\begin{equation}
\label{eq:continuity}
T^{\mu\nu}{}_{\!;\nu} = \dot\rho + 3 \left(\frac{\dot a}{a}\right) \left(\rho + p\right) = 0.
\end{equation}

The Einstein field equations then reduce to a set of two coupled, non-linear ordinary differential equations,
\begin{eqnarray}
\label{eq:generalFRW}
&&\left(\frac{\dot a}{a}\right)^2 + \frac{k}{a^2} = \frac{8 \pi}{3 \mpl^2} \rho,\cr
&&\left(\frac{\ddot a}{a}\right) = - \frac{4 \pi}{3 \mpl^2} \left(\rho + 3 p\right).
\end{eqnarray}
The first is called the {\it Friedmann Equation}, and the second is called the {\it Raychaudhuri Equation}. Note that the equations for the evolution of the scale factor depend not only on the energy density $\rho$, but also the pressure $p$: pressure gravitates! 
The continuity equation (\ref{eq:continuity}) is {\it not} independent of the Einstein Field Equations (\ref{eq:generalFRW}), but can be derived directly from the Friedmann and Raychaudhuri Equations. The expansion rate $\dot a / a$ is called the {\it Hubble parameter} $H$:
\begin{equation}
H \equiv \frac{\dot a}{a},
\end{equation}
and has units of inverse time. A positive Hubble parameter $H > 0$ corresponds to an expanding universe, and a negative Hubble parameter $H < 0$ corresponds to a collapsing universe. (Since our actual universe is expanding, we will specialize to that case.) Minkowski Space can be recovered by assuming a flat geometry $k = 0$, and no expansion, $\dot a = 0$. The Hubble parameter sets the fundamental scale of the spacetime, {\it i.e.} a characteristic time is the {\it Hubble time} $t \sim H^{-1}$, and likewise the {\it Hubble length} is $d \sim H^{-1}$. We will see later that the Hubble time sets the scale for the age of the universe, and the Hubble length sets the scale for the size of the observable universe. 

The coordinate system $\left(t,{\bf x}\right)$ is called a {\it comoving} coordinate system, because observers with constant comoving coordinates are at rest relative to the expansion, {\it i.e.} two observers with constant separation in comoving coordinates $\Delta{\bf x}$ have a physical, or {\it proper}, separation which increases in proportion to the scale factor
\begin{equation}
\Delta{\bf x}_{\rm prop} = a\left(t\right) \Delta{\bf x}_{\rm com}.
\end{equation}
An important kinematic effect of cosmological expansion is the phenomenon of {\it cosmological redshift}: we will see later that solutions to the wave equation in an FRW space have constant wavelength in {\it comoving} coordinates, so that the proper wavelength of (for example) a photon increases in time in proportion to the scale factor
\begin{equation}
\lambda \propto a\left(t\right).
\end{equation}
For a photon emitted at time $t_{\rm em}$ and detected at time $t_0$, the redshift $z$ is defined by:
\begin{equation}
\left(1 + z\right) \equiv \frac{\lambda_0}{\lambda_{\rm em}} = \frac{a\left(t_0\right)}{a\left(t_{\rm em}\right)}.
\end{equation}
(Here we introduce the convention used frequently in cosmology that a subscript $0$ refers to the {\it current} time, not an initial time.) Note that the cosmological redshift is {\it not} a Doppler shift caused by the relative velocity of the source and detector, but is an expansion effect: the wavelength of a photon traveling through the spacetime increases because the underlying spacetime is expanding. Another way to look at this is that a photon traveling through an FRW spacetime loses momentum with time,
\begin{equation}
\label{eq:freqredshift}
p = h \nu \propto a^{-1}(t).
\end{equation} 
By the equivalence principle, this momentum loss must apply to massive particles as well as photons: {\it any} particle moving in an expanding FRW spacetime will lose momentum as $p \propto a^{-1}$. For massless particles like photons, this is manifest as a redshift in the wavelength, but it means that a massive particle will asymptotically come to rest relative to the comoving coordinate system. Thus, comoving coordinates represent a preferred reference frame reminiscent of Aristotelian physics: any free body with a ``peculiar'' velocity relative to the comoving frame will eventually come to rest in that frame. 

There are three possibilities for the curvature of the universe: flat ($k = 0$), positively curved ($k = +1$), or negatively curved ($k = -1$). The current value of the Hubble parameter is (from the Hubble Space Telescope Key Project \cite{Freedman:2000cf}),
\begin{equation}
H_0 = 72 \pm 8\ {\rm km/s/Mpc}.
\end{equation}
Therefore, we can see from the Friedmann Equation (\ref{eq:generalFRW}) that, given the expansion rate $H$, the curvature is determined by the density:
\begin{equation}
k = a^2 \left(\frac{8 \pi}{3 \mpl^2} \rho - H^2\right).
\end{equation}
Note that only the {\it sign} of $k$ is physically important, since any rescaling of $k$ is equivalent to a rescaling of the scale factor $a$. We define the {\it critical density} as the density for which $k = 0$, corresponding to a geometrically flat universe,
\begin{equation}
\rho_c \equiv \frac{3 \mpl^2}{8 \pi} H^2\ \Rightarrow\ k = 0.
\end{equation}
For $\rho > \rho_c$, the universe is positively curved and {\it closed}, with finite volume, and for $\rho < \rho_c$, the universe is negatively curved and {\it open}, with infinite volume. We express the ratio of the actual density $\rho$ to the critical density $\rho_c$ as the parameter $\Omega$:
\begin{equation}
\label{eq:defOmega}
\Omega \equiv \left(\frac{\rho}{\rho_c}\right) = \frac{8 \pi}{3 \mpl^2} \frac{\rho}{H^2}. 
\end{equation}
(Do not confuse the density parameter $\Omega$ with the solid angle $d\Omega$ in Eq. \ref{eq:generalFRWmetric}!) Table \ref{tab:omega} summarizes the relation between density, curvature, and geometry. 
\begin{table}[htbp]
\centering
\caption{Cosmological density and curvature}\label{tab:omega}
\begin{tabular}{|c|c|c|} \hline
density & curvature & geometry   \\ \hline \hline
$\Omega = 1$ & $k = 0$ & flat \\ \hline
$\Omega > 1$ & $k = 1$ & closed  \\ \hline
$\Omega < 1$ & $k = 0$ & open \\ \hline
\end{tabular}
\end{table}
The density parameter $\Omega$ is not in general constant in time, and we can re-write the Friedmann Equation as
\begin{equation}
\label{eq:FRWOmega}
\Omega(t) = 1 + \frac{k}{\left(a H\right)^2}.
\end{equation}
Since the Hubble parameter is proportional to the inverse time $H \propto t^{-1}$, we see that the time-dependence of $\Omega$ is determined by the time dependence of the scale factor $a\left(t\right)$. In the next section, we tackle the problem of solving for $a\left(t\right)$. 

\subsection{Solving the Friedmann Equation}
\label{sec:SolvingTheFriedmannEquation}

In the previous section, we considered the form and kinematics of FRW spaces, but not the {\it dynamics}, that is, how does the stress-energy of the universe determine the expansion history? The answer to this question depends on what kind of matter dominates the cosmological stress-energy. In this section, we consider three basic types of cosmological stress-energy: matter, radiation, and vacuum. 

\begin{figure}
\begin{center}
%--%\psfig{file=matterbox.eps,width=4.5in}
\includegraphics[width=4.5in]{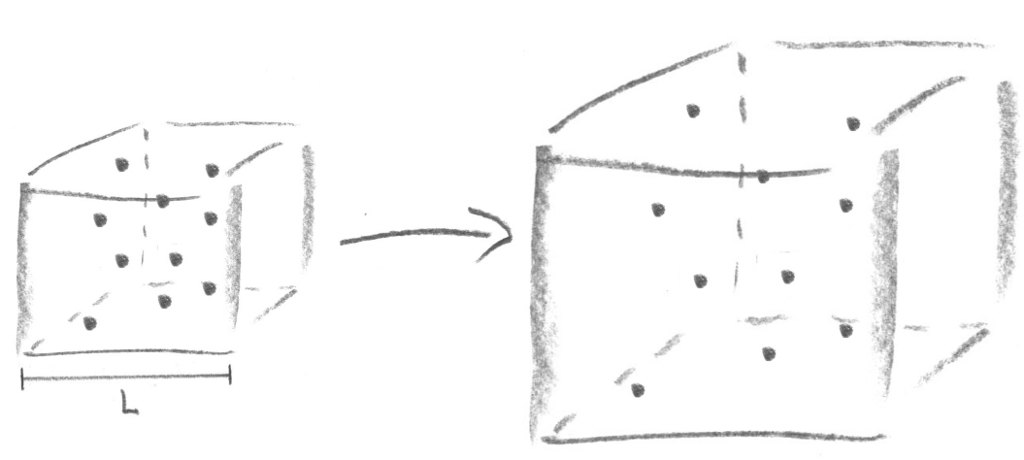}
\end{center}
\caption{A comoving box full of matter. The energy density in matter scales inversely with the volume of the box.}
\label{fig:boxofmatter}
\end{figure}

The simplest kind of cosmological stress-energy is generically referred to as {\it matter}. Imagine a comoving box with sides of length $L$. By {\it comoving} box, we mean a box whose corners are at rest in a comoving coordinate system, and whose proper dimension is therefore increasing proportional to the scale factor, $L_{\rm prop} \propto a$. That is, the box is growing with the expansion of the universe. Now imagine the box filled with $N$ particles of mass $m$, also at rest in the comoving reference frame (Fig. \ref{fig:boxofmatter}). 
%%%%%%%%%%%%%%%%%%%%%%%%%%%%%%%%%%%%%%%%%%%%%%%%%%%%%%%%%%%%%%%%%%%%%%%%%%%%%%%%%%%%%%%%%%%%%%%%%
%%%%%%%%%%%%%%%%%%%%%%%%%%%%%%%%%%%%%%%%%%%%%%%%%%%%%%%%%%%%%%%%%%%%%%%%%%%%%%%%%%%%%%%%%%%%%%%%%
%%%%%%%%%%%%%%%%%%%%%%%%%%%%%%%%%%%%%%%%%%%%%%%%%%%%%%%%%%%%%%%%%%%%%%%%%%%%%%%%%%%%%%%%%%%%%%%%%
\begin{figure}
\begin{center}
%--%\psfig{file=radbox.eps,width=4.5in}
\includegraphics[width=4.5in]{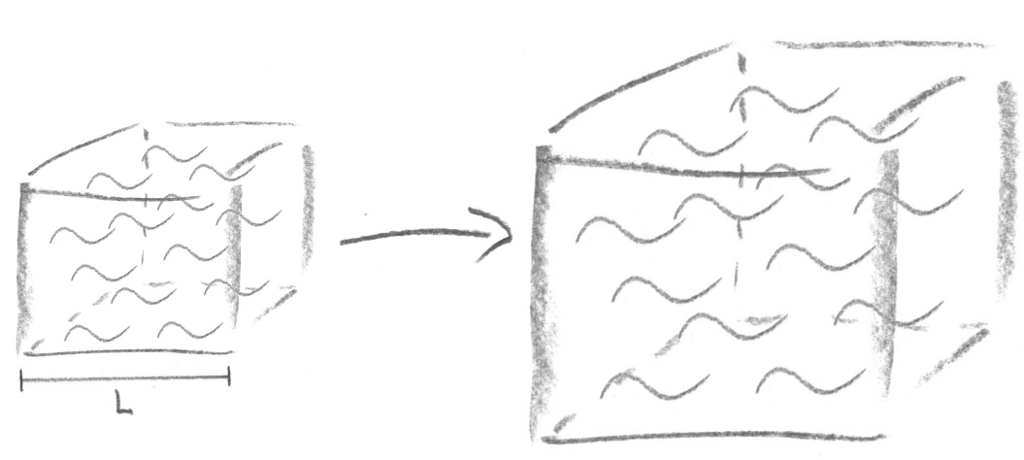}
\end{center}
\caption{A comoving box full of radiation. The number density of photons scales inversely with the volume of the box, but the photons also increase in wavelength.}
\label{fig:boxofradiation}
\end{figure}
%%%%%%%%%%%%%%%%%%%%%%%%%%%%%%%%%%%%%%%%%%%%%%%%%%%%%%%%%%%%%%%%%%%%%%%%%%%%%%%%%%%%%%%%%%%%%%%%%
%%%%%%%%%%%%%%%%%%%%%%%%%%%%%%%%%%%%%%%%%%%%%%%%%%%%%%%%%%%%%%%%%%%%%%%%%%%%%%%%%%%%%%%%%%%%%%%%%
\begin{figure}
\begin{center}
%--%\psfig{file=vacbox.eps,width=4.5in}
\includegraphics[width=4.5in]{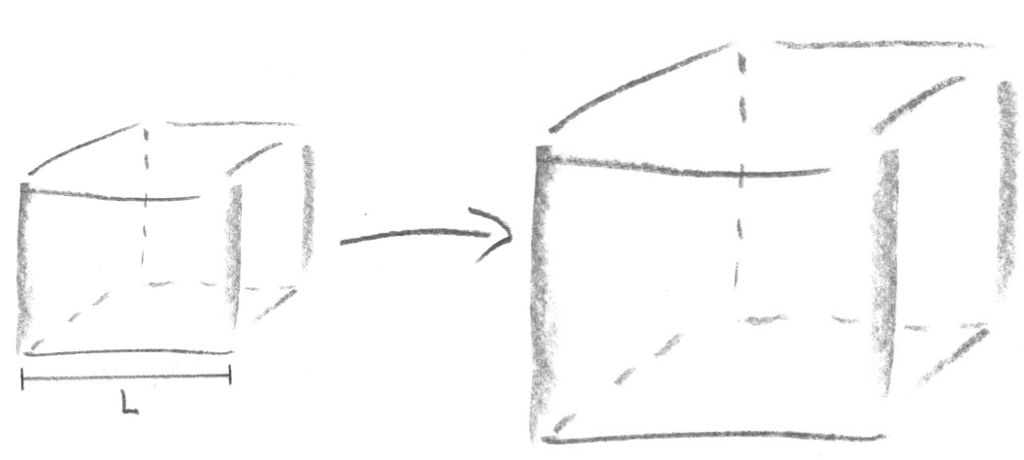}
\end{center}
\caption{A comoving box full of vacuum. The energy density of vacuum does not scale at all!}
\label{fig:boxofvac}
\end{figure}
%%%%%%%%%%%%%%%%%%%%%%%%%%%%%%%%%%%%%%%%%%%%%%%%%%%%%%%%%%%%%%%%%%%%%%%%%%%%%%%%%%%%%%%%%%%%%%%%%
In units where $c = 1$, the relativistic energy density of such a system of particles is given by
\begin{equation}
\rho_{\rm m} = \frac{M N}{V},
\end{equation}
where $V$ is the {\it proper} volume of the box, $V = L_{\rm prop}^3 \propto a^3$. Since neither $M$ nor $N$ change with expansion, we have immediately that
\begin{equation}
\rho_{\rm m} = \frac{M N}{L^3 a^3} \propto a^{-3},
\end{equation}
where $L$ is the comoving size of the box. So the proper energy density of massive particles at rest in a comoving volume evolves as the inverse cube of the scale factor. Now imagine the same box filled with $N$ photons with frequency $\nu$ (Fig. \ref{fig:boxofradiation}). The energy per photon is $h \nu$, so that the energy density in the box is then 
\begin{equation}
\rho_{\gamma} = \frac{N h \nu}{V}.
\end{equation}

As in the case of massive particles, the number density of photons in the box redshifts inversely with the proper volume of the box $n = N / V \propto a^{-3}$. But each photon also loses energy through cosmological redshift, $\nu \propto a^{-1}$ (\ref{eq:freqredshift}), so that the total energy density in photons or other massless degrees of freedom, which we generically refer to as {\it radiation}, redshifts as
\begin{equation}
\rho_{\gamma} \propto a^{-4}.
\end{equation}
Note also that cosmological redshift immediately gives us a rule for the behavior of a black-body spectrum of radiation with temperature $T$. Since all photons redshift at exactly the same rate, a system with starts out as a black-body {\it stays} a black-body, with a temperature that decreases with expansion,
\begin{equation}
T_{\gamma} \propto a^{-1}.
\end{equation}

The third type of stress-energy which is important in cosmology dates back to Einstein's introduction of a ``cosmological constant'' to his field equations. If we take the stress-energy $T_{\mu \nu}$ and add a term proportional to the metric, the identity $D_\nu g^{\mu\nu} = 0$ means the stress-energy conservation equation (\ref{eq:SEconservation}) is unchanged:
\begin{equation}
D_{\nu} T^{\mu \nu} \rightarrow D_{\nu} \left(T^{\mu \nu} + \Lambda g^{\mu \nu}\right) = 0.
\end{equation}
In our analogy with electromagnetism, this is like adding a constant to the 
electromagnetic potential, $V'(x) = V(x) + \Lambda$. The constant $\Lambda$ does not
affect local dynamics in any way, but it does affect the cosmology. From Eq. (\ref{eq:fluidstressenergy}), stress-energy of the form $T^{\mu\nu} = \Lambda g^{\mu\nu}$ corresponds to an equation of state
\begin{equation}
p_{\Lambda} = -\rho_{\Lambda}.
\end{equation}
The continuity equation (\ref{eq:continuity}) then reduces to
\begin{equation}
\dot\rho + 3 \left(\frac{\dot a}{a}\right) \left(\rho + p\right) = \dot\rho = 0,
\end{equation}
so that vacuum has a constant energy density, $\rho_\Lambda = {\rm const.}$ A cosmological constant is also frequently referred to as {\it vacuum energy}, since it is as if we are assigning an energy density to empty space. With this interpretation, a comoving box full of vacuum contains a total amount of energy which {\it grows} with the expansion of the universe (Fig. \ref{fig:boxofvac}). This highlights the curious property of General Relativity that, while energy is conserved in a local sense, it is {\it not} conserved globally. We are creating energy  out of nothing!

It is straightforward to solve the Einstein Field Equations for the three basic types of stress-energy. Consider first a matter-dominated universe. We can write the time derivative of the energy density as:
\begin{equation}
\rho_{\rm m} \propto a^{-3}\ \Rightarrow \dot\rho_{\rm m} = - 3 \left(\frac{\dot a}{a}\right) \rho.
\end{equation}
From the continuity equation (\ref{eq:continuity}), we have
\begin{equation}
\dot\rho + 3 \left(\frac{\dot a}{a}\right) \left(\rho + p\right) = 3 \left(\frac{\dot a}{a}\right) p = 0.
\end{equation}
We then have that the pressure of matter vanishes, $p_{\rm m} = 0$. The matter-dominated Friedmann Equation becomes
\begin{equation}
\left(\frac{\dot a}{a}\right)^2 + \frac{k}{a^2} = \frac{8 \pi}{3 \mpl^2} \rho \propto a^{-3}. 
\end{equation}
In the case of a flat universe, $k = 0$, the solution is especially simple:
\begin{equation}
\left(\frac{\dot a}{a}\right)^2 \propto a^{-3}\ \Rightarrow a\left(t\right) \propto t^{2/3}. 
\end{equation}
Similarly, for a radiation dominated universe, the continuity equation implies that
\begin{equation}
\rho_{\gamma} \propto a^{-4}\ \Rightarrow p_{\gamma} = \rho_{\gamma} / 3.
\end{equation}
Again assuming a flat geometry,
\begin{equation}
\left(\frac{\dot a}{a}\right)^2 \propto a^{-4}\ \Rightarrow a\left(t\right) \propto t^{1/2}. 
\end{equation}
Finally, solving the the Friedmann Equation for the vacuum case gives
\begin{equation}
\left(\frac{\dot a}{a}\right)^2 \propto \rho_{\Lambda} = {\rm const.}\ \Rightarrow a\left(t\right) \propto e^{H t},
\end{equation}
so that the universe expands exponentially quickly, with a time constant given by the Hubble parameter
\begin{equation}
H = \sqrt{\frac{8 \pi}{3 \mpl}^2 \rho_{\Lambda}} = {\rm const.}
\end{equation} 
Such a spacetime is called {\it de Sitter space}. 

\begin{figure}
\begin{center}
%--%\psfig{file=rhovsz.eps,width=4.5in}
\includegraphics[width=4.5in]{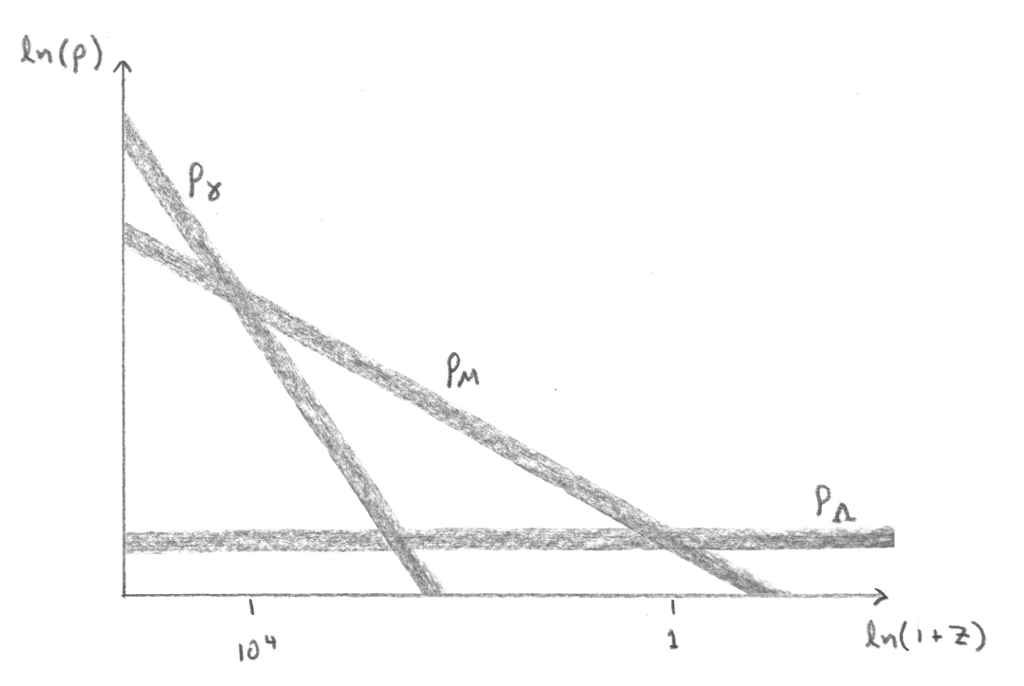}
\end{center}
\caption{Schematic diagram of how the three types of stress-energy scale with redshift $1 + z \propto a$: at early time, radiation dominates, followed by matter, and finally the universe is dominated by vacuum energy.}
\label{fig:rhovsz}
\end{figure}

Note in particular that the energy density in radiation redshifts away more quickly than the energy density in matter, and vacuum energy does not redshift at all, so that a universe with a mix of radiation, matter and vacuum will be radiation-dominated at early times, matter-dominated at later times, and eventually vacuum-dominated (Fig. \ref{fig:rhovsz}). 
%%%%%%%%%%%%%%%%%%%%%%%%%%%%%%%%%%%%%%%%%%%%%%%%%%%%%%%%%%%%%%%%%%%%%%%%%%%%%%%%%%%%%%%%%%%%%%%%%
%%%%%%%%%%%%%%%%%%%%%%%%%%%%%%%%%%%%%%%%%%%%%%%%%%%%%%%%%%%%%%%%%%%%%%%%%%%%%%%%%%%%%%%%%%%%%%%%%
Note also that for either matter- or radiation-domination, the universe is singular as $t \rightarrow 0$: the universe has finite age! Since the scale factor vanishes at $t = 0$, and the density scales as an inverse power of $a$, the initial singularity consists of infinite density. Likewise, since temperature also scales inversely with $a$, the initial singularity is also a point of infinite temperature. We therefore arrive at the standard hot Big Bang picture of the universe: a cosmological singularity at finite time in the past, followed by a hot, radiation-dominated expansion, during which the universe gradually cools as $T \propto a^{-1}$ and the radiation dilutes, followed by a period of matter-dominated expansion during which galaxies and stars and planets form. Finally, if the vacuum energy is nonzero, it will inevitably dominate, and the universe will enter a state of exponential expansion. Current evidence indicates that the real universe made a transition from matter-domination to vacuum-domination at a redshift of around $z = 1$, or about a billion years ago, so that the densities of the three types of matter today are of order
\begin{eqnarray}
\Omega_\Lambda &&\simeq 0.7,\cr
\Omega_{\rm m} &&\simeq 0.3,\cr
\Omega_{\gamma} &&\simeq 10^{-4}.
\end{eqnarray}
In the next section, we discuss one important prediction of the hot Big Bang: the presence of a background of relic photons from the early universe, called the {\it Cosmic Microwave Background}. 

\subsection{The Hot Big Bang and the Cosmic Microwave Background}
\label{sec:CMB}

The basic picture of an expanding, cooling universe leads to a number of 
startling predictions: the formation of nuclei and the resulting primordial 
abundances of elements, and the later formation of neutral atoms and the 
consequent presence of a cosmic background of photons, the Cosmic Microwave 
Background (CMB) \cite{White:1994sx,Hu:2001bc,Kosowsky:2001ue,Samtleben:2007zz,Hu:2008hd}. A rough history of the universe can be 
given as a time line of
increasing time and decreasing temperature \cite{Kolb:1990Ch3}:
\begin{itemize}
\item{$T = \infty$, $t = 0$: Big Bang.}
\item{$T \sim 10^{15}\ K$, $t \sim 10^{-12}\ {\rm sec}$: Primordial soup of
fundamental particles.}
\item{$T \sim 10^{13}\ K$, $t \sim 10^{-6}\ {\rm sec}$: Protons and neutrons form.}
\item{$T \sim 10^{10}\ K$, $t \sim 3\ {\rm min}$: Nucleosynthesis: nuclei form.}
\item{$T \sim 3000\ K$, $t \sim 300,000\ {\rm years}$: Atoms form.}
\item{$T \sim 10\ K$, $t \sim 10^{9}\ {\rm years}$: Galaxies form.}
\item{$T \sim 3\ K$, $t \sim 10^{10}\ {\rm years}$: Today.}
\end{itemize}
The epoch at which atoms form, when the universe was at an age of 300,000 years 
and at a temperature of around $3000\ {\rm K}$  is oxymoronically referred to as 
``recombination'', despite the fact that electrons and nuclei had never before 
``combined'' into atoms. The physics is simple: at a temperature of greater 
than about $3000\ {\rm K}$, the universe consisted of an ionized plasma of mostly 
protons, electrons, and photons, with a few helium nuclei and a tiny trace of 
lithium. The important characteristic of this plasma is that it was {\it 
opaque}, or, more precisely, the mean free path of a photon was a great deal 
smaller than the Hubble length. As the universe cooled and 
expanded, the plasma ``recombined'' into neutral atoms, first the helium, then a 
little later the hydrogen. 
%%%%%%%%%%%%%%%%%%%%%%%%%%%%%%%%%%%%%%%%%%%%%%%%%%%%%%%%%%%%%%%%%%%%%%%%%%%%%%%%%%%%%%%%%%%%%%%%%
\begin{figure}
\begin{center}
%--%\psfig{file=recombination.eps,width=4.5in}
\includegraphics[width=4.5in]{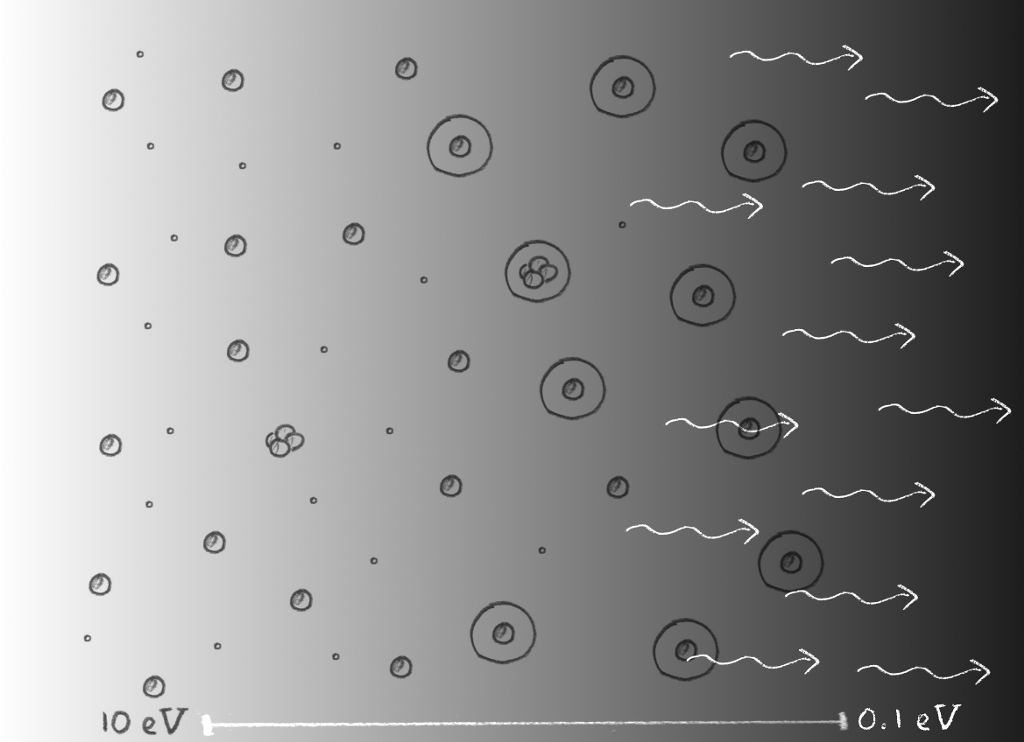}
\end{center}
\caption{Schematic diagram of recombination. At early time, the temperature of the universe is above the ionization energy of hydrogen and helium, so that the universe is full of an ionized plasma, and the mean free path for photons is short compared to the Hubble length. At late time, the temperature drops and the nuclei capture the electrons and form neutral atoms. Once this happens, the universe becomes transparent to photons, which free stream from the surface of last scattering. }
\label{fig:recombination}
\end{figure}
%%%%%%%%%%%%%%%%%%%%%%%%%%%%%%%%%%%%%%%%%%%%%%%%%%%%%%%%%%%%%%%%%%%%%%%%%%%%%%%%%%%%%%%%%%%%%%%%%

If we consider hydrogen alone, the process of recombination can be            
described by the {Saha equation} for the  equilibrium ionization fraction 
$X_{\rm e}$ of the hydrogen \cite{Kolb:1990Ch3}:
\begin{equation}
\frac{1 - X_{\rm e}}{X_{\rm e}^2} = \frac{4 \sqrt{2} \zeta(3)}{\sqrt{\pi}} \eta 
\left(\frac{T}{m_{\rm e}}\right)^{3/2} \exp\left(\frac{13.6\ {\rm eV}}{
T}\right).\label{eqsahaequation}
\end{equation}
Here $m_{\rm e}$ is the electron mass and $13.6\ {\rm eV}$ is the ionization 
energy of hydrogen. The physically important parameter affecting recombination 
is the density of protons and electrons compared to photons. This is determined
by the {\it baryon asymmetry},\footnote{If there were no excess of baryons over antibaryons, there would be no 
protons and
electrons to recombine, and the universe would be just a gas of photons and neutrinos!} 
which is described as the ratio of baryons to photons:
\begin{equation}
\eta \equiv \frac{n_{\rm b} - n_{\rm \bar b} }{ n_\gamma} = 2.68 \times 10^{-8} 
\left(\Omega_{\rm b} h^2\right).
\end{equation}
Here $\Omega_{\rm b}$ is the baryon density and $h$ is the Hubble constant in units
of $100\ {\rm km/s/Mpc}$,
\begin{equation}
h \equiv H_0 / (100\ {\rm km /s/Mpc}).
\end{equation}
The most recent result from the WMAP satellite gives $\Omega_{\rm b} h^2 = 0.02273 \pm 0.00062$ \cite{Dunkley:2008ie}. 
Recombination happens quickly (i.e., in much less than a Hubble time $t \sim H^{-1}$), but it is 
not instantaneous. The universe goes from a completely ionized state to
a neutral state over a range of redshifts $\Delta z \sim 200$. 
If we define recombination as an ionization fraction $X_{\rm e} = 0.1$, we have
that the temperature at recombination $T_{\rm R} = 0.3\ {\rm eV}$.

\begin{figure}
\begin{center}
%--%\psfig{file=lastscattering.eps,width=3.5in}
\includegraphics[width=3.5in]{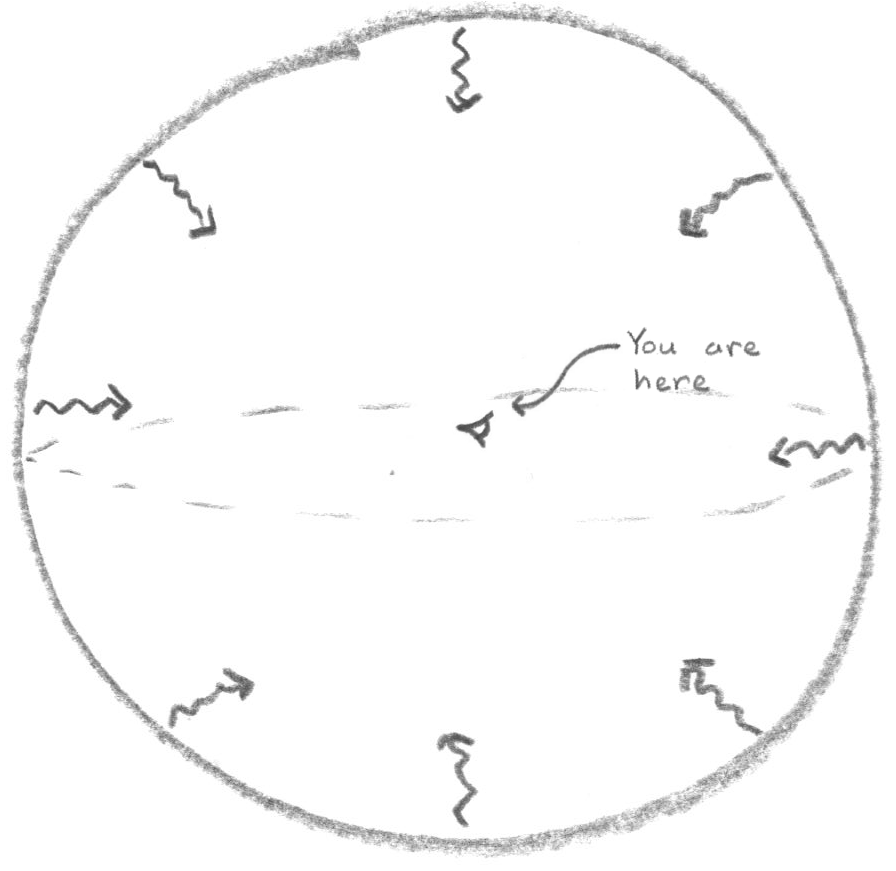}
\end{center}
\caption{Cartoon of the last scattering surface. From earth, we see blackbody radiation emitted
uniformly from all directions, forming a ``sphere'' at 
redshift $z = 1100$.}
\label{fig:lss}
\end{figure}

What happens to the photons after recombination? Once the gas in the universe
is in a neutral state, the mean free path for a photon becomes much larger than
the Hubble length. The universe is then full of a background of freely propagating
photons with a blackbody distribution of frequencies. At the time of recombination,
the background radiation has a temperature of $T = T_{\rm R} = 3000\ {\rm K}$, and as the
universe expands the photons redshift, so that the temperature of the photons drops
with the increase of the scale factor, $T \propto a(t)^{-1}$. We can detect these
photons today. Looking at the sky, this background of photons comes to us evenly from
all directions, with an observed temperature of $T_0 \simeq 2.73\ {\rm K}$. 
This allows us to determine the redshift of recombination,
\begin{equation}
1 + z_{\rm R} = \frac{a\left(t_0\right) }{ a\left(t_{\rm R}\right)} = \frac{T_{\rm R} }{ T_0} \simeq 1100.
\end{equation}
This is the cosmic microwave
background. Since by looking at higher and higher redshift objects, we are looking
further and further back in time, we can view the observation of CMB photons as imaging
a uniform ``surface of last scattering'' at a redshift of 1100 (Fig. \ref{fig:lss}).
%%%%%%%%%%%%%%%%%%%%%%%%%%%%%%%%%%%%%%%%%%%%%%%%%%%%%%%%%%%%%%%%%%%%%%%%%%%%%%%%%%%%%%%%%%%%%%%%%
%%%%%%%%%%%%%%%%%%%%%%%%%%%%%%%%%%%%%%%%%%%%%%%%%%%%%%%%%%%%%%%%%%%%%%%%%%%%%%%%%%%%%%%%%%%%%%%%%

\begin{figure}
\begin{center}
%--%\psfig{file=COBE.eps,width=4.5in}
\includegraphics[width=4.5in]{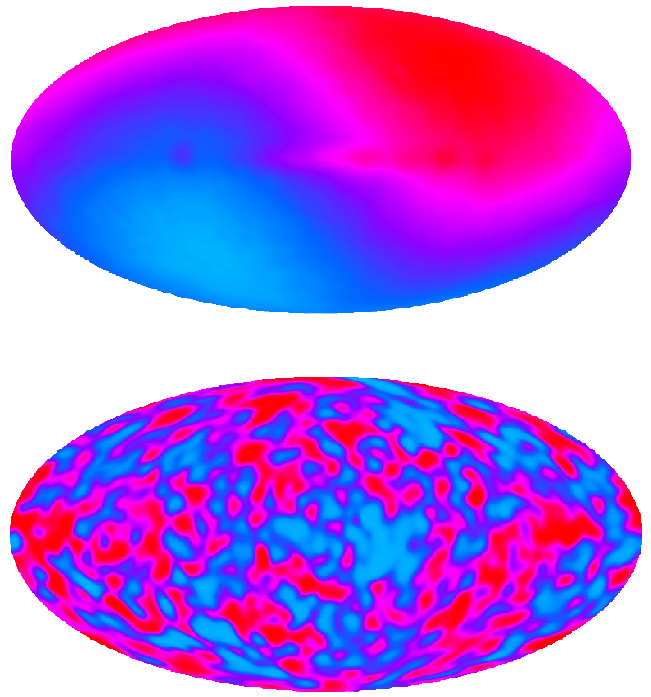}
\end{center}
\caption{The COBE measurement of the CMB anisotropy \cite{Bennett:1996ce}. The top oval is a 
map of the sky showing the dipole anisotropy $\Delta T / T \sim 10^{-3}$. The bottom oval is a similar
map with the dipole contribution and emission from our own galaxy subtracted, showing the anisotropy for $\ell > 1$,
$\Delta T / T \sim 10^{-5}$. (Figure courtesy of the COBE Science Working Group.)}
\label{fig:COBE}
\end{figure}

To the extent that recombination happens at the same time and in the same way everywhere,
the CMB will be of precisely uniform temperature.  While the observed CMB is highly isotropic, it is not perfectly so. The largest 
contribution to the anisotropy of the CMB as seen from earth is simply Doppler shift
due to the earth's motion through space. (Put more technically, the motion is the
earth's motion relative to a comoving cosmological reference frame.) CMB photons 
are slightly blueshifted in the direction of our motion and slightly redshifted 
opposite the direction of our motion. This blueshift/redshift shifts the temperature
of the CMB so the effect has the characteristic form of a ``dipole'' temperature
anisotropy (Fig. \ref{fig:COBE}).
The dipole anisotropy, however, is a {\em local} phenomenon. Any intrinsic, or primordial,
anisotropy of the CMB is potentially of much greater cosmological interest. To describe 
the anisotropy of the CMB, we remember that the surface of last scattering appears to us
as a spherical surface at a redshift of $1100$. Therefore the natural parameters to
use to describe the anisotropy of the CMB sky is as an expansion in spherical harmonics 
$Y_{\ell m}$:
\begin{equation}
\frac{\Delta T }{ T} = \sum_{\ell = 1}^{\infty} \sum_{m = -\ell}^{\ell}{a_{\ell m} 
Y_{\ell m}\left(\theta,\phi\right)}.
\end{equation}
If we assume isotropy, there is no preferred direction in the universe, and we expect the physics to be independent of the index $m$. We can then define
\begin{equation}
C_{\ell} \equiv \frac{1}{2 \ell + 1} \sum_{m}{\left\vert a_{\ell m} \right\vert^2}.
\end{equation}
The $\ell = 1$ contribution is just the dipole anisotropy, 
\begin{equation}
\left(\frac{\Delta T }{ T}\right)_{\ell = 1} \sim 10^{-3}.
\end{equation}

The dipole was first measured in the 1970's by several groups \cite{Henry:1971,Corey:1976,Smoot:1977}. 
It was not until more than a decade after the discovery of the dipole anisotropy
that the first observation was made of anisotropy for $\ell \geq 2$, by the 
differential microwave radiometer aboard the Cosmic Background Explorer (COBE) 
satellite \cite{Bennett:1996ce}, launched in in 1990. COBE observed that the anisotropy at the quadrupole
and higher $\ell$ was two orders of magnitude smaller than the dipole:
\begin{equation}
\left(\frac{\Delta T }{ T}\right)_{\ell > 1} \simeq 10^{-5}.
\end{equation}
Fig. \ref{fig:COBE} shows the dipole and higher-order CMB anisotropy as measured by
COBE. 
%%%%%%%%%%%%%%%%%%%%%%%%%%%%%%%%%%%%%%%%%%%%%%%%%%%%%%%%%%%%%%%%%%%%%%%%%%%%%%%%%%%%%%%%%%%%%%%%%
%%%%%%%%%%%%%%%%%%%%%%%%%%%%%%%%%%%%%%%%%%%%%%%%%%%%%%%%%%%%%%%%%%%%%%%%%%%%%%%%%%%%%%%%%%%%%%%%%
%%%%%%%%%%%%%%%%%%%%%%%%%%%%%%%%%%%%%%%%%%%%%%%%%%%%%%%%%%%%%%%%%%%%%%%%%%%%%%%%%%%%%%%%%%%%%%%%%
\begin{figure}
\begin{center}
%--%\psfig{file=WMAP.eps,width=4.5in}
\includegraphics[width=4.5in]{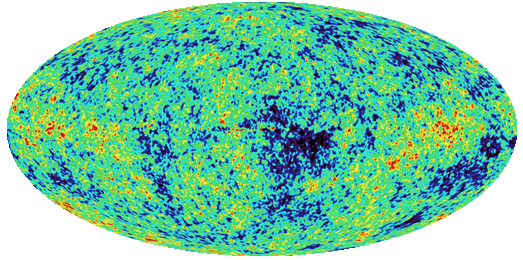}
\end{center}
\caption{The WMAP measurement of the CMB anisotropy \cite{Hinshaw:2008kr}. (Figure courtesy of the WMAP Science Working Group.) WMAP measured the anisotropy with much higher sensitivity and resolution than COBE.}
\label{fig:WMAP}
\end{figure}
%%%%%%%%%%%%%%%%%%%%%%%%%%%%%%%%%%%%%%%%%%%%%%%%%%%%%%%%%%%%%%%%%%%%%%%%%%%%%%%%%%%%%%%%%%%%%%%%%
This anisotropy represents intrinsic fluctuations in the CMB
itself, due to the presence of tiny primordial density fluctuations in the cosmological
matter present at the time of recombination. These density fluctuations are of great
physical interest, since these are the fluctuations which later collapsed
to form all of the structure in the universe, from superclusters to planets to
graduate students.  While the physics of recombination in the homogeneous case is quite simple, the presence of inhomogeneities in the universe makes the situation much more complicated. 
I describe some of the major effects in a qualitative way here, and refer the reader to the literature for a more detailed technical explanation of the relevant physics \cite{White:1994sx,Hu:2001bc,Kosowsky:2001ue,Samtleben:2007zz,Hu:2008hd}. In these lectures, I primarily focus on the current status of the CMB as a probe of inflation, but there is much more to the story.

\begin{figure}
\begin{center}
%--%\psfig{file=WMAPCl.eps,width=4.5in}
\includegraphics[width=4.5in]{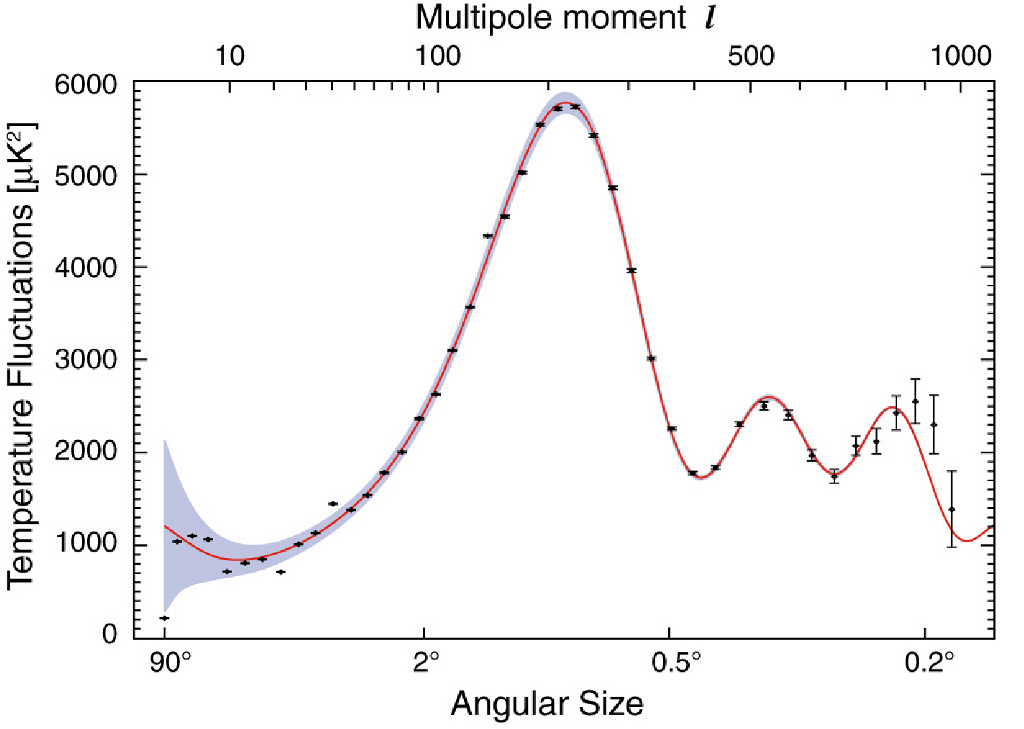}
\end{center}
\caption{The $C_\ell$ spectrum for the CMB as measured by WMAP, showing the peaks characteristic of acoustic oscillations. The gray shaded region represents the uncertainty due to cosmic variance. (Figure courtesy of the WMAP Science Working Group.)}
\label{fig:Cl}
\end{figure}

The simplest contribution to the CMB anisotropy from density fluctuations is just
a gravitational redshift, known as the {\em Sachs-Wolfe effect} \cite{Sachs:1967}. 
A photon coming from a region which is slightly denser than the average will have a slightly larger
redshift due to the deeper gravitational well at the surface of last scattering. 
Conversely, a photon coming from an underdense region will have a slightly smaller
redshift. Thus we can calculate the CMB temperature anisotropy due to the slightly
varying Newtonian potential $\Phi$ from density fluctuations at the surface
of last scattering:
\begin{equation}
\label{eq:SachsWolfe}
\frac{\delta T}{T} = \frac{1}{3} \left[\Phi_{\rm em} - \Phi_{\rm obs}\right],
\end{equation}
where $\Phi_{\rm em}$ is the potential at the point the photon was emitted on the surface of last scattering, and $\Phi_{\rm obs}$ is the potential at the point of observation, which can be treated as a constant. The factor $1/3$ is a General Relativistic correction. This simple kinematic contribution to the CMB 
anisotropy is dominant on large angular scales, corresponding to multipoles $\ell < 100$.  However, the amount of information we can gain from these multipoles is limited by an intrinsic source of error called {\it cosmic variance}. Cosmic variance is a result of the statistical nature of the primordial power spectra: since we have only one universe to measure, we have only one realization of the random field of density perturbations, and therefore there is an inescapable $1 / \sqrt{N}$ uncertainty in our ability to reconstruct the primordial power spectrum, where $N$ is the number of independent wave modes which will fit inside the horizon of the universe! On very large angular scales, this problem becomes acute, and we can write the cosmic variance error on any given $C_\ell$ as
\begin{equation}
\frac{\Delta C_\ell}{C_\ell} = \frac{1}{\sqrt{2 \ell + 1}},
\end{equation}
which comes from the fact that any $C_\ell$ is represented by $2 \ell + 1$ independent amplitudes $a_{\ell m}$. Even a perfect observation of the CMB can only approximately measure the true power spectrum --- the errors in the WMAP data, for example, are dominated by cosmic variance out to $\ell \sim 400$ (Fig. \ref{fig:Cl}).

For fluctuation modes on smaller angular scales, more complicated physics comes into
play. The dominant process that occurs on short wavelengths is {\em acoustic oscillations} in the baryon/photon plasma. The idea is simple: matter tends to collapse due to gravity  onto regions where the density is higher than average, so the baryons ``fall'' into overdense regions. However, since the baryons and the photons are still strongly coupled, the photons tend to resist this collapse and push the baryons outward. The result is ``ringing'', or oscillatory modes of compression and rarefaction in the gas due to density fluctuations. The gas heats as it compresses and cools as it expands, which creates fluctuations in the temperature of the CMB. This manifests itself in the $C_\ell$ spectrum as a series of peaks and valleys (Fig. \ref{fig:Cl}). The specific shape and location of the acoustic peaks is created by complicated but well-understood physics, involving a large number of cosmological parameters. The presence of acoustic peaks in the CMB was first suggested by Sakharov \cite{Sakharov:1965}, and later calculated by Sunyaev and Zel'dovich \cite{Zeldovich:1969ff,Sunyaev:1970eu} and Peebles and Yu \cite{Peebles:1970ag}. The complete linear theory of CMB fluctuations was worked out by Ma and Bertschinger in 1995 \cite{Ma:1995ey}. The shape of the CMB multipole spectrum depends, for example, on the baryon density $\Omega_{\rm b}$, the Hubble constant $H_0$, the densities of matter $\Omega_{\rm m}$ and cosmological constant $\Omega_{\Lambda}$, the amplitude of primordial gravitational waves, and the redshift $z_{\rm ri}$ at which the first generation of stars ionized the intergalactic medium. This makes interpretation of the spectrum something of a complex undertaking, but it also makes it a sensitive probe of cosmological models. 

In addition to anisotropy in the temperature of the CMB, the photons coming from the surface of last scattering are expected to be weakly polarized due to the presence of perturbations \cite{Kosowsky:1998mb,Zaldarriaga:2003bb}. This polarization is much less well measured than the temperature anisotropy, but it has been detected by WMAP and by a number of ground- and balloon-based measurements \cite{Leitch:2004gd,Sievers:2005gj,Montroy:2005yx,Wu:2006ji,Nolta:2008ih}. Measurement of polarization promises to greatly increase the amount of information it is possible to extract from the CMB.  Of particular interest is the odd-parity, or {\it B-mode} component of the polarization, the only primordial source of which is gravitational waves, and thus provides a clean signal for detection of these perturbations. The B-mode has yet to be detected by any measurement. 

%%%%%%%%%%%%%%%%%%%%%%%%%%%%%%%%%%%%%%%%%%%%%%%%%%%%%%%%%%%%%%%%%%%%%%%%%%%%%%%%%%%%%%%%%%%%%%%%%
%%%%%%%%%%%%%%%%%%%%%%%%%%%%%%%%%%%%%%%%%%%%%%%%%%%%%%%%%%%%%%%%%%%%%%%%%%%%%%%%%%%%%%%%%%%%%%%%%

\section{The Flatness and Horizon Problems}
\label{sec:Flatness}

We have so far considered two types of cosmological mass-energy -- matter and radiation -- and solved the Friedmann Equation for the simple case of a flat universe. What about the more general case? In this section, we consider non-flat universes with general contents. We introduce two related questions which are not explained by the standard Big Bang cosmology: why is the universe so close to flat today, and why is it so large? 

We can describe a general homogeneous, isotropic mass-energy by its equation of state
\begin{equation}
p = w \rho,
\end{equation}
so that pressureless matter corresponds to $w = 0$, and radiation corresponds to $w = 1/3$. We will consider only the case of constant equation of state, $w = {\rm const.}$ From the continuity equation, we have
\begin{equation}
\dot\rho + 3 \left(1 + w\right) \frac{\dot a}{a} \rho = 0,
\end{equation}
with solution
\begin{equation}
\rho \propto a^{-3 \left( 1 + w \right)}.
\end{equation}
The Friedmann Equation for a flat universe is then
\begin{equation}
\left(\frac{\dot a}{a}\right)^2 \propto a^{-3 \left(1 + w\right)},
\end{equation}
so that the scale factor increases as a power-law in time,
\begin{equation}
a\left(t\right) \propto t^{2 / 3 \left(1 + w\right)}.
\end{equation}
What about the evolution of a non-flat universe? Analytic solutions for $a(t)$ in the $k \neq 0$ case can be found in cosmology textbooks. For our purposes, it is sufficient to consider the time-dependence of the density parameter $\Omega$. From Eqs. (\ref{eq:generalFRW}, \ref{eq:defOmega}, \ref{eq:FRWOmega}) it is not too difficult to show that the density parameter evolves with the scale factor $a$ as:
\begin{equation}
\label{eq:Omegaevolution}
\frac{d \Omega}{d \ln{a}} = \left(1 + 3 w\right) \Omega \left(\Omega - 1\right).
\end{equation}
Proof is left as an exercise for the reader. Note that a flat universe, $\Omega = 1$ remains flat at all times, but in a non-flat universe, the density parameter $\Omega$ is a time-dependent quantity, with the evolution determined by the equation of state parameter $w$. For matter ($w = 0$) or radiation ($w = 1/3$), the prefactor in Eq. (\ref{eq:Omegaevolution}) is positive,
\begin{equation}
1 + 3 w > 0,
\end{equation}
which means a flat universe is an {\it unstable} fixed point:
\begin{equation}
\frac{d \left\vert\Omega - 1\right\vert}{d \ln{a}} > 0,\ \left(1 + 3 w\right) > 0.
\end{equation}
Any deviation from a flat geometry is amplified by the subsequent cosmological expansion, so a nearly flat universe today is a highly fine-tuned situation. The WMAP5 CMB measurement tells us the universe is flat to within a few percent, $\left\vert \Omega_0 - 1\right\vert < 0.02$ \cite{Dunkley:2008ie,Komatsu:2008hk}. If we are very conservative and take a limit on the density today as $\Omega_0 = 1 \pm 0.05$, that means that at recombination, when the CMB was emitted, $\Omega_{\rm rec} = 1 \pm 0.0004$, and at the time of primordial nucleosynthesis, $\Omega_{\rm nuc} = 1 \pm 10^{-12}$. Why did the universe start out so incredibly close to flat? The standard Big Bang cosmology provides no answer to this question, which we call the {\it flatness problem}. 

There is a second, related problem with the standard Big Bang picture, arising from the finite age of the universe. Because the universe has a finite age, photons can only have traveled a finite distance in the time since the Big Bang. Therefore, the universe has a {\it horizon}: the further out in space we look, the further back in time we see. If we could look far enough out in any direction, past the surface of last scattering, we would be able to see the Big Bang itself, and beyond that we can see no further. Every observer in an FRW spacetime sees herself at the center of a spherical horizon which defines her observable universe. To calculate the size of our horizon, we use the fact that photons travel on paths of zero proper length:
\begin{equation}
ds^2 = dt^2 - a^2\left(t\right) \left\vert d{\bf x}\right\vert^2 = 0,
\end{equation}
so that the comoving distance $\left\vert d{\bf x}\right\vert$ traversed by a photon in time $dt$ is
\begin{equation}
\left\vert d{\bf x}\right\vert = \frac{dt}{a\left(t\right)}.
\end{equation}
Therefore, the size of the cosmological horizon at time $t$ after the Big Bang is
\begin{equation}
d_{\rm H}\left(t\right) = \int_{0}^{t}{\frac{dt'}{a\left(t'\right)}}.
\end{equation}
To convert comoving length to proper length, we just multiply by $a\left(t\right)$, so that the proper horizon size is
\begin{equation}
d_{\rm H}^{\rm prop}\left(t\right) = a\left(t\right) d_{\rm H}^{\rm com}\left(t\right).
\end{equation}
Normalizing $a\left(t_0\right) = 1$, the horizon size of a 14-billion year-old flat, matter-dominated universe is $d_{\rm H} = 3 t_0 \sim 13\ {\rm Gpc}$. 

To see why the presence of a horizon is a problem for the standard Big Bang, we examine the causal structure of an FRW universe. Take the FRW metric
\begin{equation}
ds^2 =  dt^2 - a^2\left(t\right) \left\vert d{\bf x}\right\vert^2,
\end{equation}
and re-write it in terms of a redefined clock, the {\it conformal time} $\tau$:
\begin{equation}
\label{eq:conformalmetric}
ds^2 = a^2\left(\tau\right) \left[d\tau^2 - \left\vert d{\bf x}\right\vert^2\right].
\end{equation}
Conformal time is a ``clock'' which slows down with the expansion of the universe,
\begin{equation}
d\tau = \frac{dt}{a\left(t\right)},
\end{equation}
so that the comoving horizon size is just the age of the universe in conformal time
\begin{equation}
\label{eq:conformalhorizon}
d_{\rm H}\left(t\right) = \int_{0}^{t}{\frac{dt'}{a\left(t'\right)}} = \int_{0}^{\tau}{d\tau'} = \tau.
\end{equation}
The conformal metric is useful because the expansion of the spacetime is factored into a static metric multiplied by a time-dependent conformal factor $a\left(\tau\right)$, so that photon geodesics are simply described by $d\left\vert{\bf x}\right\vert = d\tau$. In a diagram of $\tau$ versus $\left\vert{\bf x}\right\vert$, photons travel on $45^{\circ}$ angles. (Note that this is true even for curved spacetimes!) We can draw light cones and infer causal relationships with the expansion factored out, in a manner identical to the usual case of Minkowski Space. 

\begin{figure}
\begin{center}
%--%\psfig{file=lightcone.eps,width=4.5in}
\includegraphics[width=4.5in]{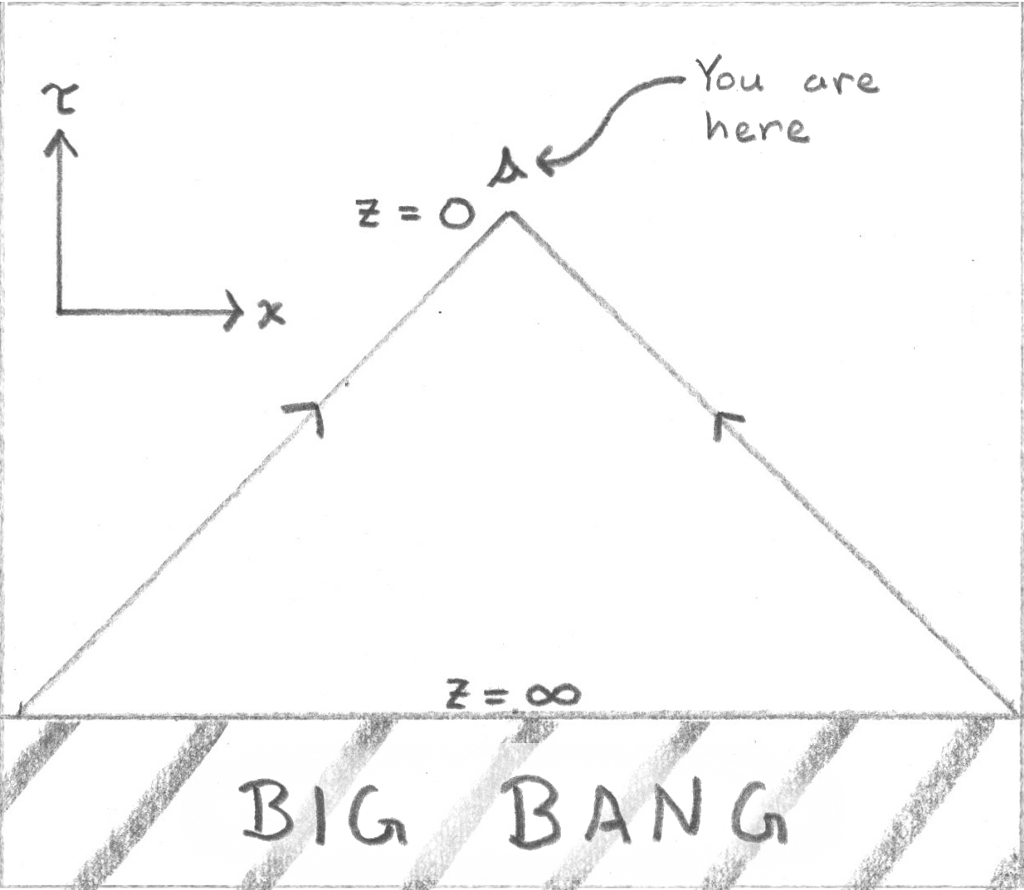}
\end{center}
\caption{A conformal diagram of a Friedmann-Robertson-Walker space. The FRW space is causally identical to Minkowski Space, except that it is not past-infinite, so that past light cones are ``cut off'' at the Big Bang, which is a spatially infinite surface at time $t = 0$. }
\label{fig:FRWdiagram}
\end{figure}

There is one major difference between FRW and Minkowski: an FRW spacetime has a {\it finite age}. Therefore, unlike the case of Minkowski Space, which has an infinite past, an FRW spacetime is ``chopped off'' at some finite past time $\tau = 0$ (Fig.\ref{fig:FRWdiagram}).
%%%%%%%%%%%%%%%%%%%%%%%%%%%%%%%%%%%%%%%%%%%%%%%%%%%%%%%%%%%%%%%%%%%%%%%%%%%%%%%%%%%%%%%%%%%%
%%%%%%%%%%%%%%%%%%%%%%%%%%%%%%%%%%%%%%%%%%%%%%%%%%%%%%%%%%%%%%%%%%%%%%%%%%%%%%%%%%%%%%%%%%%%
The initial singularity is a surface of constant conformal time, and it is easy to see from Eq. (\ref{eq:conformalhorizon}) that our horizon size is the width of our past light cone projected on the surface defined by the initial singularity. This is a very different picture from the notion many people (even scientists) have of the Big Bang, which is something akin to an explosion, with the universe initially a cosmic ``egg'' of zero size. On the contrary, in the case of a flat or open universe, the universe is spatially infinite an infinitesimal amount of time after the initial singularity: the Big Bang happens everywhere at once in an infinite space! Our {\it observable} universe is finite because we can only see a small patch of the much larger cosmos.\footnote{Of course, this is an idealization, and the actual universe could well have a nontrivial global topology, even if it is locally flat, as long as the scale of the overall manifold is much larger than our horizon size \cite{Kamionkowski:1998qx,Levin:2001fg}. } Closed universes are spatially finite, but are still much larger in extent than our observable patch. The key point is that two events on the conformal spacetime diagram are causally connected only if they share a causal past: that is, if their past light cones overlap. 

\begin{figure}
\begin{center}
%--%\psfig{file=lightconecmb.eps,width=4.5in}
\includegraphics[width=4.5in]{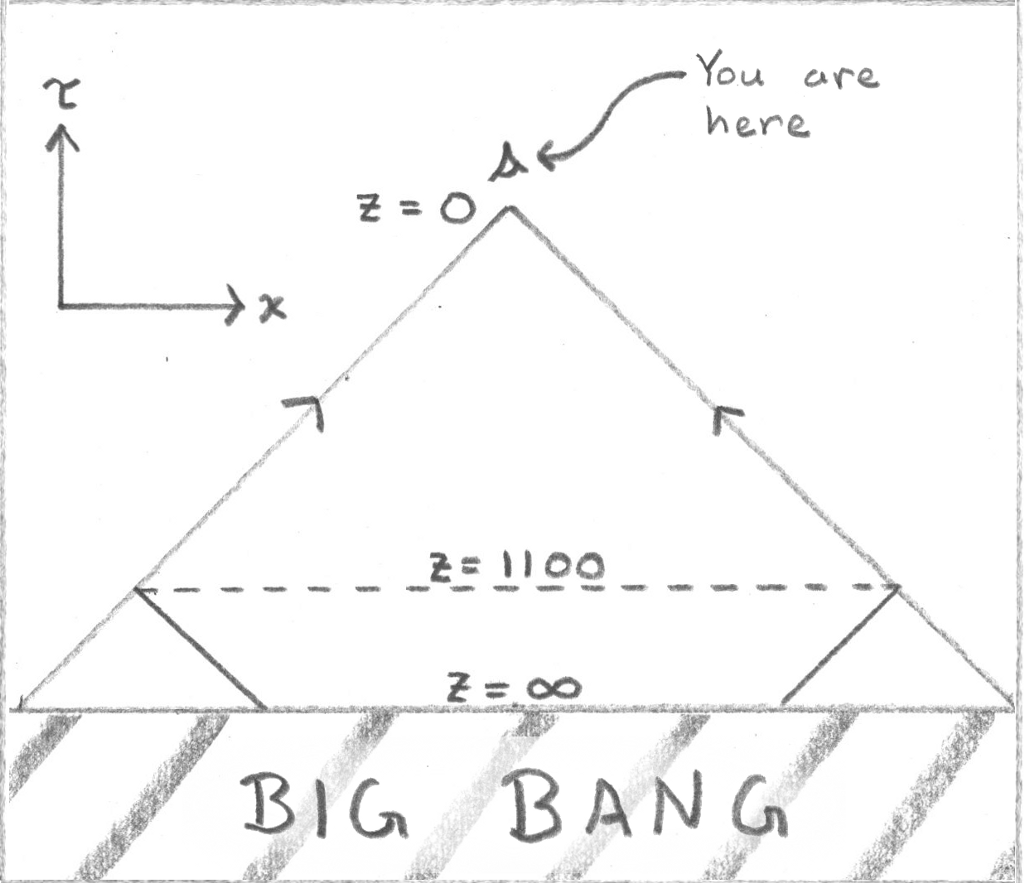}
\end{center}
\caption{A conformal diagram of the Cosmic Microwave Background. Two points on opposite sides of the sky are causally separate, since their past light cones do not intersect.}
\label{fig:CMBcausaldiagram}
\end{figure}

Consider two points on the CMB sky $180^{\circ}$ degrees apart (Fig. \ref{fig:CMBcausaldiagram}).
%%%%%%%%%%%%%%%%%%%%%%%%%%%%%%%%%%%%%%%%%%%%%%%%%%%%%%%%%%%%%%%%%%%%%%%%%%%%%%%%%%%%%%%%%%%%%%%%%
%%%%%%%%%%%%%%%%%%%%%%%%%%%%%%%%%%%%%%%%%%%%%%%%%%%%%%%%%%%%%%%%%%%%%%%%%%%%%%%%%%%%%%%%%%%%%%%%%
Their past light cones do not overlap, and the two points are causally {\it disconnected}. Those two points on the surface of last scattering occupy completely separate, disconnected observable universes. How did these points reach the observed thermal equilibrium to a few parts in $10^{5}$ if they never shared a causal past? This apparent paradox is called the {\it horizon problem}: the universe somehow reached nearly perfect equilibrium on scales much larger than the size of any local horizon. From the Friedmann Equation, it is easy to show that the horizon problem and the flatness problem are related: consider a comoving length scale $\lambda$. It is is easy to show that for $w = {\rm const.}$, the ratio of $\lambda$ to the horizon size $d_{\rm H}$ is related to the curvature by a conservation law
\begin{equation}
\label{eq:flatnesshorizoncons}
\left(\frac{\lambda}{d_{\rm H}}\right)^2 \left\vert \Omega - 1\right\vert = {\rm const.}
\end{equation}
Proof is left as an exercise for the reader. Therefore, for a universe evolving away from flatness,
\begin{equation}
\frac{d\left\vert \Omega - 1\right\vert}{d \ln{a}} > 0,
\end{equation}
the horizon size gets bigger in comoving units
\begin{equation}
\frac{d}{d\ln{a}}\left(\frac{\lambda}{d_{\rm H}}\right) < 0.
\end{equation}
That is, more and more space ``falls into'' the horizon, or becomes causally connected, at late times. 

What would be required to have a universe which evolves {\it toward} flatness, rather than away from it? From Eq. (\ref{eq:Omegaevolution}), we see that having  $1 + 3 w$ negative will do the trick,
\begin{equation}
\frac{d \left\vert\Omega - 1\right\vert}{d \ln{a}} < 0,\ \left(1 + 3 w\right) < 0.
\end{equation}
Therefore, if the energy density of the universe is dominated not by matter or radiation, but by something with sufficiently negative pressure, $p < -\rho / 3$, a curved universe will become flatter with time. From the Raychaudhuri Equation (\ref{eq:generalFRW}), we see that the case of $p < -\rho/3$ is exactly equivalent to an accelerating expansion:
\begin{equation}
\frac{\ddot a}{a} \propto - (1 + 3 w) > 0,\ \left(1 + 3 w\right) < 0. 
\end{equation}
If the expansion of the universe is slowing down, as is the case for matter- or radiation-domination, the curvature evolves away from flatness. But if the expansion is speeding up, the universe gets flatter. From Eq. (\ref{eq:flatnesshorizoncons}), we see that this negative pressure solution also solves the horizon problem, since accelerating expansion means that the horizon size is shrinking in comoving units:
\begin{equation}
\frac{d}{d\ln{a}}\left(\frac{\lambda}{d_{\rm H}}\right) > 0,\ \left(1 + 3 w\right) < 0.
\end{equation}
When the expansion accelerates, distances initially smaller than the horizon size are ``redshifted'' to scales larger than the horizon at late times. Accelerating cosmological expansion is called {\it inflation}.

\begin{figure}
\begin{center}
%--%\psfig{file=lightconeinfl.eps,width=4.5in}
\includegraphics[width=4.5in]{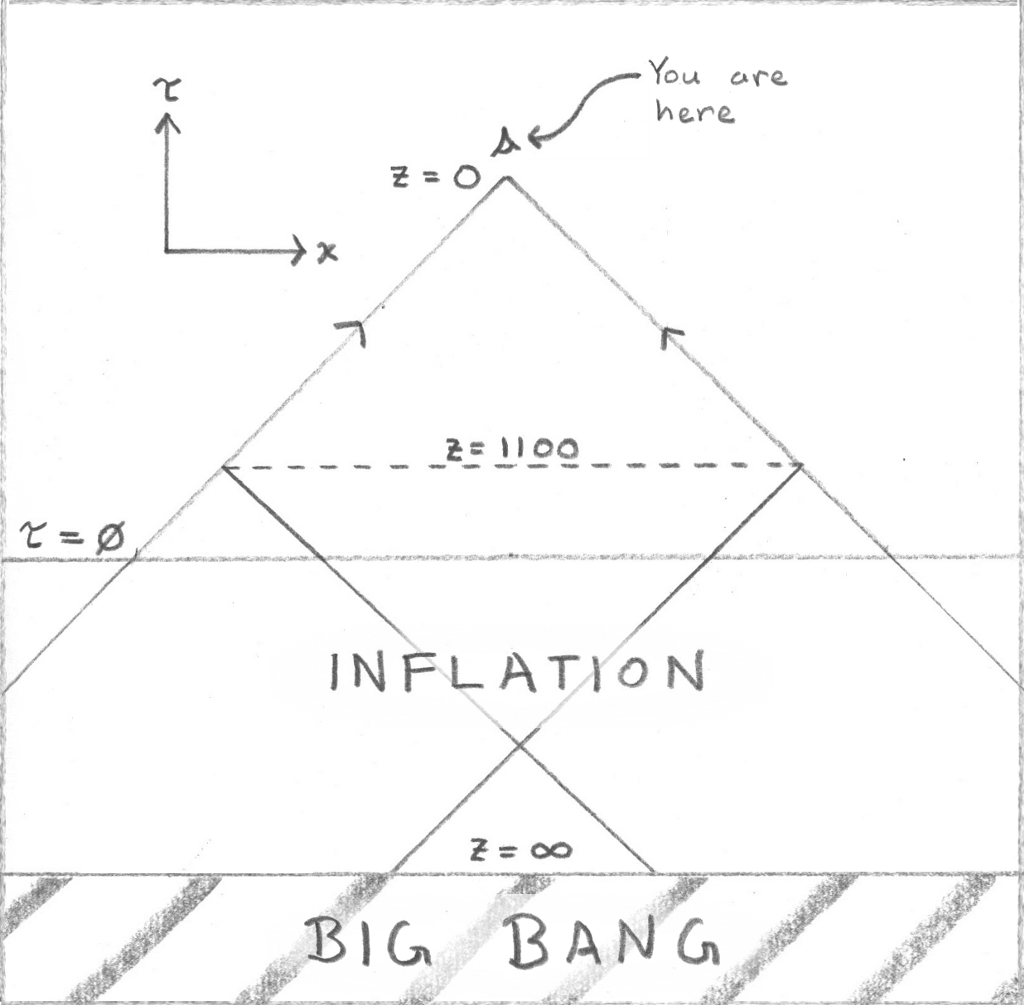}
\end{center}
\caption{A conformal diagram of light cones in an inflationary universe. Inflation ends in reheating at conformal time $\tau = 0$, which is the onset of the radiation-dominated expansion of the hot Big Bang. However, inflation provides a ``sea'' of negative conformal time, which allows the past light cones of events at the last scattering surface to overlap.}
\label{fig:INFLdiagram}
\end{figure}

The simplest example of an accelerating expansion from a negative pressure fluid is the case of vacuum energy we considered in Section (\ref{sec:SolvingTheFriedmannEquation}), for which the scale factor increases exponentially,
\begin{equation}
a \propto e^{H t}. 
\end{equation}
For such expansion, the universe is driven exponentially toward a flat geometry,
\begin{equation}
\frac{d \ln{\Omega}}{d\ln{a}} = 2 \left(1 - \Omega\right).
\end{equation}
We can see that the horizon problem is also solved by looking at the conformal time:
\begin{equation}
d\tau = \frac{dt}{a\left(t\right)} = e^{-H t} dt,
\end{equation}
so that
\begin{equation}
\label{eq:deSittertau}
\tau = - \frac{1}{H} e^{-H t} = - \frac{1}{a H}.
\end{equation}
The conformal time during the inflationary period is {\it negative}, tending toward zero at late time. Therefore, if we have a period of inflationary expansion prior to the early epoch of radiation-dominated expansion, inflation takes place in negative conformal time, and conformal time $\tau = 0$ represents not the initial singularity but the transition from the inflationary expansion to radiation domination. The initial singularity is pushed back into negative conformal time, and can be pushed arbitrarily far depending on the duration of inflation. Figure \ref{fig:INFLdiagram} shows the causal structure of an inflationary spacetime. The past light cones of two points on the CMB sky do not intersect at $\tau = 0$, but inflation provides a ``sea'' of negative conformal time, which allows those points to share a causal past. In this way, inflation solves the horizon problem. 
%%%%%%%%%%%%%%%%%%%%%%%%%%%%%%%%%%%%%%%%%%%%%%%%%%%%%%%%%%%%%%%%%%%%%%%%%%%%%%%%%%%%%%%%%%%%
%%%%%%%%%%%%%%%%%%%%%%%%%%%%%%%%%%%%%%%%%%%%%%%%%%%%%%%%%%%%%%%%%%%%%%%%%%%%%%%%%%%%%%%%%%%%

In more realistic models of inflation in the early universe, the energy density is approximately, but not exactly, constant, and the expansion is approximately, but not exactly, exponential. In such quasi-de Sitter spaces, the qualitative picture above still holds, and inflation provides a clean and compelling explanation for the peculiar boundary conditions for our universe. In the next section, we discuss how to construct more detailed models of inflation in field theory. 

\section{Inflation from Scalar Fields}
\label{sec:InflationFromScalarFields}

The example of de Sitter evolution we considered in Section \ref{sec:Flatness} gives a good qualitative picture of how inflation, or accelerated expansion, solves the horizon and flatness problems of the standard Big Bang cosmology. However, this leaves open the question: what physics is responsible for the accelerated expansion at early times? It cannot be Einstein's cosmological constant, simply because a universe dominated by vacuum energy {\it stays} dominated by vacuum energy for the infinite future, since in a de Sitter background matter ($\rho \propto a^{-3}$) and radiation ($\rho \propto a^{-4}$) are diluted exponentially quickly. Therefore, we will never reach a radiation-dominated phase, and we will never see a hot Big Bang. In order to transition from an inflating phase to a thermal equilibrium, radiation-dominated phase, the vacuum-like energy during inflation must be time-dependent. We model this dynamics with a scalar field $\phi$, for which we assume the following action:
\begin{equation}
\label{eq:scalaraction}
S = \int{d^4 x \sqrt{- g} {\mathcal L}_\phi},
\end{equation}
where $g \equiv Det\left(g_{\mu\nu}\right)$ is the determinant of the metric and the Lagrangian for the field $\phi$ is 
\begin{equation}
\label{eq:canonicalL}
{\mathcal L}_\phi = \frac{1}{2} g^{\mu\nu} \partial_\mu \phi \partial_\nu \phi - V\left(\phi\right).
\end{equation}
Comparing the action (\ref{eq:scalaraction}) and the Lagrangian (\ref{eq:canonicalL}) with their Minkowski counterparts illustrates how we generalize a classical field theory to curved spacetime:
\begin{equation}
S_{\rm Minkowski} = \int{d^4 x \left[\frac{1}{2} \eta^{\mu\nu} \partial_\mu \phi \partial_\nu \phi - V\left(\phi\right)\right]}.
\end{equation}
The metric appears in two places in the curved-spacetime action: First, it appears in the measure of volume in the four-space, $d^4 x$, where the determinant of the metric takes the role of the Jacobian for arbitrary coordinate transformations, $x \rightarrow x'$. Second, the metric appears in the kinetic term for the scalar field, where we replace the Minkowski metric $\eta^{\mu\nu}$ with the general metric $g^{\mu\nu}$. 

The action (\ref{eq:scalaraction}) is not the most general assumption we could make, as we can see by writing the full action including gravity,
\begin{equation}
S_{\rm tot} = \int{d^4 x \sqrt{- g} \left[\frac{\mpl^2}{16 \pi} R + {\mathcal L}_\phi\right]}.
\end{equation}
Here $R$ is the Ricci Scalar, composed of the metric and its derivatives. Variation of the first term in the action results in the Einstein Field Equation (\ref{eq:EFE}). Such a {\it minimally coupled} theory assumes that there is no direct coupling between the field and the metric, which would be represented in a more general action by terms which mix $R$ and  $\phi$. In practice, many such non-minimally coupled theories can be transformed to a minimally coupled form by a field redefinition. We could also write a more general theory by modifying the scalar field Lagrangian (\ref{eq:canonicalL}) to contain non-canonical kinetic terms,
\begin{equation}
{\mathcal L}_\phi = F\left(\phi, g^{\mu\nu} \partial_\mu \phi \partial_\nu\phi\right) - V\left(\phi\right).
\end{equation}
where $F()$ is some function of the field and its derivatives. Such Lagrangians appear frequently in models of inflation based on string theory, and are a topic of considerable current research interest. We could also complicate the gravitational sector by replacing the Ricci scalar $R$ with a more complicated function $f\left(R\right)$. An example of such a model is the inflation model of Starobinsky \cite{Starobinsky:1980te}, which can be reduced to the form (\ref{eq:scalaraction}) through a field redefinition. We could also introduce multiple scalar fields. 

Here we will confine ourselves for simplicity to a canonical Lagrangian (\ref{eq:canonicalL}) of a single scalar field, for which the only adjustable quantity is the choice of potential $V\left(\phi\right)$. For simplicity, we assume a flat spacetime,
\begin{equation}
\label{eq:coordinatemetric}
g_{\mu\nu} = \left(
\begin{array}{cccc}
1& & & \\
 &-a^2(t)& & \\
 & &-a^2(t)& \\
 & & &-a^2(t)
\end{array}
\right),
\end{equation}
and the equation of motion for the field $\phi$ with a Lagrangian given by Eq. (\ref{eq:canonicalL}) is:
\begin{equation}
\ddot\phi + 3 H \dot\phi - \nabla^2 \phi + \frac{\delta V}{\delta \phi} = 0,
\end{equation}
where an overdot indicates a derivative with respect to the coordinate time $t$, and $H = \dot a / a$ is the Hubble parameter.  We will be particularly interested in the homogeneous mode of the field, for which the gradient term vanishes, $\nabla \phi = 0$, so that the the functional derivative $\delta V / \delta\phi$ simplifies to an ordinary derivative, and the equation of motion simplifies to\footnote{The astute reader may well ask: if we are claiming inflation is a solution to the problems of flatness and homogeneity in the universe, why are we assuming flatness and homogeneity from the outset? The answer is that, as long as inflation gets started {\it somehow} and goes on for long enough, the late-time behavior of the field $\phi$ will always be described by Eq. (\ref{eq:inflatoneom}). We will see later that we only have observational access to the {\it end} of the inflationary period, and therefore a consistent theory of initial conditions  is not required for investigating the observational consequences of inflation.}
\begin{equation}
\label{eq:inflatoneom}
\ddot\phi + 3 H \dot\phi + V'\left(\phi\right) = 0.
\end{equation}
The stress-energy for a scalar field is given by
\begin{equation}
T_{\mu\nu} = \partial_\mu \phi \partial_\nu \phi - g_{\mu \nu} {\cal L}_\phi,
\end{equation}
and, for a homogeneous field, it takes the form of a perfect fluid with energy density $\rho$ and pressure $p$, with
\begin{eqnarray}
\label{eq:infleqofstate}
\rho &=& \frac{1}{2} \dot\phi^2 + V\left(\phi\right),\cr
p &=& \frac{1}{2} \dot\phi^2 - V\left(\phi\right).
\end{eqnarray}
We see that the de Sitter limit, $p \simeq - \rho$, is just the limit in which the potential energy of the field dominates the kinetic energy, $\dot\phi^2 \ll V\left(\phi\right)$. This limit is referred to as {\it slow roll}, and under such conditions the universe expands quasi-exponentially, 
\begin{equation}
a\left(t\right) \propto \exp{\left(\int{H dt}\right)} \equiv e^{-N},
\end{equation}
where it is conventional to define the number of e-folds $N$ with the sign convention
\begin{equation}
\label{eq:numefolds}
d N \equiv - H dt,
\end{equation}
so that $N$ is large in the far past and decreases as we go forward in time and as the scale factor $a$ increases.

This can be made quantitative by plugging the energy and pressure (\ref{eq:infleqofstate}) into the Friedmann Equation
\begin{equation}
\label{eq:scalarFriedmann}
H^2 = \left(\frac{\dot a}{a}\right)^2 = \frac{8 \pi}{3 \mpl^2} \left[\frac{1}{2} \dot\phi^2 + V\left(\phi\right)\right],
\end{equation}
and the Raychaudhuri Equation, which we write in the convenient form
\begin{equation}
\left(\frac{\ddot a}{a}\right) = - \frac{4 \pi}{3 \mpl^2} \left(\rho + 3 p\right) = H^2 \left(1 - \epsilon\right).
\end{equation}
Here $H^2$ is given in terms of $\phi$ by the Friedmann Equation (\ref{eq:scalarFriedmann}), and the parameter $\epsilon$ specifies the equation of state, 
\begin{equation}
\label{eq:defepsilon}
\epsilon \equiv \frac{3}{2} \left(\frac{p}{\rho} + 1\right) = \frac{4 \pi}{\mpl^2} \left(\frac{\dot\phi}{H}\right)^2.
\end{equation}
It is a straightforward exercise to show that $\epsilon$ is related to the evolution of the Hubble parameter by
\begin{equation}
\epsilon = - \frac{d \ln{H}}{d \ln{a}} = \frac{1}{H}\frac{d H}{d N},
\end{equation}
where $N$ is the number of e-folds (\ref{eq:numefolds}). This is a useful parameterization because the condition for accelerated expansion $\ddot a > 0$ is simply equivalent to $\epsilon < 1$. The de Sitter limit $p \rightarrow -\rho$ is equivalent to $\epsilon \rightarrow 0$, so that the potential $V\left(\phi\right)$ dominates the energy density, and 
\begin{equation}
\label{eq:srHubble}
H^2 \simeq  \frac{8 \pi}{3 \mpl^2} V\left(\phi\right).
\end{equation}
We make the additional approximation that the friction term in the equation of motion (\ref{eq:inflatoneom}) dominates, 
\begin{equation}
\label{eq:secondsrcondition}
\ddot \phi \ll 3 H \dot\phi,
\end{equation}
so that the equation of motion for the scalar field is approximately
\begin{equation}
\label{eq:sreom}
3 H \dot\phi + V'\left(\phi\right) \simeq 0.
\end{equation}
Equation (\ref{eq:sreom}) together with the Friedmann Equation (\ref{eq:srHubble}) are together referred to as the {\it slow roll approximation}. The condition (\ref{eq:secondsrcondition}) can be expressed in terms of a second dimensionless parameter, conventionally defined as
\begin{equation}
\label{eq:defeta}
\eta \equiv -\frac{\ddot \phi}{H \dot\phi} = \epsilon + \frac{1}{2 \epsilon}\frac{d \epsilon}{d N}.
\end{equation}

The parameters $\epsilon$ and $\eta$ are referred to as {\it slow roll parameters}, and the slow roll approximation is valid as long as both are small, $\epsilon,\ \left\vert\eta\right\vert \ll 1$. It is not obvious that this will be a valid approximation for situations of physical interest: $\eta$ need {\it not} be small for inflation to take place. Inflation takes place when $\epsilon < 1$, regardless of the value of $\eta$.  We later demonstrate explicitly that slow roll does in fact hold for interesting choices of inflationary potential. In the limit of slow roll, we can use Eqs. (\ref{eq:srHubble}, \ref{eq:sreom}) to write the parameter $\epsilon$ approximately as
\begin{equation}
\label{eq:srepsilon}
\epsilon = \frac{4 \pi}{\mpl^2} \left(\frac{\dot\phi}{H}\right)^2 \simeq \frac{\mpl^2}{16 \pi} \left(\frac{V'\left(\phi\right)}{V\left(\phi\right)}\right)^2.
\end{equation}
The inflationary limit, $\epsilon \ll 1$ is then just equivalent to a field evolving on a flat potential, $V'\left(\phi\right) \ll V\left(\phi\right)$. The second slow roll parameter $\eta$ can likewise be written approximately as:
\begin{eqnarray}
\label{eq:sreta}
\eta &=& - \frac{\ddot \phi}{H \dot\phi}\cr
&\simeq& \frac{\mpl^2}{8 \pi} \left[\frac{V''\left(\phi\right)}{V\left(\phi\right)} - \frac{1}{2} \left(\frac{V'\left(\phi\right)}{V\left(\phi\right)}\right)^2\right],
\end{eqnarray}
so that the curvature $V''$ of the potential must also be small for slow roll to be a valid approximation.  Similarly, we can write number of e-folds as a function $N\left(\phi\right)$ of the field as:
\begin{eqnarray}
\label{eq:srN}
N &=& - \int{H dt} = - \int{\frac{H}{\dot \phi} d\phi}
= \frac{2 \sqrt{\pi}}{\mpl} \int{\frac{d \phi}{\sqrt{\epsilon}}}\cr
&\simeq& \frac{8 \pi}{\mpl^2} \int_{\phi_e}^{\phi} \frac{V\left(\phi\right)}{V'\left(\phi\right)} d\phi,
\end{eqnarray}
The limits on the last integral are defined such that $\phi_e$ is a fixed field value, which we will later take to be the end of inflation, and $N$ increases as we go {\it backward} in time, representing the number of e-folds of expansion which take place between field value $\phi$ and $\phi_e$. 

\begin{figure}
\begin{center}
%--%\psfig{file=potential.eps,width=4.5in}
\includegraphics[width=4.5in]{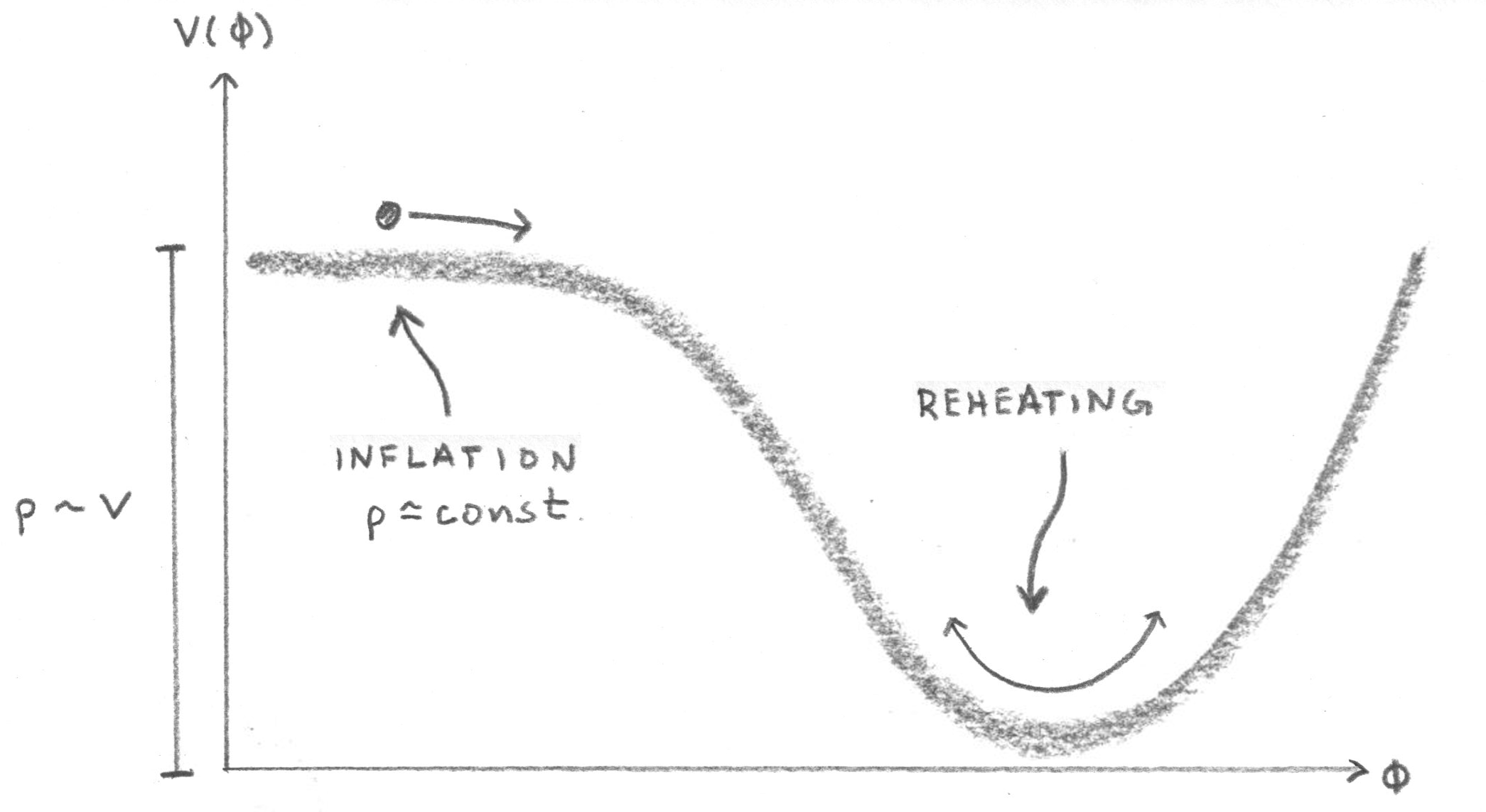}
\end{center}
\caption{A schematic of the potential for inflation. Inflation takes place on the region of the potential which is sufficiently ``flat'', and reheating takes place near the true vacuum for the field.}
\label{fig:inflationschematic}
\end{figure}

The qualitative picture of scalar field-driven inflation is that of a phase transition with order parameter given by the field $\phi$. At early times, the energy density of the universe is dominated by the field $\phi$ which is slowly evolving on a nearly constant potential, so that it approximates a cosmological constant (Fig. \ref{fig:inflationschematic}). During this period, the universe is exponentially driven toward flatness and homogeneity. Inflation ends as the potential steepens and the field begins to oscillate about its vacuum state at the minimum of the potential. At this point, we have an effectively zero-temperature scalar in a state of coherent oscillation about the minimum of the potential, and the universe is a huge Bose-Einstein condensate: hardly a hot Big Bang! In order to transition to a radiation-dominated hot Big Bang cosmology, the energy in the inflaton field must decay into Standard Model particles, a process generically termed {\it reheating}. This process is model-dependent, but it typically happens very rapidly. 
%%%%%%%%%%%%%%%%%%%%%%%%%%%%%%%%%%%%%%%%%%%%%%%%%%%%%%%%%%%%%%%%%%%%%%%%%%%%%%%%%%%%%%%%%%%%
%%%%%%%%%%%%%%%%%%%%%%%%%%%%%%%%%%%%%%%%%%%%%%%%%%%%%%%%%%%%%%%%%%%%%%%%%%%%%%%%%%%%%%%%%%%%
Note that the field $\phi$ need not be a fundamental field like a Higgs boson (although it could in fact be fundamental). {\it Any} order parameter for a phase transition will do, as long as it has the quantum numbers of vacuum, and the effective potential has the correct properties. The inflaton $\phi$ could well be a scalar composite of more fundamental degrees of freedom, the coordinate of a brane in a higher-dimensional compactification from string theory, a supersymmetric modulus, or something even more exotic. The simple single-field picture we discuss here is therefore an effective representation of a large variety of underlying fundamental theories. All of the physics important to inflation is contained in the shape of the potential $V\left(\phi\right)$. (The details of the underlying theory {\it are} important for understanding the epoch of reheating, since the reheating process depends crucially on the specific couplings of the inflaton to the other degrees of freedom in the theory.)

How long does inflation need to go on in order to solve the flatness and horizon problems? We use a thermodynamic argument, which rests on a simple fact about cosmological expansion: as long as there are no decays or annihilations of massive particles, all other interactions conserve photon number, so that the number of photons in a comoving volume is {\it constant}. Since the entropy of photons is proportional to the number density, that means the entropy per comoving volume is also constant. Therefore, the total entropy in the Cosmic Microwave Background (or, equivalently, the total number of photons) is a convenient measure of spatial volume in the universe. Since the entropy per photon $s$ is (up to a few constants) given by the cube of the temperature,
\begin{equation}
s \sim T^3,
\end{equation}
the total photon entropy $S$ in our current horizon volume is of order
\begin{equation}
S_{\rm hor} \sim T_{\rm CMB}^3 d_{\rm H}^3 \sim \left(\frac{T_{\rm CMB}}{H_0}\right)^3 \sim 10^{88},
\end{equation}
where we have taken the CMB temperature to be $2.7\ K$ and the current Hubble parameter $H_0$ to be $70\ {\rm km/s/MpC}$. (The interesting unit conversion from km/s/MpC to Kelvin is left as an exercise for the reader.) 

Let us consider a highly over-simplified picture of the universe, in which no particle decays or annihilations occur between the end of inflation and today. In that case, the only time when the photon number (and therefore the entropy) in the universe changes is during the reheating process itself, when the inflation $\phi$ decays into radiation and sets the initial state for the hot Big Bang. Therefore, we must {\it at minimum} create an entropy of $10^{88}$ during reheating. Let us say that the energy density during inflation is 
\begin{equation}
\rho \sim V\left(\phi\right) \sim \Lambda^4,
\end{equation}
where $\Lambda$ is some energy scale. Therefore, the horizon size during inflation is then
\begin{equation}
d_{\rm H} \sim H^{-1} \sim \frac{\mpl}{\Lambda^2},
\end{equation}
so that the initial volume of the inflationary ``patch'' which undergoes exponential expansion is 
\begin{equation}
V_i \sim d_{\rm H}^3 \sim \frac{\mpl^3}{\Lambda^{6}}.
\end{equation}
Suppose inflation continues for $N$ e-folds of expansion, so that the scale factor $a$ increases by a factor of $e^{N}$ during inflation. The {\it proper} volume of the initial inflationary patch increases by the cube of the scale factor
\begin{equation}
V_f \sim e^{3 N} d_{\rm H}^3 \sim  e^{3 N} \frac{\mpl^3}{\Lambda^{6}}.
\end{equation}
Inflation takes a tiny patch of the universe and blows it up exponentially large, but in such a way that the energy {\it density} remains approximately constant: we have created an exponential amount of energy out of nothing! During reheating, this huge store of energy in the coherently oscillating field $\phi$ decays into radiation and the temperature and entropy of the universe undergo an explosive increase. If reheating is highly efficient, then all or most of the energy stored in the inflaton field will be transformed into radiation, and the temperature of the universe after reheating will be of order the energy density of the inflaton field,
\begin{equation}
T_{\rm RH} \sim \Lambda.
\end{equation}
The entropy per comoving volume after reheating will then be $s_{\rm RH} \sim T_{\rm RH}^3 \sim \Lambda^3$, and the {\it total} entropy in our inflating patch will be 
\begin{equation}
S_{\rm RH} \sim V_f T_{\rm RH}^3 \sim e^{3 N} \frac{\mpl^3}{\Lambda^{3}}.
\end{equation}
Since this is our only source of entropy in our toy-model universe, this entropy must be at least as large as the entropy in our current horizon volume, $S_{\rm RH} \geq 10^{88}$. The only adjustable parameter is the number of e-folds of inflation. Taking the logarithm of both sides gives a lower bound on $N$,
\begin{equation}
\label{eq:lowerboundN}
N \geq 68 + \ln{\left(\frac{\Lambda}{\mpl}\right)}.
\end{equation}
We will see later that the amplitude of primordial density fluctuations $\delta \rho / \rho \sim 10^{-5}$ typically constrains the inflationary energy scale to be of order $\Lambda \sim 10^{-4} m_{\rm Pl}$, so that we have a lower limit on the number of e-folds of inflation of
\begin{equation}
N > N_{\rm min} \sim 60.
\end{equation}

Most inflation models hugely oversaturate this bound, with $N_{\rm tot} \gg N_{\rm min}$. There is in fact no {\it upper} bound on the number of e-folds of inflation, an idea which is central to Linde's idea of ``eternal'' inflation \cite{Linde:1986fc,Guth:2000ka,Aguirre:2007gy,Winitzki:2008zz}, in which inflation, once initiated, never completely ends, with reheating occurring only in isolated patches of the cosmos. Furthermore, it is easy to see that our oversimplified toy model of the universe gives a remarkably accurate estimate of $N_{\rm min}$. In the real universe, all sorts of particle decays and annihilations happen between the end of inflation and today, which create additional entropy. However, our lower bound (\ref{eq:lowerboundN}) is only logarithmically sensitive to these processes. The dominant uncertainty is in the reheat temperature: it is possible that the energy scale of inflation is very low, or that the reheating process is very inefficient, and there are very few {\it observational} bounds on these scales. We do know that the universe has to be radiation dominated and in equilibrium by the time primordial nucleosynthesis happens at temperatures of order MeV. Furthermore, the baryon asymmetry of the universe is at least a good hint that the Big Bang was hot to at least the scale of electroweak unification. A typical assumption is that the reheat temperature is something between $1\ {\rm TeV}$ and  $10^{16}\ {\rm GeV}$, which translates into a range for $N_{\rm min}$ of order \cite{Liddle:2003as,Kinney:2005in}
\begin{equation}
N_{\rm min} \simeq \left[46,60\right].
\end{equation}

\subsection{Example: the $\lambda \phi^4$ potential}

We are now in a position to apply this to a specific case. We use the simple case of a quartic potential,
\begin{equation}
V\left(\phi\right) = \lambda \phi^4.
\end{equation}
The slow roll equations (\ref{eq:sreom}, \ref{eq:srHubble}) imply that the field evolves as:
\begin{equation}
\dot\phi = - \frac{V'\left(\phi\right)}{3 H} = - \sqrt{\frac{\mpl^2}{24 \pi}} \frac{V'\left(\phi\right)}{\sqrt{V\left(\phi\right)}} \propto \phi. 
\end{equation}
Note that this potential does not much qualitatively resemble the schematic in Fig. \ref{fig:INFLdiagram}: the ``flatness'' of the potential arises because the energy density $V\left(\phi\right) \propto \phi^4$ rises much more quickly than the kinetic energy, $\dot\phi^2 \propto \phi^2$, so that if the field is far enough out on the potential, the slow roll approximation is self-consistent. The field rolls down to the potential toward the vacuum at the origin, and the equation of state is determined by the parameter $\epsilon$,
\begin{equation}
\epsilon\left(\phi\right) \simeq \frac{\mpl^2}{16 \pi} \left(\frac{V'\left(\phi\right)}{V\left(\phi\right)}\right)^2 = \frac{1}{\pi} \left(\frac{\mpl}{\phi}\right)^2.
\end{equation}
The field value $\phi_e$ at the end of inflation is when $\epsilon\left(\phi_e\right) = 1$, or
\begin{equation}
\phi_e = \frac{m_{\rm Pl}}{\sqrt{\pi}}. 
\end{equation}
For $\phi > \phi_e$, $\epsilon < 1$ and the universe is inflating, and for $\phi < \phi_e$, $\epsilon > 1$ and the expansion enters a decelerating phase. Therefore, even this simple potential has the necessary characteristics to support a period of early-universe inflation followed by reheating and a hot Big Bang cosmology. What about the requirement that the universe inflate for at least 60 e-folds? Using Eq. (\ref{eq:srepsilon}), we can express the number of e-folds before the end of inflation (\ref{eq:srN}) as
\begin{equation}
N = \frac{2 \sqrt{\pi}}{\mpl} \int_{\phi_e}^{\phi}{\frac{dx}{\sqrt{\epsilon\left(x\right)}}} = \pi \left(\frac{\phi}{\mpl}\right)^2 - 1,
\end{equation}
where we integrate {\it backward} from $\phi_e$ to $\phi$ to be consistent with the sign convention (\ref{eq:numefolds}). Therefore the field value $N$ e-folds before the end of inflation is
\begin{equation}
\label{eq:phi4N}
\phi_N = \mpl\sqrt{\frac{N + 1}{\pi}},
\end{equation}
so that
\begin{equation}
\phi_{60} = 4.4 \mpl.
\end{equation}
We obtain sufficient inflation, but at a price: the field must be a long way (several times the Planck scale) out on the potential. However, we do {\it not} necessarily have to invoke quantum gravity, since for small enough coupling $\lambda$, the energy density in the field can much less than the Planck density, and the energy density is the physically important quantity. 

In this section, we have seen that the basic picture of an early epoch in the universe dominated by vacuum-like energy, leading to nearly exponential expansion, can be realized within the context of a simple scalar field theory. The equation of state for the field approximates a cosmological constant $p = -\rho$ when the energy density is dominated by the field potential $V\left(\phi\right)$, and inflation ends when the potential becomes steep enough that the kinetic energy $\dot\phi^2 / 2$ dominates over the potential. To solve the horizon and flatness problems and create a universe consistent with observation, we must have {\it at least} 60 or so e-folds of inflation, although in principle inflation could continue for much longer than this minimum amount. This dynamical explanation for the flatness and homogeneity of the universe is an interesting, but hardly compelling scenario. It could be that the universe started out homogeneous and flat because of initial conditions, either through some symmetry we do not yet understand, or because there are many universes, and we just happen to find ourselves in a highly unlikely realization which is homogeneous and geometrically flat. In the absence of any other observational handles on the physics of the very early universe, it is impossible to tell. However, flatness and homogeneity are not the whole story: inflation provides an elegant mechanism for explaining the {\it inhomogeneity} of the universe as well, which we discuss in Section \ref{sec:PerturbationsInInflation}.

\section{Perturbations in Inflation}
\label{sec:PerturbationsInInflation}

The universe we live in today is homogeneous, but only when averaged over very large scales. On small scales, the size of people or solar systems or galaxies or even clusters of galaxies, the universe we see is highly inhomogeneous. Our world is full of complex structure, created by gravitational instability acting on tiny ``seed'' perturbations in the early universe. If we look as far back in time as the epoch of recombination, the universe on all scales was homogeneous to a high degree of precision, a few parts in $10^5$. Recent observational efforts such as the WMAP satellite have made exquisitely precise measurements of the first tiny inhomogeneities in the universe, which later collapsed to form the structure we see today. (We discuss the WMAP observation in more detail in Section \ref{sec:ObservationalConstraints}.) Therefore, another mystery of Big Bang cosmology is: what created the primordial perturbations? This mystery is compounded by the fact that the perturbations we observe in the CMB exhibit correlations on scales much larger than the horizon size at the time of recombination, which corresponds to an angular multipole of $\ell \simeq 100$, or about $1^\circ$ as observed on the sky today. This is another version of the horizon problem: not only is the universe homogeneous on scales larger than the horizon, but whatever created the primordial perturbations must also have been capable of generating fluctuations on scales larger than the horizon. Inflation provides just such a mechanism \cite{Starobinsky:1979ty,Mukhanov:1981xt,Hawking:1982cz,Hawking:1982my,Starobinsky:1982ee,Guth:1982ec,Bardeen:1983qw}. 

\begin{figure}
\begin{center}
%--%\psfig{file=horizoninfl.eps,width=4.5in}
\includegraphics[width=4.5in]{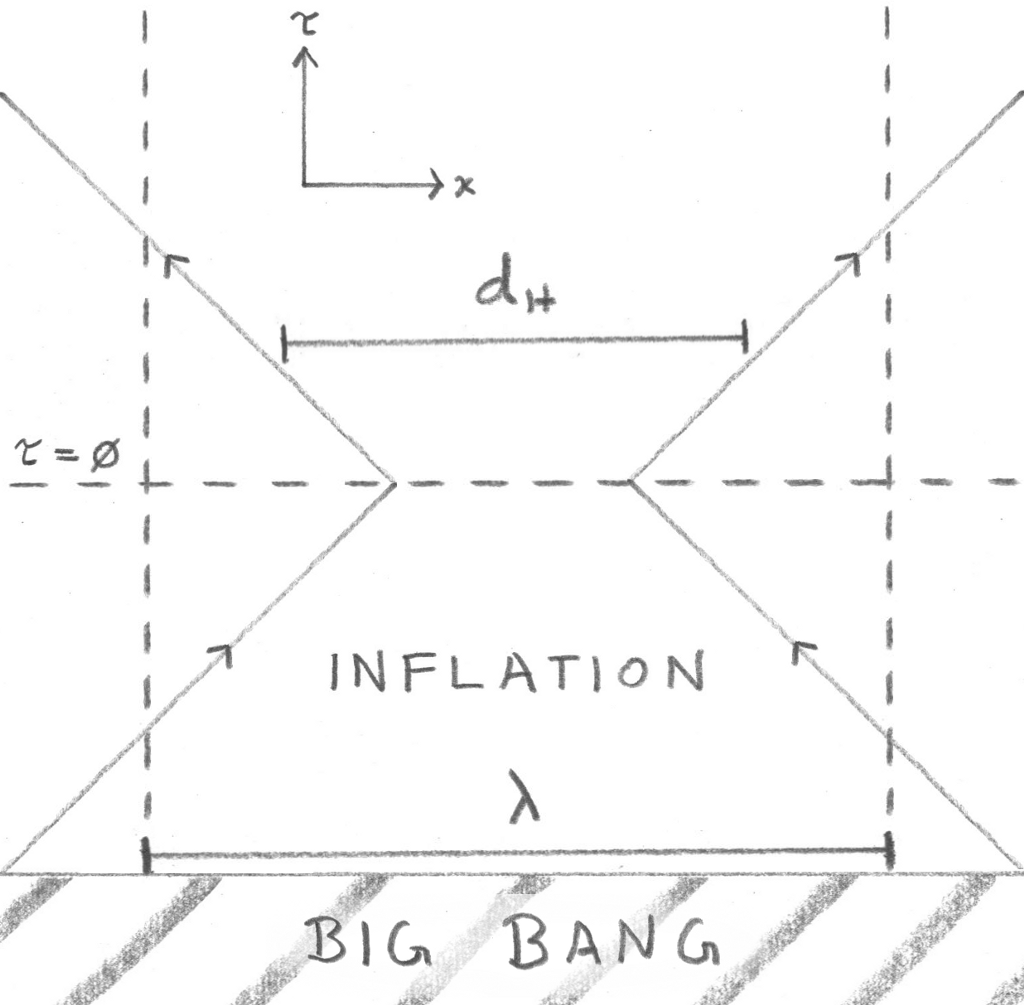}
\end{center}
\caption{A conformal diagram of the horizon in an inflationary universe. The comoving horizon shrinks during inflation, and grows during the radiation- and matter-dominated expansion, while the comoving wavelengths of perturbations remain constant. This drives comoving perturbations to ``superhorizon'' scales.}
\label{fig:horizon}
\end{figure}

Consider a perturbation in the cosmological fluid with wavelength $\lambda$. Since the proper wavelength redshifts with expansion, $\lambda_{\rm prop} \propto a\left(t\right)$, the {\it comoving} wavelength of the perturbation is a constant, $\lambda_{\rm com} = {\rm const.}$ This is true of photons or density perturbations or gravitational waves or any other wave propagating in the cosmological background. Now consider this wavelength relative to the size of the horizon: We have seen that in general the horizon as measured in comoving units is proportional to the conformal time, $d_H \propto \tau$. Therefore, for matter- or radiation-dominated expansion, the horizon size {\it grows} in comoving units, so that a comoving length which is larger than the horizon at early times is smaller than the horizon at late times: modes ``fall into'' the horizon. The opposite is true during inflation, where the conformal time is negative and evolving toward zero: the comoving horizon size is still proportional to $\tau$, but it now {\it shrinks} with cosmological expansion, and comoving perturbations which are initially smaller than the horizon are ``redshifted'' to scales larger than the horizon at late times (Fig. \ref{fig:horizon}).
%%%%%%%%%%%%%%%%%%%%%%%%%%%%%%%%%%%%%%%%%%%%%%%%%%%%%%%%%%%%%%%%%%%%%%%%%%%%%%%%%%%%%%%%%%%%
%%%%%%%%%%%%%%%%%%%%%%%%%%%%%%%%%%%%%%%%%%%%%%%%%%%%%%%%%%%%%%%%%%%%%%%%%%%%%%%%%%%%%%%%%%%%

If the universe is inflating at early times, and radiation- or matter-dominated at late times, perturbations in the density of the universe which are initially smaller than the horizon are redshifted during inflation to superhorizon scales. Later, as the horizon begins to grow in comoving coordinates, the perturbations fall back into the horizon, where they act as a source for structure formation. In this way inflation explains the observed properties of perturbations in the universe, which exist at both super- and sub-horizon scales at the time of recombination. Furthermore, an important consequence of this process is that the last perturbations to exit the horizon are the {\it first} to fall back in. Therefore, the shortest wavelength perturbations are the ones which exited the horizon just at the end of inflation, $N = 0$, and longer wavelength perturbations exited the horizon earlier. Perturbations about the same size as our horizon today exited the horizon during inflation at around $N = 60$. Perturbations which exited the horizon earlier than that, $N > 60$, are still larger than our horizon today. Therefore, it is only possible to place observational constraints on the {\it end} of inflation, about the last 60 e-folds. Everything that happened before that, including anything that might tell us about the initial conditions which led to inflation, is most probably inaccessible to us. 

This kinematic picture, however, does not itself explain the physical origin of the perturbations. Inflation driven by a scalar field provides a natural explanation for this as well. The inflaton field $\phi$ evolving on the potential $V\left(\phi\right)$ will not evolve completely classically, but will also be subject to small quantum fluctuations about its classical trajectory, which will in general be {\it inhomogeneous}. Since the energy density of the universe during inflation is dominated by the inflaton field, quantum fluctuations in $\phi$ couple to the spacetime curvature and result in fluctuations in the density of the universe. Therefore, in the same way that the classical behavior of the field $\phi$ provides a description of the background evolution of the universe, the quantum behavior of $\phi$ provides a description of the inhomogeneous perturbations about that background. We defer a full treatment of inflaton perturbations to Appendix A, and in the next section focus on the much simpler case of quantizing a {\it decoupled} scalar $\varphi$ in an inflationary spacetime. In addition to its relative simplicity, this case has direct relevance to the generation of gravitational waves in inflation. 

\subsection{The Klein-Gordon Equation in Curved Spacetime}

Consider an arbitrary free scalar field, which we denote $\varphi$ to distinguish it from the inflaton field $\phi$. The Lagrangian for the field is
\begin{equation}
\label{eq:freeaction}
{\mathcal L} = \frac{1}{2} g^{\mu\nu} \partial_\mu \varphi \partial_\nu \varphi,
\end{equation}
and varying the action (\ref{eq:freeaction}) gives the Euler-Lagrange equation of motion 
\begin{equation}
\label{eq:freescalareom}
\frac{1}{\sqrt{-g}} \partial_\nu \left(g^{\mu\nu} \sqrt{-g} \partial_\mu \varphi\right) = 0.
\end{equation}
It will prove convenient to express the background FRW metric in conformal coordinates 
\begin{equation}
g_{\mu\nu} = a^2\left(\tau\right) \eta_{\mu\nu}
\end{equation}
instead of the coordinate-time metric (\ref{eq:coordinatemetric}) we used in Section (\ref{sec:InflationFromScalarFields}). Here $\tau$ is the conformal time and $\eta_{\mu\nu} = {\rm diag.}\left(1,-1,-1,-1\right)$ is the Minkowski metric. In conformal coordinates, the free scalar equation of motion (\ref{eq:freescalareom}) is
\begin{equation}
\label{eq:FRWfreescalareom}
\varphi'' + 2 \left(\frac{a'}{a}\right) \varphi' - \nabla^2 \varphi = 0,
\end{equation}
where $' = d/d\tau$ is a derivative with respect to {\it conformal} time. Note that unlike the case of the inflaton $\phi$, we are solving for perturbations and therefore retain the gradient term $\nabla^2 \varphi$. The field $\varphi$ is a decoupled spectator field evolving in a {\it fixed} cosmological background, and does not effect the time evolution of the scale factor $a\left(\tau\right)$. An example of such a field is gravitational waves. If we express the spacetime metric as an FRW background $g^{\rm FRW}_{\mu\nu}$ plus perturbation $\delta g_{\mu\nu}$, we can express the tensorial portion of the perturbation in general as a sum of two scalar degrees of freedom
\begin{eqnarray}
\label{eq:tensormetric}
&&\delta g_{0i} = \delta g_{i 0} = 0\cr
&&\delta g_{ij} = \frac{32 \pi}{\mpl} \left( \varphi_+ \hat e^{+}_{ij}  + \varphi_\times \hat e^{\times}_{ij}\right),
\end{eqnarray}
where $i, j = 1,2,3$, and $\hat e^{+,\times}_{ij}$ are longitudinal and transverse polarization tensors, respectively. It is left as an exercise for the reader to show that the scalars $\varphi_{+,\times}$ behave to linear order as free scalars, with equation of motion (\ref{eq:FRWfreescalareom}).

To solve the equation of motion (\ref{eq:FRWfreescalareom}), we first Fourier expand the field into momentum states $\varphi_k$,
\begin{equation}
\label{eq:classicalmodeexpansion}
\varphi\left(\tau,{\bf x}\right) = \int{\frac{d^3 k}{\left(2 \pi\right)^{3/2}} \left[\varphi_{\bf k}\left(\tau\right) b_{\bf k} e^{i {\bf k} \cdot {\bf x}} + \varphi_{\bf k}^* \left(\tau\right) b_{\bf k}^* e^{-i {\bf k} \cdot {\bf x}}\right]}.
\end{equation}
Note that the coordinates ${\bf x}$ are comoving coordinates, and the wavevector ${\bf k}$ is a comoving wavevector, which does not redshift with expansion. The proper wavevector is 
\begin{equation}
{\bf k}_{\rm prop} = {\bf k} / a\left(\tau\right).
\end{equation}
Therefore, the comoving wavenumber ${\bf k}$ is not itself dynamical, but is just a set of constants labeling a particular Fourier component.  The equation of motion for a single mode $\varphi_{\bf k}$ is 
\begin{equation}
\varphi_{\bf k}'' + 2 \left(\frac{a'}{a}\right) \varphi_{\bf k}' + k^2 \varphi_{\bf k} = 0.
\end{equation}
It is convenient to introduce a field redefinition
\begin{equation}
u_k \equiv a\left(\tau\right) \varphi_{\bf k}\left(\tau\right),
\end{equation}
and the mode function $u_k$ obeys a generalization of the Klein-Gordon equation to an expanding spacetime,
\begin{equation}
\label{eq:generaltensormode}
u_k'' + \left[k^2 - \frac{a''}{a}\right] u_k = 0.
\end{equation}
(We have dropped the vector notation ${\bf k}$ on the subscript, since the Klein-Gordon equation depends only on the magnitude of $k$.) 

Any mode with a fixed comoving wavenumber $k$ redshifts with time, so that early time corresponds to short wavelength (ultraviolet) and late time corresponds to long wavelength (infrared). The solutions to the mode equation show qualitatively different behaviors in the ultraviolet and infrared limits:
\begin{itemize}
\item{
{\it Short wavelength limit}, $k \gg a''/a$. In this case, the equation of motion is that for a conformally Minkowski Klein-Gordon field,
\begin{equation}
u_k'' + k^2 u_k = 0,
\end{equation}
with solution
\begin{equation}
\label{eq:UVmode}
u_k\left(\tau\right) = \frac{1}{\sqrt{2 k}}\left( A_k e^{-i k \tau} + B_k e^{i k \tau}\right).
\end{equation}
Note that this is in terms of {\it conformal} time and {\it comoving} wavenumber, and can only be identified with an exactly Minkowski spacetime in the ultraviolet limit.
}
\item{{\it Long wavelength limit}, $k \ll a''/a$. In the infrared limit, the mode equation becomes
\begin{equation}
a'' u_k = a u_k'',
\end{equation}
with the trivial solution 
\begin{equation}
\label{eq:genericIRsolution}
u_k \propto a\ \Rightarrow \varphi_k = {\rm const.}
\end{equation}
This illustrates the phenomenon of {\it mode freezing}: field modes $\varphi_k$ with wavelength longer than the horizon size cease to be dynamical, and asymptote to a constant, {\it nonzero} amplitude.\footnote{The second solution to this equation is a decaying mode, which is always subdominant in the infrared limit.} This is a quantitative expression of our earlier qualitative notion of particle creation at the cosmological horizon. The amplitude of the field at long wavelength is determined by the boundary condition on the mode, {\it i.e.} the integration constants $A_k$ and $B_k$. 
}
\end{itemize}
Therefore, all of the physics boils down to the question of how we set the boundary condition on field perturbations in the ultraviolet limit. This is fortunate, since in that limit the field theory describing the modes becomes approximately Minkowskian, and we know how to quantize fields in Minkowski Space. Once the integration constants are fixed, the behavior of the mode function $u_k$ is completely determined, and the long-wavelength amplitude of the perturbation can then be calculated without ambiguity. We next discuss quantization. 

\subsection{Quantization}
We have seen that the equation of motion for field perturbations approaches the usual Minkowski Space Klein-Gordon equation in the ultraviolet limit, which  corresponds to the limit of early time for a mode redshifting with expansion. We determine the boundary conditions for the mode function via canonical quantization. To quantize the field $\varphi_k$, we promote the Fourier coefficients in the classical mode expansion (\ref{eq:classicalmodeexpansion}) to annihilation and creation operators
\begin{equation}
b_{\bf k}\ \rightarrow \hat b_{\bf k},\ \ \ b^*_{\bf k}\ \rightarrow \hat b^\dag_{\bf k},
\end{equation}
with commutation relation
\begin{equation}
\label{eq:canonicalcommutator}
\left[\hat b_{\bf k}, \hat b^\dag_{\bf k'}\right] \equiv \delta^3\left({\bf k} - {\bf k}'\right).
\end{equation}
Note that the commutator in an FRW background is given in terms of {\it comoving} wavenumber, and holds whether we are in the short wavelength limit or not. In the short wavelength limit, this becomes equivalent to a Minkowski Space commutator. The quantum field $\varphi$ is then given by the usual expansion in operators $\hat b_{\bf k}$, $\hat b^\dag_{\bf k}$
\begin{equation}
\label{eq:quantummodeexpansion}
\varphi\left(\tau,{\bf x}\right) = \int{\frac{d^3 k}{\left(2 \pi\right)^{3/2}} \left[\varphi_{\bf k}\left(\tau\right) b_{\bf k} e^{i {\bf k} \cdot {\bf x}} + {\rm H.C.}\right]}
\end{equation}

The corresponding canonical momentum is
\begin{equation}
\Pi\left(\tau,{\bf x}\right) \equiv \frac{\delta {\mathcal L}}{\delta\left(\partial_0 \varphi\right)} = a^2\left(\tau\right) \frac{\partial \varphi}{\partial \tau}.
\end{equation}
It is left as an exercise for the reader to show that the canonical commutation relation
\begin{equation}
\label{eq:quantization}
\left[\varphi\left(\tau,{\bf x}\right), \Pi\left(\tau,{\bf x}'\right)\right] = i \delta^3\left({\bf x} - {\bf x}'\right)
\end{equation}
corresponds to a Wronskian condition on the mode $u_k$,
\begin{equation}
u_k \frac{\partial u^*_k}{\partial \tau} - u^*_k \frac{\partial u_k}{\partial \tau} = i,
\end{equation}
which for the ultraviolet mode function (\ref{eq:UVmode}) results in a condition on the integration constants
\begin{equation}
\label{eq:quantizationAB}
\left\vert A_k\right\vert^2 - \left\vert B_k\right\vert^2 = 1.
\end{equation}
This quantization condition corresponds to one of the two boundary conditions which are necessary to completely determine the solution. The second boundary condition comes from vacuum selection, {\it i.e.} our definition of which state corresponds to a zero-particle state for the system. In the next section, we discuss the issue of vacuum selection in detail.

\subsection{Vacuum Selection}
Consider a quantum field in Minkowski Space. The state space for a quantum field theory is a set
of states $\left\vert n({\bf k}_1),\ldots,n({\bf k}_i)\right\rangle$ representing
the number of particles with momenta ${\bf k}_1,\ldots,{\bf k}_i$. The creation and annihilation
operators $\hat a^{\dagger}_{\bf k}$ and ${\hat a_{\bf k}}$ act on these states by adding or subtracting a particle from the state:
\begin{eqnarray}
\hat a^{\dagger}_{\bf k} \left\vert n({\bf k})\right\rangle &&= \sqrt{n + 1} \left\vert n({\bf k}) + 1\right\rangle\cr
{\hat a_{\bf k}} \left\vert n({\bf k})\right\rangle &&= \sqrt{n} \left\vert n({\bf k}) - 1\right\rangle.
\end{eqnarray}
The ground state, or vacuum state of the space, is just the zero particle state:
\begin{equation}
{\hat a_{\bf k}} \left\vert 0 \right\rangle = 0.
\end{equation}
Note in particular that the vacuum state $\left\vert 0 \right\rangle$ is {\em not} equivalent
to zero. The vacuum is not nothing:
\begin{equation}
\left\vert 0 \right\rangle \neq 0.
\end{equation}

To construct a quantum field, we look at the familiar classical wave equation for a scalar field,
\begin{equation}
\label{eqminkowskiwaveequation}
\frac{\partial^2 \phi }{ \partial t^2} - \nabla^2 \phi = 0.
\end{equation}
To solve this equation, we decompose into Fourier modes $u_{\rm k}$,
\begin{equation}
\label{eqfourierexpansion}
\phi =  \int{d^3 k \left[a_{\bf k} u_{\bf k}(t) e^{i {\bf k}\cdot{\bf x}} + a^*_{\bf k} u^*_{\bf k}(t) e^{- i {\bf k}\cdot{\bf x}}\right]},
\end{equation}
where the mode functions $u_{\bf k}(t)$ satisfy the ordinary differential equation
\begin{equation}
\label{eqminkowskimode}
\ddot u_{\bf k} + k^2 u_{\bf k} = 0.
\end{equation}
This is a classical wave equation with a classical solution, and the Fourier coefficients
$a_{\rm k}$ are just complex numbers. The solution for the mode function is 
\begin{equation}
\label{eqminkowskimodefunction}
u_{\bf k} \propto e^{-i \omega_k t},
\end{equation}
where $\omega_k$ satisfies the dispersion relation
\begin{equation}
\omega_k^2 - {\bf k}^2 = 0.
\end{equation}
To turn this into a quantum field, we identify the
Fourier coefficients with creation and annihilation operators
\begin{equation}
a_{\bf k} \rightarrow \hat a_{\bf k},\ a^*_{\bf k} \rightarrow \hat a^{\dagger}_{\bf k},
\end{equation}
and enforce the commutation relations
\begin{equation}
\left[\hat a_{\bf k}, \hat a^{\dagger}_{\bf k'}\right] = \delta^3\left({\bf k} - {\bf k'}\right).
\end{equation}

This is the standard quantization of a scalar field in 
Minkowski Space, which should be familiar. But what probably is not familiar is that this solution has an interesting symmetry. Suppose we define a new mode function $u_{\bf k}$ which 
is a rotation of the solution (\ref{eqminkowskimodefunction}):
\begin{equation}
\label{eqrotatedmodefunction}
u_{\bf k} = A(k) e^{-i \omega t + i {\bf k} \cdot {\bf x}} + B(k) e^{i \omega t - i {\bf k} \cdot {\bf x}}.
\end{equation}
This is {\em also} a perfectly valid solution to the original wave equation 
(\ref{eqminkowskiwaveequation}), since it is just a superposition of the Fourier modes. But we 
can then re-write the quantum
field in terms of our original Fourier modes and new {\em operators} $\hat b_{\bf k}$ and 
$\hat b^{\dagger}_{\bf k}$ and the original Fourier modes $e^{i {\bf k} \cdot {\bf x}}$ as:
\begin{equation}
\phi =  \int{d^3 k \left[{\hat b_{\bf k}} e^{-i \omega t + i {\bf k}\cdot{\bf x}} + \hat b^{\dagger}_{\bf k}  e^{+ i \omega t - i {\bf k}\cdot{\bf x}}\right]},
\end{equation}
where the new operators $\hat b_{\bf k}$ are given in terms of the old operators ${\hat a_{\bf k}}$ by
\begin{equation}
\hat b_{\bf k} = A(k) \hat a_{\bf k} + B^*(k) \hat a^{\dagger}_{\bf k}.
\end{equation}
This is completely equivalent to our original solution (\ref{eqfourierexpansion}) as long as the
new operators satisfy the same commutation relation as the original operators,
\begin{equation}
\left[\hat b_{\bf k}, \hat b^{\dagger}_{\bf k'}\right] = \delta^3\left({\bf k} - {\bf k'}\right).
\end{equation}
This can be shown to place a condition on the coefficients $A$ and $B$,
\begin{equation}
\left\vert A\right\vert^2 - \left\vert B\right\vert^2 = 1.
\end{equation}
Otherwise, we are free to choose $A$ and $B$ as we please. 

This is just a standard property of linear differential equations: any linear combination of
solutions is itself a solution. But what does it mean physically? In one case, we have an
annihilation operator ${\hat a_{\bf k}}$ which gives zero when acting on a particular state which
we call the vacuum state:
\begin{equation}
{\hat a_{\bf k}} \left\vert 0_a \right\rangle = 0.
\end{equation}
Similarly, our rotated operator $\hat b_{\bf k}$ gives zero when acting on some state
\begin{equation}
\hat b_{\bf k} \left\vert 0_b\right\rangle = 0.
\end{equation}
The point is that the two ``vacuum'' states are not the same
\begin{equation}
\left\vert 0_a \right\rangle \neq \left\vert 0_b\right\rangle.
\end{equation}
From this point of view, we can define any state we wish to be the ``vacuum'' and build a
completely consistent quantum field theory based on this assumption. From another equally 
valid point of view this state will contain particles. How do we tell which is the 
{\em physical} vacuum state? To define the real vacuum, we have to consider the spacetime
the field is living in. For example, in regular special relativistic quantum field theory,
the ``true'' vacuum is the zero-particle state as seen by an inertial observer. Another
more formal way to state this is that we require the vacuum to be Lorentz symmetric. This
fixes our choice of vacuum $\left\vert 0\right\rangle$ and defines unambiguously our set
of creation and annihilation operators $\hat a$ and $\hat a^{\dagger}$. A consequence of this is that an {\em accelerated} observer in the Minkowski vacuum will think that the space is full
of particles, a phenomenon known as the Unruh effect \cite{Unruh:1976db}. The zero-particle
state for an accelerated observer is different than for an inertial observer. The case of an FRW spacetime is exactly analogous, except that the FRW equivalent of an inertial observer is an observer at rest in comoving coordinates. Since an FRW spacetime is asymptotically Minkowski in the ultraviolet limit, we choose the vacuum field which corresponds to the usual Minkowski vacuum in that limit,
\begin{equation}
u_k\left(\tau\right) \propto e^{-i k \tau}\ \Rightarrow A_k = 1,\ B_k = 0.
\end{equation}
This is known as the {\it Bunch-Davies} vacuum. This is not the only possible choice, although it is widely believed to be the most natural. The issue of vacuum ambiguity of inflationary perturbations is a subject which is extensively discussed in the literature, and is still the subject of controversy. It is known that the choice of vacuum is potentially sensitive to quantum-gravitational physics \cite{Hui:2001ce,Danielsson:2002kx,Easther:2002xe}, a subject which is referred to as {\it Trans-Planckian} physics \cite{Martin:2000xs,Niemeyer:2000eh,Kinney:2003xf}. For the remainder of our discussion, we will assume a Bunch-Davies vacuum. 

The key point is that quantization and vacuum selection together {\it completely} specify the mode function, up to an overall phase. This means that the amplitude of the mode once it has redshifted to long wavelength and frozen out is similarly determined. In the next section, we solve the mode equation at long wavelength for an inflationary background. 

\subsection{Exact Solutions and the Primordial Power Spectrum}

The exact form of the solution to Eq. (\ref{eq:generaltensormode}) depends on the evolution of the background spacetime, as encoded in $a\left(\tau\right)$, which in turn depends on the equation of state of the field driving inflation. We will consider the case where the equation of state is constant, which will {\it not} be the case in general for scalar field-driven inflation, but will nonetheless turn out to be a good approximation in the limit of a slowly rolling field. Generalizing Eq. (\ref{eq:deSittertau}) to the case of arbitrary equation of state parameter $\epsilon = {\rm const.}$, the conformal time can be written
\begin{equation}
\label{eq:inflconformaltime}
\tau = - \left(\frac{1}{a H}\right) \left(\frac{1}{1 - \epsilon}\right),
\end{equation}
and the Friedmann and Raychaudhuri Equations (\ref{eq:generalFRW}) give
\begin{equation}
\frac{a''}{a} = a^2 H^2 \left(2 - \epsilon\right),
\end{equation}
where a prime denotes a derivative with respect to conformal time. The conformal time, as in the case of de Sitter space, is negative and tending toward zero during inflation. (Proof of these relations is left as an exercise for the reader.) We can then write the mode equation (\ref{eq:generaltensormode}) as
\begin{equation}
u_k'' + \left[k^2 - a^2 H^2 \left(2 - \epsilon\right)\right] u_k = 0.
\end{equation}
Using Eq. (\ref{eq:inflconformaltime}) to write $a H$ in terms of the conformal time $\tau$,  the equation of motion becomes
\begin{equation}
\tau^2 \left(1 - \epsilon\right)^2 u_k'' + \left[\left(k \tau\right)^2 \left(1 - \epsilon\right)^2 - \left(2 - \epsilon\right)\right] u_k = 0. 
\end{equation}
This is a Bessel equation, with solution
\begin{equation}
\label{eq:generalmodesolution}
u_k \propto \sqrt{- k \tau} \left[J_\nu\left(-k \tau\right) \pm i Y_\nu\left(-k \tau\right)\right],
\end{equation}
where the index $\nu$ is given by:
\begin{equation}
\nu = \frac{3 - \epsilon}{2 \left(1 - \epsilon\right)}.
\end{equation}
The quantity $- k \tau$ has special physical significance, since from Eq. (\ref{eq:inflconformaltime}) we can write
\begin{equation}
\left(- k \tau\right) \left(1 - \epsilon\right) = \frac{k}{a H},
\end{equation}
where the quantity $\left(k / a H\right)$ expresses the wavenumber $k$ in units of the comoving horizon size $d_{\rm H} \sim (a H)^{-1}$. Therefore, the short wavelength limit is $-k \tau \rightarrow -\infty$, or $\left(k / a H\right) \gg 1$. The long-wavelength limit is $-k \tau \rightarrow 0$, or $\left(k / a H\right) \ll 1$. 

\begin{figure}
\begin{center}
%--%\psfig{file=modefunction.eps,width=4.5in}
\includegraphics[width=4.5in]{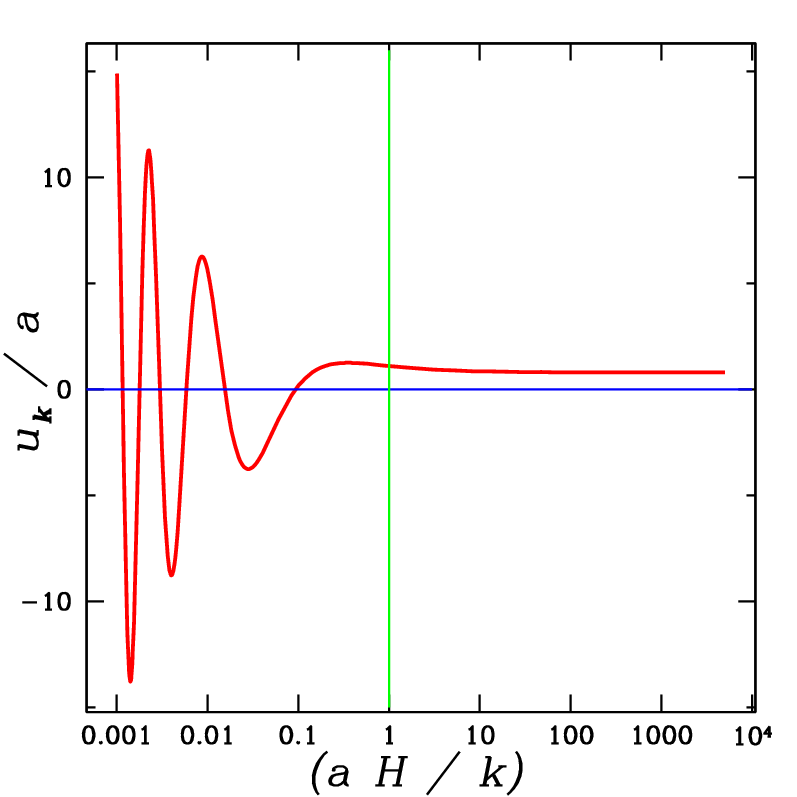}
\end{center}
\caption{The normalized mode function in de Sitter space, showing oscillatory behavior on subhorizon scales $k / a H > 1$, and mode freezing on superhorizon scales, $k / a H < 1$. }
\label{fig:modefunction}
\end{figure}

The simple case of de Sitter space ($p = - \rho$) corresponds to the limit $\epsilon = 0$, so that the Bessel index is $\nu = 3/2$ and the mode function (\ref{eq:generalmodesolution}) simplifies to
\begin{equation}
u_k \propto \left(\frac{k \tau - i}{k\tau}\right) e^{\pm i k \tau}.
\end{equation}
In the short wavelength limit, $\left( -k \tau\right) \rightarrow -\infty$, the mode function is given, as expected, by 
\begin{equation}
u_k \propto e^{\pm i k \tau}.
\end{equation}
Selecting the Bunch-Davies vacuum gives $u_k \propto e^{i k \tau}$, and canonical quantization fixes the normalization,
\begin{equation}
u_k = \frac{1}{\sqrt{2 k}} e^{-i k \tau}.
\end{equation}
Therefore, the fully normalized exact solution is
\begin{equation}
\label{eq:deSittermode}
u_k = \frac{1}{\sqrt{2 k}} \left(\frac{k \tau - i}{k\tau}\right) e^{-i k \tau}.
\end{equation}
This solution has no free parameters aside from an overall phase, and is valid at {\it all} wavelengths, including after the mode has been redshifted outside of the horizon and becomes non-dynamical, or ``frozen''. In the long wavelength limit, $-k \tau \rightarrow 0$, the mode function (\ref{eq:deSittermode}) becomes
\begin{equation}
u_k \rightarrow \frac{1}{\sqrt{2 k}} \left(\frac{i}{\left(- k \tau\right)}\right) = \frac{i}{2 k}\left(\frac{a H}{k}\right) \propto a,
\end{equation}
consistent with the qualitative result (\ref{eq:genericIRsolution}). Therefore the field amplitude $\varphi_k$ is given by
\begin{equation}
\left\vert \varphi_k\right\vert = \left\vert \frac{u_k}{a}\right\vert \rightarrow \frac{H}{\sqrt{2} k^{3/2}} = {\rm const.}
\end{equation}
The quantum mode therefore displays the freezeout behavior we noted qualitatively above (Fig. \ref{fig:modefunction}).  
%%%%%%%%%%%%%%%%%%%%%%%%%%%%%%%%%%%%%%%%%%%%%%%%%%%%%%%%%%%%%%%%%%%%%%%%%%%%%%%%%%%%%%%%%%%%
%%%%%%%%%%%%%%%%%%%%%%%%%%%%%%%%%%%%%%%%%%%%%%%%%%%%%%%%%%%%%%%%%%%%%%%%%%%%%%%%%%%%%%%%%%%%
The amplitude of quantum fluctuations is conventionally expressed in terms of the two-point correlation function of the field $\varphi$. It is left as an exercise for the reader to show that the vacuum two-point correlation function is given by
\begin{eqnarray}
\left\langle 0 \left\vert \varphi\left(\tau,{\bf x}\right)\varphi\left(\tau,{\bf x}'\right) \right\vert 0 \right\rangle &=& \int{\frac{d^3 k}{\left(2 \pi\right)^3} \left\vert\frac{u_k}{a}\right\vert^2 e^{i {\bf k}\cdot\left({\bf x} - {\bf x}'\right)}}\cr
&=& \int{\frac{d k}{k} P\left(k\right) e^{i {\bf k}\cdot\left({\bf x} - {\bf x}'\right)}},
\end{eqnarray}
where the {\it power spectrum} $P\left(k\right)$ is defined as
\begin{equation}
P\left(k\right) \equiv \left(\frac{k^3}{2 \pi^2}\right) \left\vert\frac{u_k}{a}\right\vert^2 \longrightarrow \left(\frac{H}{2 \pi}\right)^2,\ -k \tau \rightarrow 0.
\end{equation}
The power per logarithmic interval $k$ in the field fluctuation is then given in the long wavelength limit by the Hubble parameter $H = {\rm const.}$ This property of scale invariance is exact in the de Sitter limit.

In a more general model, the spacetime is only {\it approximately} de Sitter, and we expect that the power spectrum of field fluctuations will only be approximately scale invariant. It is convenient to express this dynamics in terms of the equation of state parameter $\epsilon$, 
\begin{equation}
\epsilon = \frac{1}{H}\frac{d H}{d N}.
\end{equation}
We must have $\epsilon < 1$ for inflation, and for a slowly rolling field $\left\vert \eta\right\vert \ll 1$ means that $\epsilon$ will also be slowly varying, $\epsilon \simeq {\rm const.}$ It is straightforward to show that for $\epsilon = {\rm const.} \neq 0$ that:
\begin{itemize}
\item{
The Bunch-Davies vacuum corresponds to the positive mode of Eq. (\ref{eq:generalmodesolution}),
\begin{equation}
\label{eq:generalBDmode}
u_k \propto \sqrt{- k \tau} \left[J_\nu\left(-k \tau\right) + i Y_\nu\left(-k \tau\right)\right].
\end{equation}
}
\item{
Quantization fixes the normalization as
\begin{equation}
\label{eq:tensormodeexactsoln}
u_k = \frac{1}{2}\sqrt{\frac{\pi}{k}} \sqrt{- k \tau} \left[J_\nu\left(-k \tau\right) + i Y_\nu\left(-k \tau\right)\right].
\end{equation}
}
\item{
The power spectrum in the long-wavelength limit $k / a H \rightarrow 0$ is a power law in $k$:
\begin{equation}
\label{eq:tensorinfrared}
\left[P\left(k\right)\right]^{1/2} \longrightarrow 2^{\nu - 3/2} \frac{\Gamma\left(\nu\right)}{\Gamma\left(3/2\right)} \left(1 - \epsilon\right) \left(\frac{H}{2 \pi}\right) \left(\frac{k}{a H \left(1 - \epsilon\right)}\right)^{3/2 - \nu},
\end{equation}
where $\Gamma\left(\nu\right)$ is a gamma function, and
\begin{equation}
\nu = \frac{3 - \epsilon}{2 \left(1 - \epsilon\right)}.
\end{equation}
}
\end{itemize}
Proof is left as an exercise for the reader.\footnote{Note that the quantization condition (\ref{eq:quantization}) can be applied to the solution (\ref{eq:generalmodesolution}) exactly, resulting in the normalization condition (\ref{eq:quantizationAB}), without approximating the solution in the short-wavelength limit!} Note that in the case $\epsilon = {\rm const.}$, both the background and perturbation equations are {\it exactly} solvable. 

We can use these solutions as approximate solutions in the more general slow roll case, where $\epsilon \ll 1 \simeq {\rm const.}$, so that the dependence of the power spectrum on $k$ is approximately a power-law,
\begin{equation}
P\left(k\right) \propto k^{n},
\end{equation}
with spectral index
\begin{equation}
\label{eq:freescalarn}
n = 3 - 2 \nu = 3 -  \frac{3 - \epsilon}{1 - \epsilon} \simeq -2 \epsilon.
\end{equation}
Equation (\ref{eq:tensorinfrared}) is curious, however, because it does not obviously exhibit complete mode freezing at long wavelength, since $a$ and $H$ both depend on time. We can show that $P\left(k\right)$ does in fact approach a time-dependent value at long wavelength by evaluating
\begin{eqnarray}
&&\frac{d}{d N} \left[H \left(\frac{k}{a H}\right)^{3/2 - \nu}\right] = \frac{d}{d N} \left[H \left(\frac{k}{a H}\right)^{-\epsilon / \left(1 - \epsilon\right)}\right]\cr
&&= H \epsilon \left(\frac{k}{a H}\right)^{-\epsilon / \left(1 - \epsilon\right)} - \frac{\epsilon}{1 - \epsilon} \left(\frac{k}{a H}\right)^{-\epsilon / \left(1 - \epsilon\right) - 1} \left(\frac{k}{a H} - \frac{\epsilon k}{a H}\right)\cr 
&&= 0,
\end{eqnarray}
which can be easily shown using $a \propto e^{-N}$ and $H \propto e^{\epsilon N}$. That is, the time-dependent quantities $a$ and $H$ in Eq. (\ref{eq:tensorinfrared}) are combined in such a way as to form an {\it exactly} conserved quantity. Since it is conserved, we are free to evaluate it at any time (or value of $a H$) that we wish. It is conventional to evaluate the power spectrum at {\it horizon crossing}, or at $a H = k$, so that
\begin{equation}
\label{eq:horizoncrossing}
P^{1/2}\left(k\right) \simeq \left(\frac{H}{2 \pi}\right)_{k = a H},
\end{equation}
where we have approximated the $\nu$-dependent multiplicative factor as order unity. \footnote{This is {\it not} the value of the scalar field power spectrum at the moment the mode is physically crossing outside the horizon, as is often stated in the literature: it is the value of the power spectrum in the asymptotic long-wavelength limit. It is easy to show from the exact solution (\ref{eq:tensormodeexactsoln}) that the mode function is still evolving with time as it crosses the horizon at $k = a H$, and the asymptotic amplitude differs from the amplitude at horizon crossing by about a factor of two. See Ref. \cite{Kinney:2005vj} for a more detailed discussion of this point.}  

It is straightforward to calculate the spectral index (\ref{eq:freescalarn}) directly from the horizon crossing expression (\ref{eq:horizoncrossing}) by using
\begin{equation}
a \propto e^{-N},\ H \propto e^{\epsilon N},
\end{equation}
so that we can write derivatives in $k$ at horizon crossing as derivatives in the number of e-folds $N$,
\begin{equation}
d\ln{k}\vert_{k = a H} = d\ln{\left(a H\right)} = \frac{1}{a H} \frac{d \left(a H\right)}{d N} dN = \left(\epsilon - 1\right) d N.
\end{equation}
The spectral index is then, to lowest order in slow roll
\begin{eqnarray}
n = \frac{d \ln{P\left(k\right)}}{d \ln{k}} &&= \frac{k}{H^2}\frac{d H^2}{d k}\bigg\vert_{k = a H}\cr  
&&= \frac{1}{ H^{2} \left(\epsilon - 1\right)} \frac{d H^2}{d N}\cr 
&&= \frac{2 \epsilon}{ \left(\epsilon - 1\right)}\cr
&&\simeq - 2 \epsilon,
\end{eqnarray}
in agreement with (\ref{eq:freescalarn}). Note that we are rather freely changing variables from the wavenumber $k$ to the comoving horizon size $(a H)^{-1}$ to the number of e-folds $N$. As long as the cosmological evolution is monotonic, these are all different ways of measuring time: the time when a mode with wavenumber $k$ exits the horizon, the time at which the horizon is a particular size, the number of e-folds $N$ and the field value $\phi$ are all effectively just different choices of a clock, and we can switch from one to another as is convenient. For example, in the slow roll approximation, the Hubble parameter $H$ is just a function of $\phi$, $H \propto \sqrt{V\left(\phi\right)}$. Because of this, it is convenient to define $N\left(k\right)$ to be the number of e-folds (\ref{eq:srN}) when a mode with wavenumber $k$ crosses outside the horizon, and $\phi_N\left(k\right)$ to be the field value $N\left(k\right)$ e-folds before the end of inflation. Then the power spectrum can be written equivalently as {\it either} a function of $k$ or of $\phi$:
\begin{equation}
P^{1/2}\left(k\right) = \left(\frac{H}{2 \pi}\right)_{k = a H} = \left(\frac{H}{2 \pi}\right)_{\phi = \phi_N\left(k\right)} \simeq \sqrt{\frac{2 V\left(\phi_N\right)}{3 \pi \mpl^2}}.
\end{equation}
Wavenumbers $k$ are conventionally normalized in units of $h {\rm Mpc}^{-1}$ as measured in the {\it current} universe.  We can relate $N$ to scales in the current universe by recalling that modes which are of order the horizon size in the universe today, $k \sim a_0 H_0$, exited the horizon during inflation when $N = \left[46,60\right]$, so that we can calculate the amplitude of perturbations at the scale of the CMB quadrupole today by evaluating the power spectrum for field values between $\phi_{46}$ and $\phi_{60}$. 

One example of a free scalar in inflation is gravitational wave modes, where the transverse and longitudinal polarization states of the gravity waves evolve as independent scalar fields. Using Eq. (\ref{eq:tensormetric}), we can then calculate the power spectrum in gravity waves (or {\it tensors}) as the sum of the two-point correlation functions for the separate polarizations:
\begin{equation}
P_T = \left\langle \delta g_{ij}^{2}\right\rangle = 2 \times \frac{32}{\mpl^2} \left\langle \varphi^2 \right\rangle = \frac{16 H^2}{\pi \mpl^2} \propto k^{n_T},
\end{equation}
with spectral index
\begin{equation}
n_T = - 2 \epsilon.
\end{equation}
If the amplitude is large enough, such a spectrum of primordial gravity waves will be observable in the cosmic microwave background anisotropy and polarization, or be directly detectable by proposed experiments such as Big Bang Observer \cite{Smith:2005mm,Friedman:2006zt}. 

The second type of perturbation generated during inflation is perturbations in the density of the universe, which are the dominant component of the CMB anisotropy $\delta T / T \sim \delta \rho / \rho \sim 10^{-5}$, and are responsible for structure formation. Density, or {\it scalar} perturbations are more complicated than tensor perturbations because they are generated by quantum fluctuations in the inflaton field itself: since the background energy density is dominated by the inflaton, fluctuations of the inflaton up or down the potential generate perturbations in the density. The full calculation requires self-consistent General Relativistic perturbation theory, and is presented in Appendix A. Here we simply state the result: Perturbations in the inflaton field $\delta\phi \simeq H / 2 \pi$ generate density perturbations with power spectrum
\begin{equation}
P_{\cal R}\left(k \right) = \left(\frac{\delta N}{\delta \phi} \delta\phi\right)^2 = \frac{H^2}{\pi \mpl^2 \epsilon}\bigg\vert_{k = a H} \propto k^{n_S - 1},
\end{equation}
where $N$ is the number of e-folds. Scalar perturbations are therefore enhanced relative to tensor perturbations by a factor of $1 / \epsilon$. The scalar power spectrum is also an approximate power-law, with spectral index
\begin{equation}
n_S - 1=  \frac{\epsilon}{H^{2} \left(\epsilon - 1\right)} \frac{d}{d N}\left(\frac{H^2}{\epsilon}\right) \simeq - 4 \epsilon + 2 \eta,
\end{equation}
where $\eta$ is the second slow roll parameter (\ref{eq:defeta}). Therefore, for any particular choice of inflationary potential, we have four measurable quantities: the amplitudes $P_T$ and $P_{\cal R}$ of the tensor and scalar power spectra, and their spectral indices $n_T$ and $n_S$. However, not all of these parameters are independent. In particular, the ratio $r$ between the scalar and tensor amplitudes is given by the parameter $\epsilon$, as is the tensor spectral index $n_T$:
\begin{equation}
\label{eq:consistency}
r \equiv \frac{P_T}{P_S} = 16 \epsilon = - 8 n_T.
\end{equation}
This relation is known as the {\it consistency condition} for single-field slow roll inflation, and is in principle testable by a sufficiently accurate measurement of the primordial perturbation spectra. 

In the next section, we apply these results to our example $\lambda \phi^4$ potential and calculate the inflationary power spectra. 

\subsection{Example: $\lambda \phi^4$}

For the case of our example model with $V\left(\phi\right) = \lambda \phi^4$, it is now straightforward to calculate the scalar and tensor perturbation spectra. We express the normalization of the power spectra as a function of the number of e-folds $N$ by
\begin{eqnarray}
P_{\cal R}^{1/2} &=& \frac{H}{\mpl \sqrt{\pi \epsilon}}\bigg\vert_{\phi=\phi_N}\cr
&=& \frac{4 \sqrt{24 \pi}}{3 \mpl^3} \frac{\left[V\left(\phi_N\right)\right]^{3/2}}{V'\left(\phi_N\right)}\cr
&=& \frac{24 \pi}{3} \left(\frac{N + 1}{\pi}\right) \lambda^{1/2} \sim 10^{-5},
\end{eqnarray}
where we have used the slow roll expressions for $H$ (\ref{eq:srHubble}) and $\epsilon$ (\ref{eq:srepsilon}) and Eq. (\ref{eq:phi4N}) for $\phi_N$. For perturbations about the current size of our horizon, $N = 60$, and CMB normalization forces the self-coupling to be very small,
\begin{equation}
\lambda \sim 10^{-15}.
\end{equation}
The presence of an extremely small parameter is not peculiar to the $\lambda \phi^4$ model, but is generic, and is referred to as the {\it fine tuning} problem for inflation. 

We can similarly calculate the tensor amplitude
\begin{equation}
P_T^{1/2} = \frac{4 H}{\mpl \sqrt{\pi}},
\end{equation}
which is usually expressed in terms of the tensor/scalar ratio
\begin{eqnarray}
r &=&  16 \epsilon\left(\phi_N\right) =\frac{\mpl}{\pi} \left(\frac{V'\left(\phi_N\right)}{V\left(\phi_N\right)}\right)^2 \cr
&=& \frac{16}{\pi} \left(\frac{\mpl}{\phi_N}\right)^2 = \frac{16}{N + 1} \simeq 0.26,
\end{eqnarray}
where we have again taken $N = 60$. For this particular model, the power in gravitational waves is large, about a quarter of the power in scalar perturbations. This is {\it not} generic, but is quite model-dependent. Some choices of potential predict large tensor contributions (where ``large'' means of order 10\%), and other choices of potential predict very tiny tensor contributions, well below 1\%. 

The tensor spectral index $n_T$ is fixed by the consistency condition (\ref{eq:consistency}), but the scalar spectral index $n_S$ is an independent parameter because of its dependence on $\eta$:
\begin{equation}
n = 1 - 4 \epsilon\left(\phi_N\right) + 2 \eta\left(\phi_N\right),
\end{equation}
where
\begin{equation}
\epsilon\left(\phi_N\right) = \frac{1}{N + 1},
\end{equation}
and
\begin{eqnarray}
\eta\left(\phi_N\right) &=& \frac{\mpl^2}{8 \pi} \left[\frac{V''\left(\phi_N\right)}{V\left(\phi_N\right)} - \frac{1}{2} \left(\frac{V'\left(\phi_N\right)}{V\left(\phi_N\right)}\right)^2 \right]\cr
&=& \frac{\mpl^2}{8 \pi} \left[\frac{12}{\phi_N^2} - \frac{8}{\phi_N^2}\right]\cr
&=& \frac{1}{2 \pi} \left(\frac{\mpl}{\phi_N}\right)^2 = \frac{1}{2 \left(N + 1\right)}.
\end{eqnarray}
The spectral index is then
\begin{equation}
n =  1 - \frac{3}{N + 1} \simeq 0.95.
\end{equation}
Note that we have assumed slow roll from the beginning in the calculation without {\it a priori} knowing that it is a good approximation for this choice of potential. However, at the end of the day it is clear that the slow roll ansatz was a good one, since $\epsilon$ and $\eta$ are both of order $0.01$. 

Finally, we note that the energy density during inflation is characterized by a mass scale
\begin{equation}
\rho^{1/4} \sim \Lambda \sim \lambda^{1/4} \mpl \sim 10^{15}\ {\rm GeV},
\end{equation}
about the scale for which we expect Grand Unification to be important. This interesting coincidence suggests that the physics of inflation may be found in Grand Unified Theories (GUTs). Different choices of potential $V\left(\phi\right)$ will give different values for the amplitudes and shapes of the primordial power spectra. Since the normalization is fixed by the CMB to be $P_{\cal R} \sim 10^{-5}$, the most useful observables for distinguishing among different potentials are the scalar/tensor ratio $r$ and the scalar spectral index $n_S$. In single-field inflationary models, the tensor spectral index is fixed by the consistency condition (\ref{eq:consistency}), and is therefore not an independent parameter. The consistency condition can therefore be taken to be a {\it prediction} of single-field inflation, which is in principle verifiable by observation. In practice, this is very difficult, since it involves measuring not just the amplitude of the gravitational wave power spectrum, but also its {\it shape}. We will see in Section \ref{sec:ObservationalConstraints} that current data place only a rough upper bound on the tensor/scalar ratio $r$, and it is highly unlikely that any near-future measurement of primordial gravitational waves will be accurate enough to constrain $n_T$ well enough to test the consistency condition. In the next section, we discuss current observational constraints on the form of the inflationary potential. 

\section{Observational Constraints}
\label{sec:ObservationalConstraints}

Our simple picture of inflation generated by single, minimally coupled scalar field makes a set of very definite predictions for the form of primordial cosmological fluctuations:
\begin{itemize}
\item{Gaussianity: since the two-point correlation function of a free scalar field $\left\langle \varphi^2\right\rangle$ is Gaussian, cosmological perturbations generated in a single-field inflation model will by necessity also form a Gaussian random distribution.}
\item{Adiabaticity: since there is only one order parameter $\phi$ governing the generation of density perturbations, we expect the perturbations in all the components of the cosmological fluid (baryons, dark matter, neutrinos) to be {\it in phase} with each other. Such a case is called {\it adiabatic}. If one or more components fluctuates out of phase with others, these are referred to as {\it isocurvature} modes. Single-field inflation predicts an absence of isocurvature fluctuations.}
\item{Scale invariance: In the limit of de Sitter space, fluctuations in any quantum field are exactly scale invariant, $n = 1$, as a result of the fact that the Hubble parameter is exactly constant. Since slow-roll inflation is quasi-de Sitter, we expect the perturbation spectra to be nearly, but not exactly scale invariant, with $\left\vert n_S - 1\right\vert = \left\vert  2 \eta - 4 \epsilon \right\vert \ll 1$.}
\item{Scalar perturbations dominate over tensor perturbations, $r = 16 \epsilon$.}
\end{itemize}

Furthermore, given a potential $V\left(\phi\right)$, we have a ``recipe'' for calculating the form of the primordial power spectra generated during inflation:
\begin{enumerate}
\item{
Calculate the field value at the end of inflation $\phi_e$ from 
\begin{equation}
\epsilon\left(\phi_e\right) = \frac{\mpl^2}{16 \pi} \left(\frac{V'\left(\phi_e\right)}{V\left(\phi_e\right)}\right)^2 = 1.
\end{equation}
}
\item{
Calculate the field value $N$ e-folds before the end of inflation $\phi_N$ by integrating backward on the potential from $\phi = \phi_e$,
\begin{equation}
N = \frac{2 \sqrt{\pi}}{\mpl} \int_{\phi_e}^{\phi_N}{\frac{d\phi'}{\sqrt{\epsilon\left(\phi'\right)}}}.
\end{equation}
}
\item{
Calculate the normalization of the scalar power spectrum by
\begin{equation}
P_{\cal R}^{1/2} = \frac{H}{\mpl \sqrt{\pi \epsilon}}\bigg\vert_{\phi=\phi_N} \sim 10^{-5},
\end{equation}
where the CMB quadrupole corresponds to roughly $N = 60$. A more accurate calculation includes the uncertainty in the reheat temperature, which gives a range $N \simeq \left[46,60\right]$, and a corresponding uncertainty in the observable parameters.
}
\item{
Calculate the tensor/scalar ratio $r$ and scalar spectral index $n_S$ at $N = [46,60]$ by 
\begin{equation}
r = 16 \epsilon\left(\phi_N\right),
\end{equation}
and
\begin{equation}
n_s = 1 - 4 \epsilon\left(\phi_N\right)  + 2 \eta\left(\phi_N\right),
\end{equation}
where the second slow roll parameter $\eta$ is given by:
\begin{equation}
\eta\left(\phi_N\right) = \frac{\mpl^2}{8 \pi} \left[\frac{V''\left(\phi_N\right)}{V\left(\phi_N\right)} - \frac{1}{2} \left(\frac{V'\left(\phi_N\right)}{V\left(\phi_N\right)}\right)^2 \right].
\end{equation}
}
\end{enumerate}
The key point is that the scalar power spectrum $P_{\cal R}$ and the tensor power spectrum $P_T$ are both completely determined by the choice of potential $V\left(\phi\right)$.\footnote{Strictly speaking, this is true only for scalar fields with a canonical kinetic term, where the speed of sound of perturbations is equal to the speed of light. More complicated scenarios such as DBI inflation \cite{Silverstein:2003hf} require specification of an extra free function, the speed of sound $c_S\left(\phi\right)$, to calculate the power spectra. For constraints on this more general class of models, see Ref. \cite{Agarwal:2008ah}.} Therefore, if we measure the primordial perturbations in the universe accurately enough, we can in principle constrain the form of the inflationary potential. This is extremely exciting, because it gives us a very rare window into physics at extremely high energy, perhaps as high as the GUT scale or higher, far beyond the reach of accelerator experiments such as the Large Hadron Collider. 

\begin{figure}
\begin{center}
%--%\psfig{file=largefield.eps,width=4.5in}
\includegraphics[width=4.5in]{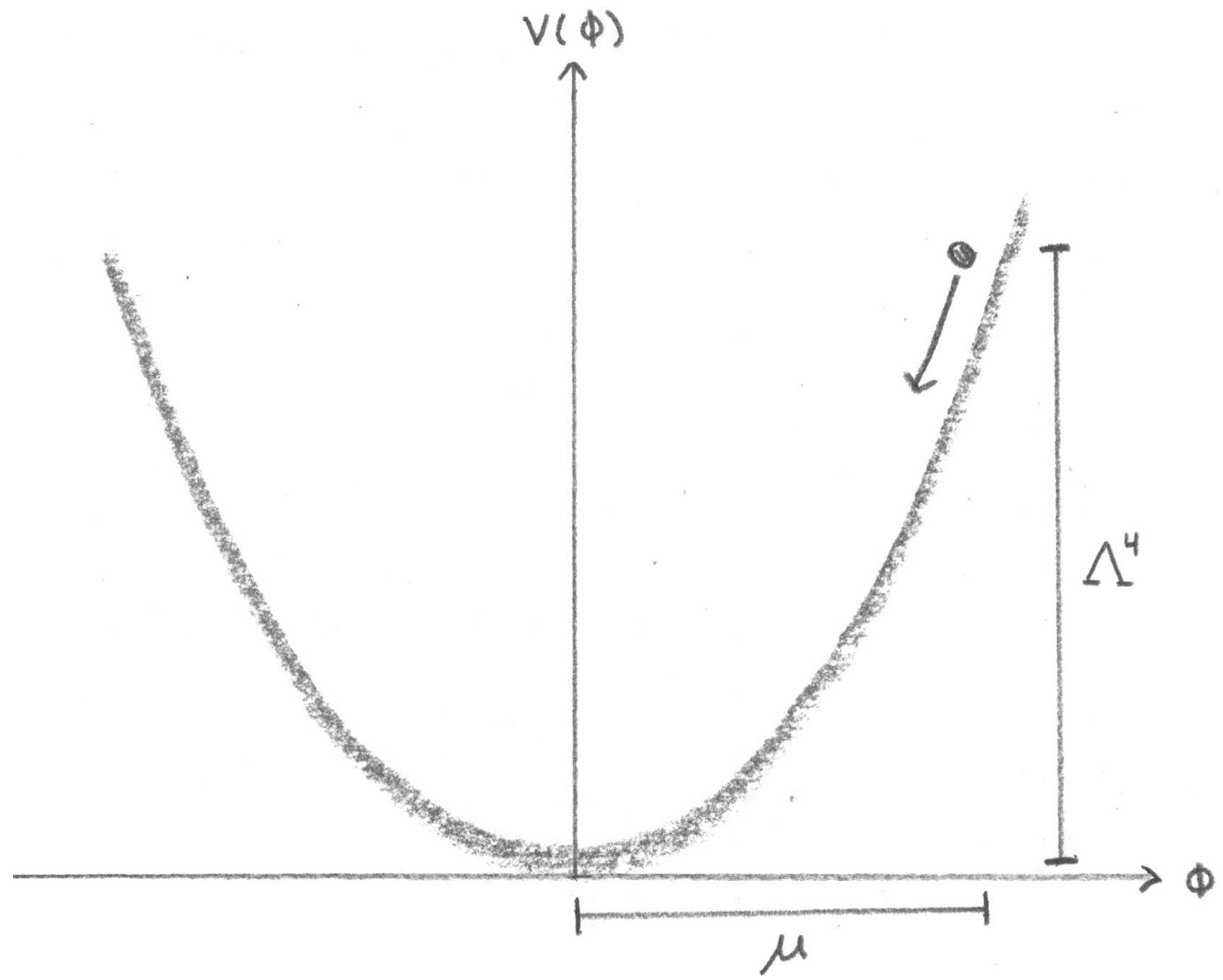}
\end{center}
\caption{A schematic of a large-field potential.}
\label{fig:largefield}
\end{figure}

\begin{figure}
\begin{center}
%--%\psfig{file=smallfield.eps,width=4.5in}
\includegraphics[width=4.5in]{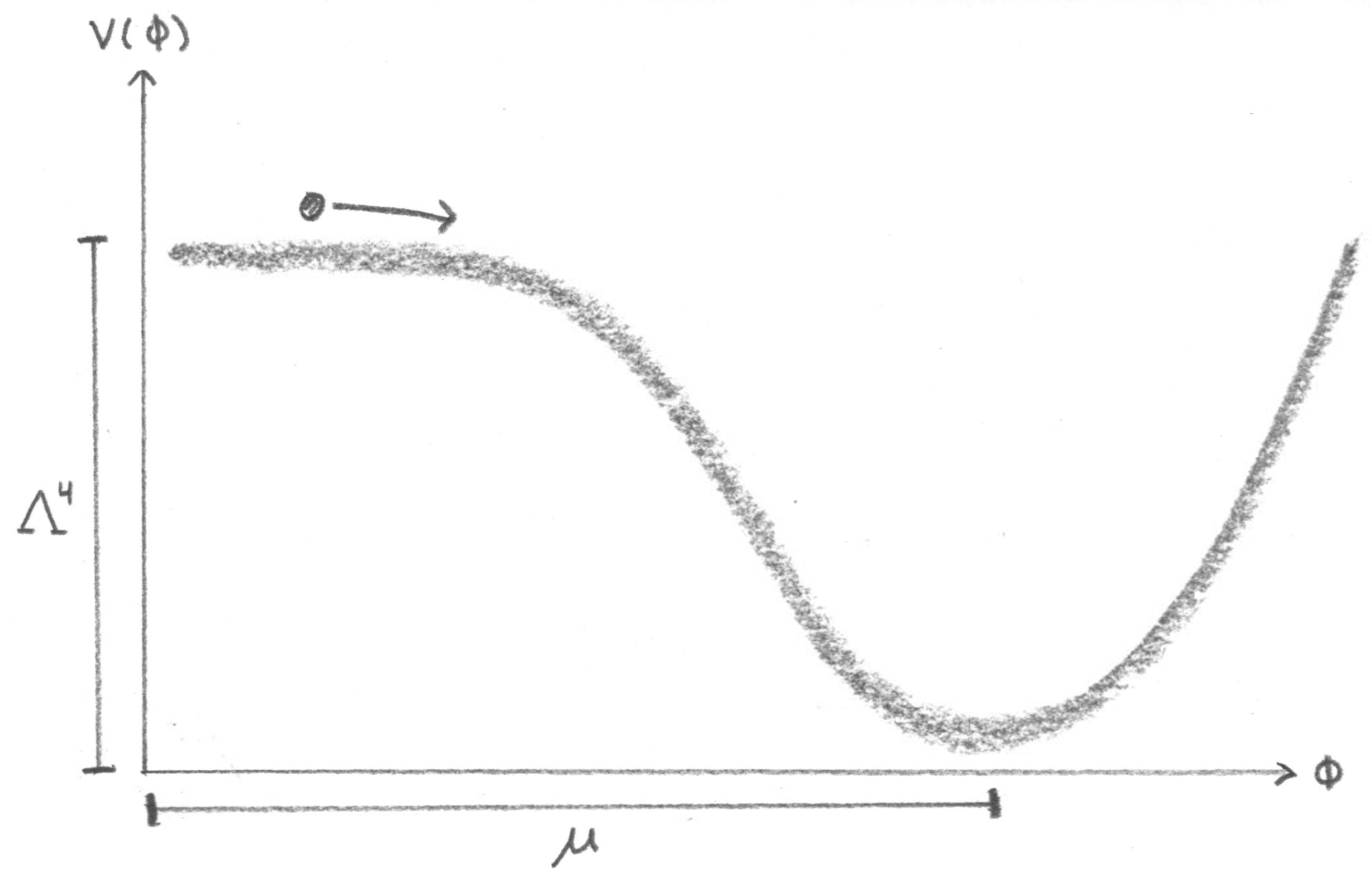}
\end{center}
\caption{A schematic of a small-field potential.}
\label{fig:smallfield}
\end{figure}

\begin{figure}
\begin{center}
%--%\psfig{file=hybrid.eps,width=4.5in}
\includegraphics[width=4.5in]{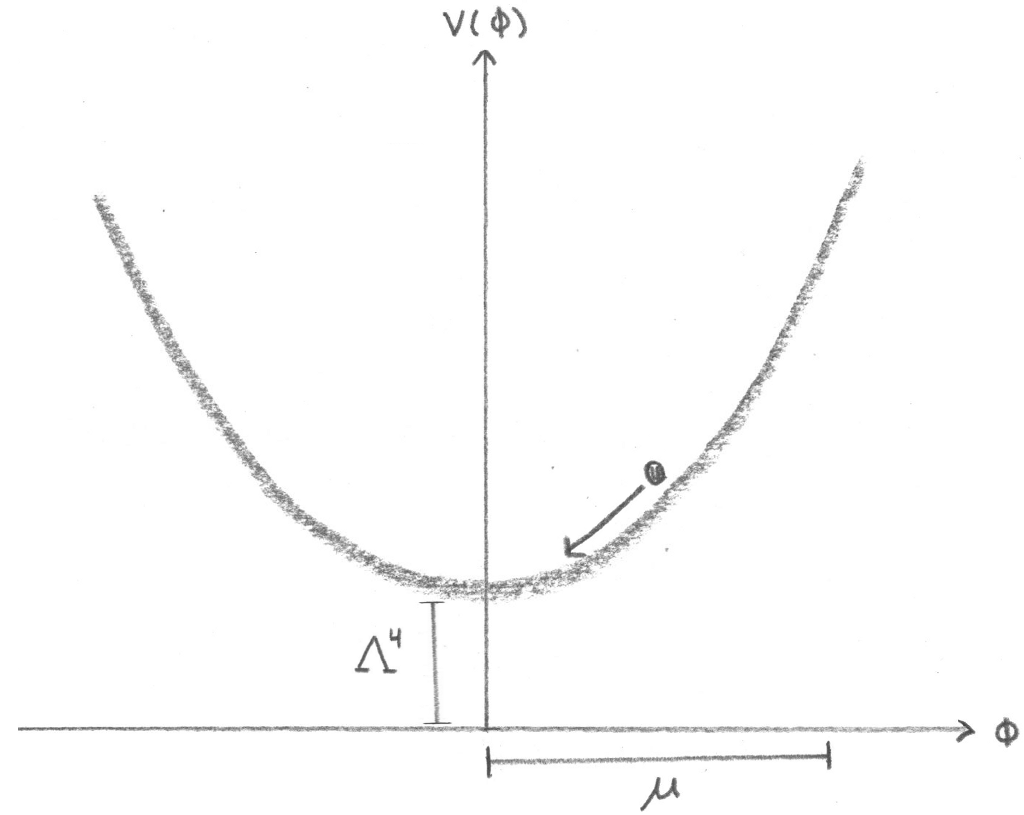}
\end{center}
\caption{A  schematic of a hybrid potential.}
\label{fig:hybrid}
\end{figure}

It is convenient to divide the set of possible single-field potentials into a few basic types \cite{Dodelson:1997hr}:
\begin{itemize}
\item{{\it Large-field potentials} (Fig. \ref{fig:largefield}). These are the simplest potentials one might imagine, with potentials of the form $V\left(\phi\right) = m^2 \phi^2$, or our example case, $V\left(\phi\right) = \lambda \phi^4$. Another widely-noted example of this type of model is inflation on an exponential potential, $V\left(\phi\right) = \Lambda^4 \exp{\left(\phi/\mu\right)}$, which has the useful property that both the background evolution and the perturbation equations are exactly solvable. In the large-field case, the field is displaced from the vacuum at the origin by an amount of order $\phi \sim \mpl$ and rolls down the potential toward the origin. Large-field models are typically characterized by a ``red'' spectral index $n_S < 1$, and a substantial gravitational wave contribution, $r \sim 0.1$.} 
\item{{\it Small-field potentials} (Fig. \ref{fig:smallfield}). These are potentials characteristic of spontaneous symmetry breaking phase transitions, where the field rolls off an unstable equilibrium with $V'\left(\phi\right) = 0$ toward a displaced vacuum. Examples of small-field inflation include a simple quadratic potential, $V\left(\phi\right) = \lambda \left(\phi^2 - \mu^2\right)^2$, inflation from a pseudo-Nambu-Goldstone boson or a shift symmetry in string theory (called {\it Natural Inflation}) with a potential typically of the form $V\left(\phi\right) = \Lambda^4 \left[1 + \cos{\left(\phi / \mu\right)}\right]$, or Coleman-Weinberg potentials, $V\left(\phi\right) = \lambda \phi^4 \ln\left(\phi\right)$. Small-field models are characterized by a red spectral index $n < 1$, and a small tensor/scalar ratio, $r \leq 0.01$.}
\item{{\it Hybrid potentials} (Fig. \ref{fig:hybrid}). A third class of models are potentials for which there is a residual vacuum energy when the field is at the minimum of the potential, for example a potential like $V\left(\phi\right) = \lambda \left(\phi^2 + \mu^2\right)^2$. In this case, inflation will continue {\it forever}, so additional physics is required to end inflation and initiate reheating. The {\it hybrid} mechanism, introduced by Linde \cite{Linde:1993cn}, solves this problem by adding a second field coupled to the inflaton which is stable for $\phi$ large, but becomes unstable at a critical field value $\phi_c$ near the minimum of $V\left(\phi\right)$ . During inflation, however, only $\phi$ is dynamical, and these models are effectively single-field. Typical models of this type predict negligible tensor modes, $r \ll 0.01$ and a ``blue'' spectrum, $n_S > 1$, which is disfavored by the data, and we will not discuss them in more detail here. (Ref. \cite{Komatsu:2008hk} contains a good discussion of current limits on general hybrid models.) Note also that such potentials will also support large-field inflation if the field is displaced far enough from its minimum. }
\end{itemize}
%%%%%%%%%%%%%%%%%%%%%%%%%%%%%%%%%%%%%%%%%%%%%%%%%%%%%%%%%%%%%%%%%%%%%%%%%%%%%%%%%%%%%%%%%%%%%%%%%
%%%%%%%%%%%%%%%%%%%%%%%%%%%%%%%%%%%%%%%%%%%%%%%%%%%%%%%%%%%%%%%%%%%%%%%%%%%%%%%%%%%%%%%%%%%%%%%%%
%%%%%%%%%%%%%%%%%%%%%%%%%%%%%%%%%%%%%%%%%%%%%%%%%%%%%%%%%%%%%%%%%%%%%%%%%%%%%%%%%%%%%%%%%%%%%%%%%
%%%%%%%%%%%%%%%%%%%%%%%%%%%%%%%%%%%%%%%%%%%%%%%%%%%%%%%%%%%%%%%%%%%%%%%%%%%%%%%%%%%%%%%%%%%%%%%%%
%%%%%%%%%%%%%%%%%%%%%%%%%%%%%%%%%%%%%%%%%%%%%%%%%%%%%%%%%%%%%%%%%%%%%%%%%%%%%%%%%%%%%%%%%%%%%%%%%
%%%%%%%%%%%%%%%%%%%%%%%%%%%%%%%%%%%%%%%%%%%%%%%%%%%%%%%%%%%%%%%%%%%%%%%%%%%%%%%%%%%%%%%%%%%%%%%%%

An important feature of all of these models is that each is characterized by two basic parameters, the ``height'' of the potential $\Lambda^4$, which governs the energy density during inflation, and the ``width'' of the potential $\mu$. (Hybrid models have a third free parameter $\phi_c$ which sets the end of inflation.) In order to have a flat potential and a slowly rolling field, there must be a hierarchy of scales such that the width is larger than the height, $\Lambda \ll \mu$. As we saw in the case of the $\lambda \phi^4$ large-field model, typical inflationary potentials have widths of order the Planck scale $\mu \sim \mpl$ and heights of order the scale of Grand Unification $\Lambda \sim M_{\rm GUT} \sim 10^{15}\ {\rm GeV},$ although models can be constructed for which inflation happens at a much lower scale \cite{Knox:1992iy,Linde:1993cn,Kinney:1995cc}. 

\begin{figure}
\begin{center}
%--%\psfig{file=WMAP5Labeled.eps,width=4.5in}
\includegraphics[width=4.5in]{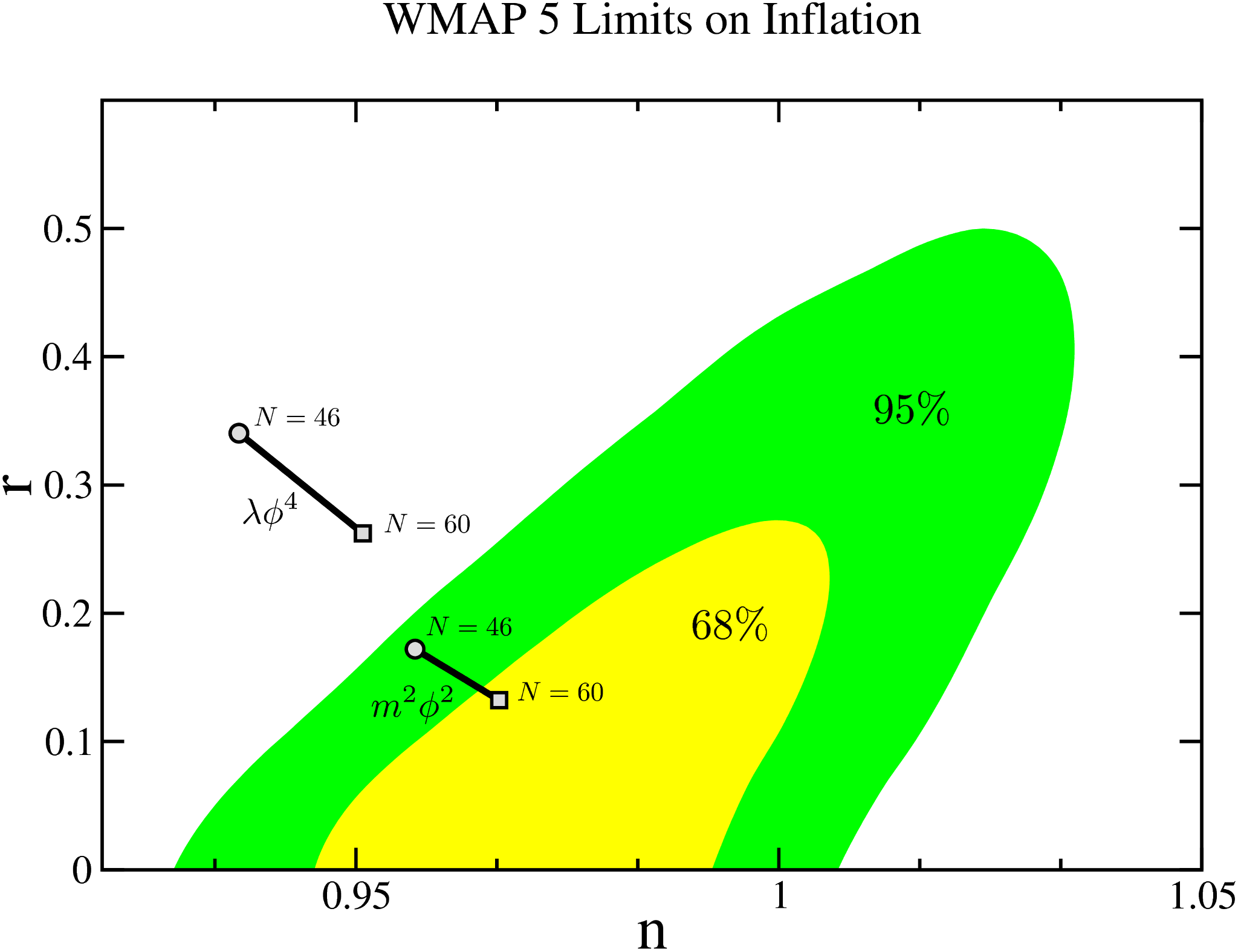}
\end{center}
\caption{Constraints on the $r$, $n$ plane from Cosmic Microwave Background measurements. Shaded regions are the regions allowed by the WMAP5 measurement to 68\% and 95\% confidence. Models plotted are ``large-field'' potentials $V\left(\phi\right) \propto \phi^2$ and $V\left(\phi\right) \propto \phi^4$.}
\label{fig:WMAPrn}
\end{figure}

\begin{figure}
\begin{center}
%--%\psfig{file=WMAP5logLabeled.eps,width=4.5in}
\includegraphics[width=4.5in]{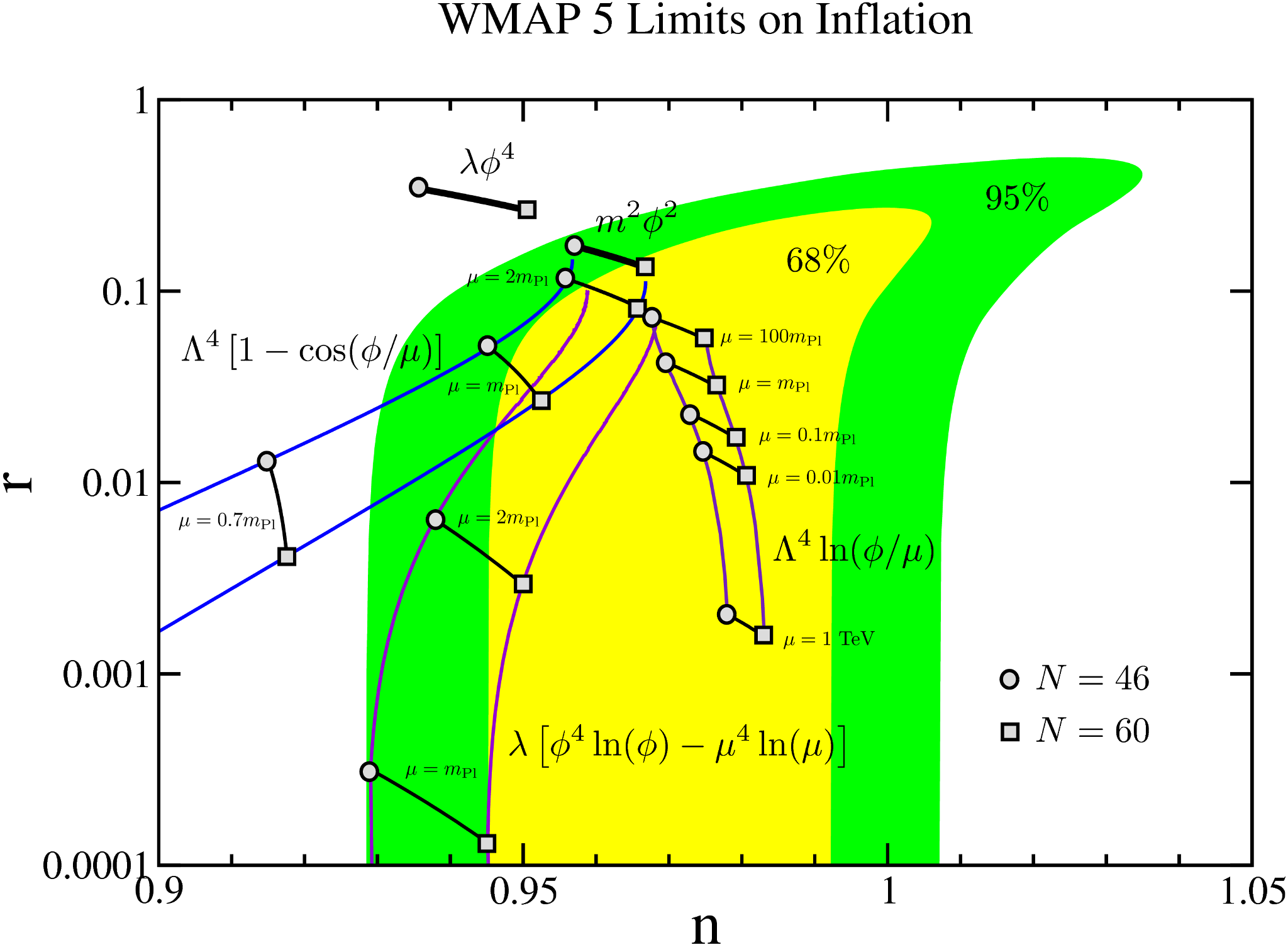}
\end{center}
\caption{Constraints on the $r$, $n$ plane from Cosmic Microwave Background measurements, with the tensor/scalar ratio plotted on a log scale. In addition to the large-field models shown in Fig. \ref{fig:WMAPrn}, three small-field models are plotted against the data: ``Natural Inflation'' from a pseudo-Nambu-Goldstone boson \cite{Freese:1990rb}, with potential $V\left(\phi\right) = \Lambda^4 \left[1 - \cos{\left(\phi / \mu\right)}\right]$, a logarithmic potential $V\left(\phi\right) \propto \ln{\left(\phi\right)}$ typical of supersymmetric models \cite{Stewart:1994ts,Dvali:1994ms,Barrow:1995xb}, and a Coleman-Weinberg potential $V\left(\phi\right) \propto \phi^4 \ln{\left(\phi\right)}$.}
\label{fig:WMAPrnlog}
\end{figure}

The quantities we are interested in for constraining models of inflation are the primordial power spectra $P_{\cal R}$ and $P_T$, which are the underlying source of the CMB temperature anisotropy and polarization. However, the observed CMB  anisotropies depend on a handful of unrelated cosmological parameters, since the primordial fluctuations are processed through the complicated physics of acoustic oscillations. This creates uncertainties due to parameter degeneracies: our best-fit values for $r$ and $n_S$ will depend on what values we choose for the other cosmological parameters such as the baryon density $\Omega_{\rm b}$ and the redshift of reionization $z_{\rm ri}$. To accurately estimate the errors on $r$ and $n_S$, we must fit all the relevant parameters {\it simultaneously}, a process which is computationally intensive, and is typically approached using Bayesian Monte Carlo Markov Chain techniques \cite{Lewis:2002ah}. Here we simply show the results: Figure \ref{fig:WMAPrn} shows the regions of the $r$, $n_S$ parameter space allowed by the WMAP 5-year data set \cite{Kinney:2006qm,Kinney:2008wy}. We have fit over the parameters $\Omega_{\rm CDM}$, $\Omega_{\rm b}$, $\Omega_{\rm Lambda}$, $H_0$, $P_{\cal R}$, $z_{\rm ri}$, $r$, and $n_s$, with a constraint that the universe must be flat, as predicted by inflation, $\Omega_{\rm b} + \Omega_{\rm CDM} + \Omega_{\rm Lambda} = 1$. We see that the data favor a red spectrum, $n_S < 1$, although the scale-invariant limit $n_S = 1$ is still within the 95\%-confidence region.  Our example inflation model $V\left(\phi\right) = \lambda \phi^4$ is convincingly ruled out by WMAP, but the simple potential $V\left(\phi\right) = m^2 \phi^2$ is nicely consistent with the data.\footnote{Liddle and Leach point out that $\lambda \phi^4$ models are special because of their reheating properties, and should be more accurately evaluated at $N = 64$ \cite{Liddle:2003as}. However, this assumes that the potential has no other terms which might become dominant during reheating, and in any case is also ruled out by WMAP5.} Figure \ref{fig:WMAPrnlog} shows the WMAP constraint with $r$ on a logarithmic scale, with the prediction of several small-field models for reference. There is no evidence in the WMAP data for a nonzero tensor/scalar ratio $r$, with a 95\%-confidence upper limit of $r < 0.5$. It is possible to improve these constraints somewhat by adding other data sets, for example the ACBAR high-resolution CMB anisotropy measurement \cite{Reichardt:2008ay} or the Sloan Digital Sky Survey \cite{Loveday:2002ax,Abazajian:2008wr}, which improve the upper limit on the tensor/scalar ratio to $r < 0.3$ or so. Current data are completely consistent with Gaussianity and adiabaticity, as expected from simple single-field inflation models. In the next section, we discuss the outlook for future observation.

%%%%%%%%%%%%%%%%%%%%%%%%%%%%%%%%%%%%%%%%%%%%%%%%%%%%%%%%%%%%%%%%%%%%%%%%%%%%%%%%%%%%%%%%%%%%%%%%%
%%%%%%%%%%%%%%%%%%%%%%%%%%%%%%%%%%%%%%%%%%%%%%%%%%%%%%%%%%%%%%%%%%%%%%%%%%%%%%%%%%%%%%%%%%%%%%%%%

%%%%%%%%%%%%%%%%%%%%%%%%%%%%%%%%%%%%%%%%%%%%%%%%%%%%%%%%%%%%%%%%%%%%%%%%%%%%%%%%%%%%%%%%%%%%%%%%%
%%%%%%%%%%%%%%%%%%%%%%%%%%%%%%%%%%%%%%%%%%%%%%%%%%%%%%%%%%%%%%%%%%%%%%%%%%%%%%%%%%%%%%%%%%%%%%%%%

\section{Outlook and Conclusion}
\label{sec:Conclusions}

 The basic hot Big Bang scenario, in which the universe arises out of a hot, dense, smooth initial state and cools through expansion, is now supported by a compelling set of observations, including the existence of the Cosmic Microwave Background, the primordial abundances of the elements, and the evolution of structure in the universe, all of which are being measured with unprecedented precision. However, this scenario leaves questions unanswered: Why is the universe so big and so old? Why is the universe so close to geometrically flat? What created the initial perturbations which later collapsed to form structure in the universe? The last of these questions is particularly interesting, because recent observations of the CMB, in particular the all-sky anisotropy map made by the landmark WMAP satellite, have directly measured the form of these primordial perturbations.  A striking property of these observed primordial perturbations is that they are correlated on scales larger than the cosmological horizon at the time of last scattering. Such apparently {\it acausal} correlations can only be produced in a few ways \cite{Spergel:1997vq}: 
\begin{itemize}
\item{Inflation.}
\item{Extra dimensions \cite{Khoury:2001wf}.} 
\item{A universe much older than $H_0^{-1}$ \cite{Khoury:2003rt,Brandenberger:2005qj}.}
\item{A varying speed of light \cite{Albrecht:1998ir}.}
\end{itemize} 

In addition, the WMAP data contain spectacular confirmation of the basic predictions of the inflationary paradigm: a geometrically flat universe with Gaussian, adiabatic, nearly scale-invariant perturbations. No other model explains these properties of the universe with such simplicity and economy, and much attention has been devoted to the implications of WMAP for inflation \cite{Spergel:2006hy,Alabidi:2006qa,Seljak:2006bg,Kinney:2006qm,Martin:2006rs,Lesgourgues:2007aa,Peiris:2008be,Alabidi:2008ej,Dunkley:2008ie,Komatsu:2008hk,Kinney:2008wy,Xia:2008ex,Hamann:2008pb,Smith:2008pf,Li:2008vf}. Inflation also makes predictions which have not been well tested by current data but {\it can} be by future experiments, most notably a deviation from a scale-invariant spectrum and the production of primordial gravitational waves. A non-scale-invariant spectrum is weakly favored by the existing data, but constraints on primordial gravity waves are still quite poor.
The outlook for improved data is promising: over the next five to ten years, there will be a continuous stream of increasingly  high-precision data made available which will allow constraint of cosmological parameters relevant for understanding the early universe. The most useful measurements for direct constraint of the inflationary parameter space are observations of the CMB, and current activity in this area is intense. The Planck satellite mission is scheduled to launch in 2009 \cite{:2006uk,Bouchet:2007zz}, and will be complemented by ground- and balloon-based measurements using a variety of technologies and strategies \cite{Leitch:2004gd,Kuo:2006ya,Sievers:2005gj,Ruhl:2004kv,Yoon:2006jc,Taylor:2006jw,Samtleben:2008sj,Crill:2008rd}. 

At the same time, cosmological parameter estimation is a well-developed field. A set of standard cosmological parameters such as the baryon density $\Omega_{\rm b} h^2$, the matter density $\Omega_{\rm m} h^2$, the expansion rate $H_0 \equiv 100 h\ {\rm km/sec}$ are being measured with increasing accuracy.  The observable quantities most meaningful for constraining models of inflation are the ratio $r$ of tensor to scalar fluctuation amplitudes, and the spectral index $n_S$ of the scalar power spectrum. This kind of simple parameterization is at the moment sufficient to describe the highest-precision cosmological data sets. Furthermore, the simplest slow-roll models of inflation predict a nearly exact power-law perturbation spectrum. In this sense, a simple concordance cosmology is well-supported by both data and by theoretical expectation. It could be that the underlying universe really is that simple. However, the simplicity of concordance cosmology is at present as much a statement about the data as about the universe itself. Only a handful of parameters are required to explain existing cosmological data. Adding more parameters to the fit does no good: any small improvement in the fit of the model to the data is offset by the statistical penalty one pays for introducing extra parameters \cite{Trotta:2005ar,Magueijo:2006we,Parkinson:2006ku,Liddle:2006tc,Liddle:2007fy,Pahud:2007gi,Linder:2007fv,Liddle:2007ez,Efstathiou:2008ed}. But the optimal parameter set is a moving target: as the data get better, we will be able to probe more parameters. It may be that a ``vanilla'' universe \cite{Easther:2003fy} of a half-dozen or so parameters will continue to be sufficient to explain observation. But it is reasonable to expect that, as  measurements improve in accuracy, we will see evidence of deviation from such a lowest-order expectation. This is where the interplay between theory and experiment gains the most leverage, because we must understand: (1) what deviations from a simple universe are predicted by models, and (2) how to look for those deviations in the data. It is of course impossible to predict which of the many possible signals (if any) will be realized in the universe in which we live. I discuss below four of the best motivated possibilities, in order of the quality of current constraints. (For a more detailed treatment of these issues, the reader is referred to the very comprehensive CMBPol Mission Concept Study \cite{Baumann:2008aq}.)

\bigskip
\noindent{\em Features in the density power spectrum}

\smallskip
Current data are consistent with a purely power-law spectrum of density perturbations, $P(k) \propto k^{n_S - 1}$ with a ``red'' spectrum ($n_S < 1$) favored by the data at about a $90\%$ confidence level, a figure which depends on the choice of parameter set and priors. Assuming it is supported by future data, the detection of a deviation from a scale-invariant ($n_S=1$) spectrum is a significant milestone, and represents a confirmation of one of the basic predictions of inflation. In slow-roll inflation, this power-law scale dependence is nearly exact, and any additional scale dependence is strongly suppressed. Therefore, detection of a nonzero ``running'' $\alpha = d{n_S}/d\ln{k}$ of the spectral index would be an indication that slow roll is a poor approximation. There is currently no evidence for scale-dependence in the spectral index, but constraints on the overall shape of the power spectrum are likely to improve dramatically through measurements of the CMB anisotropy at small angular scales, improved polarization measurements, and better mapping of large-scale structure. Planck is expected to measure the shape of the spectrum with $2 \sigma$ uncertainties of order $\Delta n \sim 0.01$ and $\Delta\alpha \sim 0.01$ \cite{Kinney:1998md,Copeland:1997mn,Colombo:2008ta,Adshead:2008vn}. Over the longer term, measurements of 21cm radiation from neutral hydrogen promises to be a precise probe of the primordial power spectrum, and would improve these constraints significantly \cite{Barger:2008ii}.

\bigskip
\noindent{\em Primordial Gravitational Waves}
 
\smallskip
In addition to a spectrum $P_{\cal R}$ of scalar perturbations, inflation generically predicts a spectrum $P_T$ of tensor perturbations. The relative amplitude of the two is determined by the equation of state of the fluid driving inflation,
\begin{equation}
r = 16 \epsilon
\end{equation}
Since the scalar amplitude is known from the COBE normalization to be $P_{\cal R} \sim H^2 / \epsilon \sim 10^{-10}$, it follows that measuring the tensor/scalar ratio $r$ determines the inflationary expansion rate $H$ and the associated energy density $\rho$.  Typical inflation models take place with an energy density of around $\rho \sim (10^{15}\ {\rm GeV})^4$, which corresponds to a tensor/scalar ratio of $r \sim 0.1$, although this figure is highly model-dependent. Single-field inflation does not make a definite prediction for the value of $r$: while many choices of potential generate a substantial tensor component, other choices of potential result in an unobservably small tensor/scalar ratio, and there is no particular reason to favor one scenario over another. 

There is at present no observational evidence for primordial gravitational waves: the current upper limit on the tensor/scalar ratio is around $r \leq 0.3$. Detection of even a large primordial tensor signal requires extreme sensitivity. The crucial observation is detection of the odd-parity, or B-mode, component of the CMB polarization signal, which is suppressed relative to the temperature fluctuations, themselves at the $10^{-4}$ level, by at least another four orders of magnitude. This signal is considerably below known foreground levels \cite{Kogut:2007tq}, severely complicating data analysis. Despite the formidable challenges, the observational community has undertaken a broad-based effort to search for the B-mode, and a detection would be a boon for inflationary cosmology. Planck will be sensitive to a tensor/scalar ratio of around $r \simeq 0.1$, and dedicated ground-based measurements can potentially reach limits of order $r \simeq 0.01$. The proposed CMBPol polarization satellite would reach $r$ of order $10^{-3}$ \cite{Baumann:2008aq,Dunkley:2008am}, and direct detection experiments such as BBO could in principle detect $r$ of order $10^{-4}$ \cite{Smith:2005mm}.  

\bigskip
\noindent{\em Primordial Non-Gaussianity}

\smallskip
In addition to a power-law power spectrum, inflation predicts that the primordial perturbations will be distributed according to Gaussian statistics. Like running of the power spectrum, non-Gaussianity is suppressed in slow-roll inflation \cite{Maldacena:2002vr}. However, detection of even moderate non-Gaussianity is considerably more difficult. If the perturbations are Gaussian, the two-point correlation function completely describes the perturbations. This is not the case for non-Gaussian fluctuations: higher-order correlations contain additional information. However, higher-order correlations require more statistics and are therefore more difficult to measure, especially at large angular scales where cosmic variance errors are significant.  Current limits are extremely weak \cite{Spergel:2006hy,Creminelli:2005hu}, and future high angular resolution CMB maps will still fall well short of being sensitive to a signal from slow-roll inflation or even weakly {\it non-}slow-roll models \cite{Liguori:2007sj}. It will take a strong deviation from the slow-roll scenario to generate observable non-Gaussianity. However, a measurement of non-Gaussianity would in one stroke rule out virtually all slow-roll inflation models and force consideration of more exotic scenarios such as DBI inflation \cite{Silverstein:2003hf}, Warm Inflation \cite{Moss:2007cv}, or curvaton scenarios \cite{Sasaki:2006kq}.

\bigskip
\noindent{\em Isocurvature perturbations}

\smallskip
In a universe where the matter consists of multiple components, there are two general classes of perturbation about a homogeneous background: adiabatic, in which the perturbations in all of the fluid components are in phase, and isocurvature, in which the perturbations have independent phases. Single-field inflation predicts purely adiabatic primordial perturbations, for the simple reason that if there is a single field $\phi$ responsible for inflation, then there is a single order parameter governing the generation of density perturbations. This is a nontrivial prediction, and the fact that current data are consistent with adiabatic perturbations is support for the idea of quantum generation of perturbations in inflation. However, current limits on the isocurvature fraction are quite weak \cite{Moodley:2004nz,Bean:2006qz}. If isocurvature modes are detected, it would rule out {\it all} single-field models of inflation. Multi-field models, on the other hand, naturally contain multiple order parameters and can generate isocurvature modes. Multi-field models are naturally motivated by the string ``landscape'', which is believed to contain an enormous number of degrees of freedom. Another possible mechanism for the generation of isocurvature modes is the curvaton mechanism, in which cosmological perturbations are generated by a field other than the inflaton \cite{Lyth:2002my,Lyth:2003ip}. 

The rich interplay between theory and observation that characterizes cosmology today is likely to continue for the foreseeable future. As measurements improve, theory will need to become more precise and complete than the simple picture of inflation that we have outlined in these lectures, and single-field inflation models could yet prove to be a poor fit to the data. However, at the moment, such models provide an elegant, compelling, and (most importantly) scientifically useful picture of the very early universe.  

\section*{Acknowledgments}
I would like to thank the organizers of the Theoretical Advanced Studies Institute (TASI) at Univ. of Colorado, Boulder for giving me the opportunity to return to my alma mater to lecture. Various versions of these lectures were also given at the Perimeter Institute Summer School on Particle Physics, Cosmology, and Strings in 2007, at the Second Annual Dirac Lectures at Florida State University in 2008, and at the Research Training Group at the University of W\"urzburg in 2008. This research is supported in part by the National Science Foundation under grants NSF-PHY-0456777 and NSF-PHY-0757693. I thank Dennis Bessada, Richard Easther, Hiranya Peiris, and Brian Powell for comments on a draft version of the manuscript. 

\begin{appendix}
%--%\label{sec:Appendix}
\section{The Curvature Perturbation in Single-Field Inflation}
\label{sec:Appendix}

\begin{figure}
\begin{center}
%--%\psfig{file=FRWFoliation.eps,width=4.5in}
\includegraphics[width=4.5in]{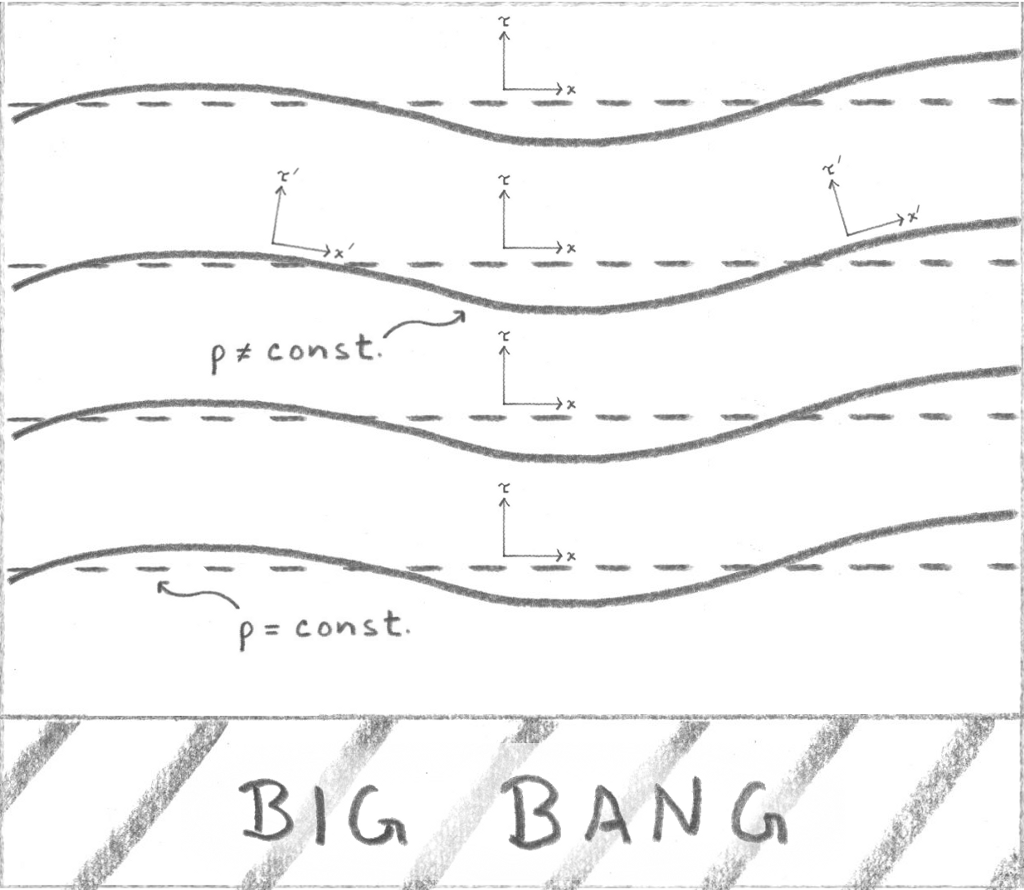}
\end{center}
\caption{Foliations of an FRW spacetime. Comoving hypersurfaces (dashed lines) have constant density, but another choice of gauge (solid lines) will have unphysical density fluctuations which are an artifact of the choice of gauge.}
\label{fig:FRWfoliation}
\end{figure}

In this section, we discuss the generation of perturbations in the density $\delta\left({\bf x}\right) \equiv \delta \rho / \rho$ generated during inflation. The process is similar to the case of a free scalar field discussed in Sec. \ref{sec:PerturbationsInInflation}: the inflaton field $\phi$, like any other scalar, will have quantum fluctuations which are stretched to superhorizon scales and subsequently freeze out as classical perturbations. The difference is that the energy density of the universe is dominated by the inflaton potential, so that quantum fluctuations in $\phi$ generate perturbations in the density $\rho$. Dealing with such density perturbations is complicated by the fact that in General Relativity, we are free to choose any coordinate system, or {\it gauge}, we wish. To see why, consider the case of an FRW spacetime evolving with scale factor $a\left(t\right)$ and uniform energy density $\rho\left(t,{\bf x}\right) = \bar\rho\left(t\right)$. What we mean here by ``uniform'' energy density, or homogeneity, is that the density is a constant in {\it comoving} coordinates. But the physics is independent of coordinate system, so we could equally well work in coordinates $t'$, ${\bf x}'$ for which constant-time hypersurfaces do {\it not} have constant density (Fig. \ref{fig:FRWfoliation}). 
%%%%%%%%%%%%%%%%%%%%%%%%%%%%%%%%%%%%%%%%%%%%%%%%%%%%%%%%%%%%%%%%%%%%%%%%%%%%%%%%%%%%%%%%%%%%%%%%%
%%%%%%%%%%%%%%%%%%%%%%%%%%%%%%%%%%%%%%%%%%%%%%%%%%%%%%%%%%%%%%%%%%%%%%%%%%%%%%%%%%%%%%%%%%%%%%%%%
Such a division of spacetime into a time coordinate and a set of orthogonal spacelike hypersurfaces is called a {\it foliation} of the spacetime, and is an arbitrary choice equivalent to a choice of coordinate system. 

For an FRW spacetime, comoving coordinates correspond to a foliation of the spacetime into spatial hypersurfaces with constant density: this is the most physically intuitive description of the spacetime. Any other choice of foliation of the spacetime would result in density ``perturbations'' which are entirely due to the choice of coordinate system. Such unphysical perturbations are referred to as {\it gauge modes}. Another way to think of this is that the division between what we call ``background'' and what we call ``perturbation'' is itself gauge-dependent. For perturbations with wavelength smaller than the horizon, it is possible to define background and perturbation without ambiguity, since all observers can agree on a definition of time coordinate $t$ and on an average density $\bar\rho\left(t\right)$. Not so for superhorizon modes: if we consider a perturbation mode with wavelength much larger than the horizon size, observers in different horizons will see themselves in independently evolving, homogeneous patches of the universe: a ``perturbation'' can be defined only by comparing causally disconnected observers, and there is an inherent gauge ambiguity in how we do this. The canonical paper on gauge issues in General Relativistic perturbation theory is by Bardeen \cite{Bardeen:1980kt}. A good pedagogical treatment with a focus on inflationary perturbations can be found in Ref. \cite{Komatsu:2002db}. 

In practice, instead of the density perturbation $\delta$, the quantity most directly relevant to CMB physics is the Newtonian potential $\Phi$ on the surface of last scattering. For example, this is the quantity that directly appears in Eq. (\ref{eq:SachsWolfe}) for the Sachs-Wolfe Effect. The Newtonian potential is related to the density perturbation $\delta$ through the Poisson Equation:
\begin{equation}
\nabla^2 \Phi =  4 \pi G \bar\rho a^2 \delta,
\end{equation} 
where the factor of $a^2$ comes from defining the gradient $\nabla$ relative to comoving coordinates. Like $\delta$, the Newtonian potential $\Phi$ is a gauge-dependent quantity: its value depends on how we foliate the spacetime. For example, we are free to choose spatial hypersurfaces such that the density is constant, and the Newtonian potential vanishes everywhere: $\Phi\left(t,{\bf x}\right) = 0$. This foliation of the spacetime is equivalent to the qualitative picture above of different horizon volumes as independently evolving homogeneous universes. Observers in different horizons use the density $\rho$ to synchronize their clocks with one another. Such a foliation is not very useful for computing the Sachs-Wolfe effect, however! Instead, we need to define a gauge which corresponds to the Newtonian limit in the present universe. To accomplish this, we describe the evolution of a scalar field dominated cosmology using the useful fluid flow approach \cite{Hawking:1966qi,Ellis:1989jt,Liddle:1993fq,Sasaki:1995aw,Challinor:1998xk}. (An alternate strategy involves the construction of gauge-invariant variables: see Refs. \cite{Kodama:1985bj,Mukhanov:1990me} for reviews.)

\begin{figure}
\begin{center}
%--%\psfig{file=TimelikeCongruence.eps,width=4.5in}
\includegraphics[width=4.5in]{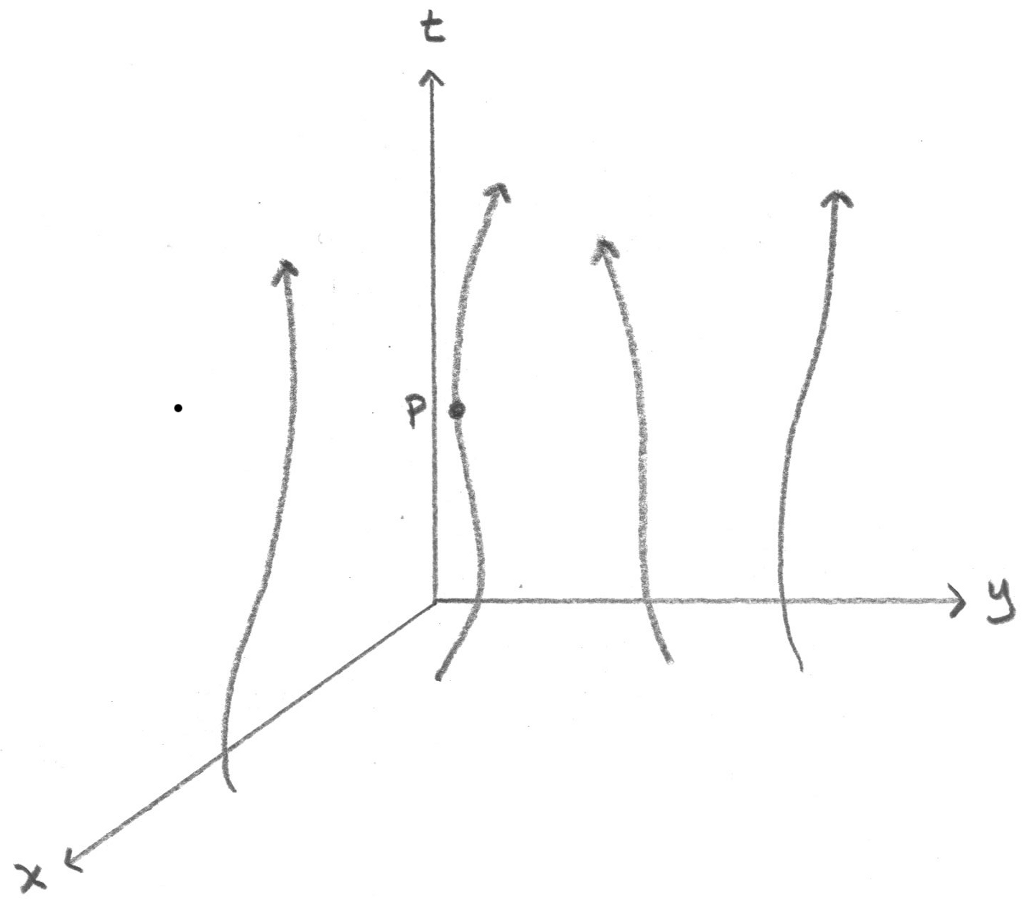}
\end{center}
\caption{A timelike congruence in spacetime. Each event $P$ is intersected by exactly one world line in the congruence.}
\label{fig:timelikecongruence}
\end{figure}

Consider a scalar field $\phi$ in an arbitrary background $g_{\mu\nu}$. The stress-energy tensor of the scalar field may be written
\begin{equation}
\label{eq:generalstressenergy}
T_{\mu \nu} = \phi_{,\mu} \phi_{,\nu} - g_{\mu \nu} \left[ \frac{1}{2} g^{\alpha\beta} \phi_{,\alpha} \phi_{,\beta} - V\left(\phi\right) \right].
\end{equation}
Note that we have not yet made any assumptions about the metric $g_{\mu \nu}$ or about the scalar field $\phi$. Equation (\ref{eq:generalstressenergy}) is a completely general expression. 
We can define a fluid four-velocity for the scalar field by
\begin{equation}
\label{eq:deffourvelocity}
u_\mu \equiv \frac{\phi_{,\mu}}{\sqrt{g^{\alpha \beta} \phi_{,\alpha} \phi_{,\beta}}}.
\end{equation}
It is not immediately obvious why this should be considered a four-velocity. Consider any perfect fluid filling spacetime. Each element of the fluid has four-velocity $u^\mu\left(x\right)$ at every point in spacetime which is everywhere timelike,
\begin{equation}
u^\mu\left(x\right) u_\mu\left(x\right) = 1\ \forall x.
\end{equation}
Such a collection of four-vectors is called a {\it timelike congruence}. We can draw the congruence defined by the fluid four-velocity as a set of flow lines in spacetime (Fig. \ref{fig:timelikecongruence}). Each event $P$ in spacetime has one and only one flow line passing through it. The fluid four-velocity is then a set of unit-normalized tangent vectors to the flow lines, $u^\mu u_\mu = 1$. 
%%%%%%%%%%%%%%%%%%%%%%%%%%%%%%%%%%%%%%%%%%%%%%%%%%%%%%%%%%%%%%%%%%%%%%%%%%%%%%%%%%%%%%%%%%%%%%%%%
%%%%%%%%%%%%%%%%%%%%%%%%%%%%%%%%%%%%%%%%%%%%%%%%%%%%%%%%%%%%%%%%%%%%%%%%%%%%%%%%%%%%%%%%%%%%%%%%%
For a scalar field, we construct a timelike congruence by Eq. (\ref{eq:deffourvelocity}), which is by construction unit normalized:
\begin{equation}
u^\mu u_\mu = \frac{g^{\mu \nu} \phi_{,\mu} \phi_{,\nu}}{g^{\alpha \beta} \phi_{,\alpha} \phi_{,\beta}} = 1.
\end{equation}

We then define the ``time'' derivative of any scalar quantity $f(x)$ by the projection of the derivative along the fluid four-velocity:
\begin{equation}
\label{eq:deftimederiv}
\dot f \equiv u^{\mu} f_{,\mu}.
\end{equation}
In particular, the time derivative of the scalar field itself is
\begin{equation}
\label{eq:defphidot}
\dot\phi \equiv u^{\mu} \phi_{,\mu} = \sqrt{g^{\alpha \beta} \phi_{,\alpha} \phi_{,\beta}}.
\end{equation}
Note that in the homogeneous case, we recover the usual time derivative,
\begin{equation}
\nabla \phi = 0 \Rightarrow \dot\phi = \sqrt{g^{00} \phi_{,0} \phi_{,0}} = \frac{d \phi}{d t}.
\end{equation}
The stress-energy tensor (\ref{eq:generalstressenergy}) in terms of $\dot\phi$ takes the form
\begin{equation}
T_{\mu\nu} = \left[\frac{1 }{ 2} \dot\phi^2 + V\left(\phi\right)\right] u_\mu u_\nu + \left[\frac{1 }{ 2} \dot\phi^2 - V\left(\phi\right)\right] \left(u_\mu u_\nu - g_{\mu\nu}\right).
\end{equation}
We can then define a generalized density $\rho$ and and pressure $p$ by
\begin{eqnarray}
\label{eq:defrhop}
\rho &&\equiv \frac{1 }{ 2} \dot\phi^2 + V\left(\phi\right),\cr
p &&\equiv \frac{1 }{ 2} \dot\phi^2 - V\left(\phi\right).
\end{eqnarray}
Note that despite the familiar form of these expressions, they are defined without any assumption of homogeneity of the scalar field or even the imposition of a particular  metric. 

In terms of the generalized density and pressure, the stress-energy (\ref{eq:generalstressenergy}) is
\begin{equation}
\label{eq:simplestressenergy}
T_{\mu\nu} = \rho u_\mu u_\nu + p h_{\mu \nu},
\end{equation}
where the tensor $h_{\mu\nu}$ is defined as:
\begin{equation}
h_{\mu\nu} \equiv u_\mu u_\nu - g_{\mu\nu}.
\end{equation}
The tensor $h_{\mu\nu}$ can be easily seen to be a projection operator onto hypersurfaces orthogonal to the four-velocity $u^\mu$. For any vector field $A^\mu$, the product $h_{\mu\nu} A^\nu$ is identically orthogonal to the four-velocity:
\begin{equation}
\left(h_{\mu\nu} A^\nu\right) u^\mu = A^\nu \left(h_{\mu\nu} u^\mu\right) = 0.
\end{equation}
Therefore, as in the case of the time derivative, we can define gradients by projecting the derivative onto surfaces orthogonal to the four-velocity
\begin{equation}
\label{eq:defgradient}
\left(\nabla f\right)^\mu \equiv h^{\mu\nu} f_{,\nu}.
\end{equation}
In the case of a scalar field fluid with four-velocity given by Eq. (\ref{eq:deffourvelocity}), the gradient of the field identically vanishes,
\begin{equation}
\left(\nabla \phi\right)^\mu = 0.
\end{equation}
Note that despite its relation to a ``spatial'' gradient, $\nabla f$ is a covariant quantity, {\it i.e.} a four-vector. 

\begin{figure}
\begin{center}
%--%\psfig{file=foliation.eps,width=4.5in}
\includegraphics[width=4.5in]{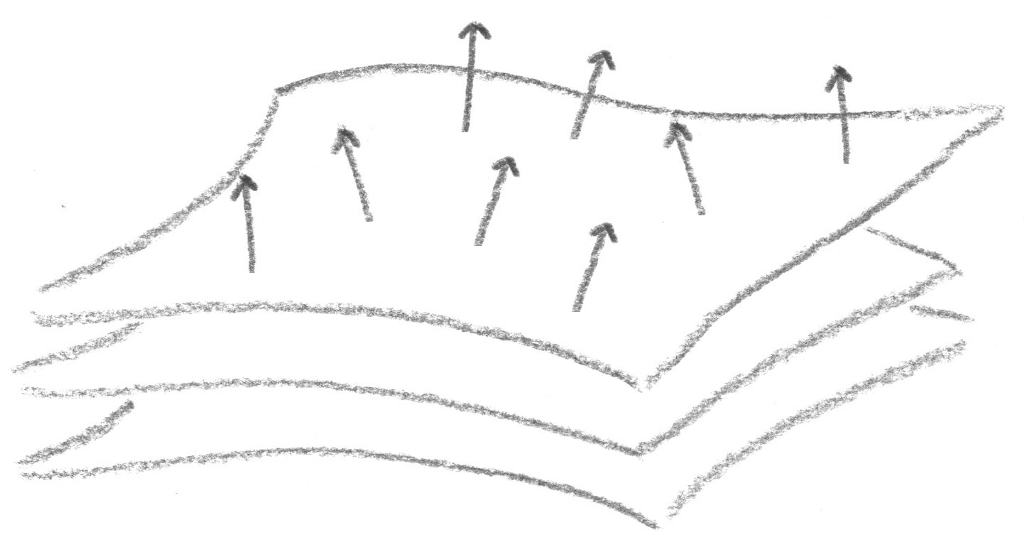}
\end{center}
\caption{A comoving foliation of spacetime. Spatial hypersurfaces are everywhere orthogonal to the fluid four-velocity $u^{\mu}$.}
\label{fig:comovingfoliation}
\end{figure}

Our fully covariant definitions of ``time'' derivatives and ``spatial'' gradients suggest a natural foliation of the spacetime into spacelike hypersurfaces, with time coordinate orthogonal to those hypersurfaces. We can define spatial hypersurfaces to be everywhere orthogonal to the fluid flow (Fig. \ref{fig:comovingfoliation}). This is equivalent to choosing a coordinate system for which $u^i = 0$ everywhere. Such a gauge choice is called {\it comoving} gauge. In the case of a scalar field, we can equivalently define comoving gauge as a coordinate system in which spatial gradients of the scalar field $\phi_{,i}$ are defined to vanish. Therefore the time derivative (\ref{eq:deftimederiv}) is just the derivative with respect to the coordinate time in comoving gauge
\begin{equation}
\dot\phi = \left(\frac{\partial \phi }{ \partial t}\right)_{\rm c}.
\end{equation}
Similarly, the generalized density and pressure (\ref{eq:defrhop}) are just defined to be those quantities as measured in comoving gauge.
%%%%%%%%%%%%%%%%%%%%%%%%%%%%%%%%%%%%%%%%%%%%%%%%%%%%%%%%%%%%%%%%%%%%%%%%%%%%%%%%%%%%%%%%%%%%%%%%%
%%%%%%%%%%%%%%%%%%%%%%%%%%%%%%%%%%%%%%%%%%%%%%%%%%%%%%%%%%%%%%%%%%%%%%%%%%%%%%%%%%%%%%%%%%%%%%%%%

The equations of motion for the fluid can be derived from stress-energy conservation,
\begin{equation}
T^{\mu\nu}{}_{\!;\nu} = 0 = \dot\rho u^\mu + (\nabla p)^\mu + \left(\rho + p\right) \left(\dot u^\mu + u^\mu \Theta\right),
\end{equation}
where the quantity $\Theta$ is defined as the divergence of the four-velocity,
\begin{equation}
\Theta \equiv u^\mu{}_{\!;\mu}.
\end{equation}
We can group the terms multiplied by $u^\mu$ separately, resulting in familiar-looking equations for the generalized density and pressure
\begin{eqnarray}
\dot\rho + \Theta \left(\rho + p\right) = 0,&&\cr
(\nabla p)^\mu + \left(\rho + p\right) \dot u^\mu = 0.&&
\end{eqnarray}
The first of these equations, similar to the usual continuity equation in the homogeneous case, can be rewritten using the definitions of the generalized density and pressure (\ref{eq:defrhop}) in terms of the field as
\begin{equation}
\label{eq:generalizedeqofmotion}
\ddot\phi + \Theta \dot\phi + V'\left(\phi\right) = 0.
\end{equation}
This suggests identifying the divergence $\Theta$ as a generalization of the Hubble parameter $H$ in the homogeneous case. In fact, if we take $g_{\mu\nu}$ to be a flat Friedmann-Robertson-Walker (FRW) metric and take comoving gauge, $u^\mu = (1,0,0,0)$,
we have
\begin{equation}
u^\mu{}_{\!;\mu} = 3 H,
\end{equation}
and the generalized equation of motion (\ref{eq:generalizedeqofmotion}) becomes the familiar equation of motion for a homogeneous scalar,
\begin{equation}
\ddot\phi + 3 H \dot\phi + V'\left(\phi\right) = 0.
\end{equation}

Now consider perturbations $\delta g_{\mu\nu}$ about a flat FRW metric,
\begin{equation}
g_{\mu\nu} = a^2\left(\tau\right) \left[\eta_{\mu\nu} + \delta g_{\mu\nu}\right],
\end{equation}
where $\tau$ is the conformal time and $\eta$ is the Minkowski metric $\eta = {\rm diag}\left(1,-1,-1,-1\right)$. A general metric perturbation $\delta g_{\mu\nu}$ can be separated into components which transform independently under coordinate transformations \cite{Bardeen:1980kt}, 
\begin{equation}
\delta g_{\mu\nu} = \delta g_{\mu\nu}^{\rm scalar} + \delta g_{\mu\nu}^{\rm vector} + \delta g_{\mu\nu}^{\rm tensor}.
\end{equation}
The tensor component is just the transverse-traceless gravitational wave perturbation, discussed in Section \ref{sec:PerturbationsInInflation}, and vector perturbations are not sourced by single-field inflation.
We therefore specialize to the case of scalar perturbations, for which the metric perturbations can be written generally in terms of four scalar functions of space and time $A$, $B$, ${\cal R}$, and $H_T$:
\begin{eqnarray}
&&\delta g_{00} = 2 A\cr
&&\delta g_{0i} = \partial_i B\cr
&&\delta g_{ij} = 2 \left[{\cal R} \delta_{ij} + \partial_i \partial_j H_T\right].
\end{eqnarray}
We are interested in calculating $\cal R$. Recall that in the Newtonian limit of General Relativity, we can write perturbations about the Minkowski metric in terms of the Newtonian potential $\Phi$ as:
\begin{equation}
ds^2 = \left( 1 + 2 \Phi\right) dt^2 - \left(1 - 2 \Phi\right) \delta_{ij} dx^i dx^j.
\end{equation}
Similarly, we can write Newtonian perturbations about a flat FRW metric as
\begin{equation}
ds^2 = a^2\left(\tau\right) \left[\left( 1 + 2 \Phi\right) d\tau^2 - \left(1 - 2 \Phi\right) \delta_{ij} dx^i dx^j\right].
\end{equation}
We therefore expect $\Phi \propto {\cal R}$ in the Newtonian limit. A careful calculation \cite{Kodama:1985bj,Liddle:1993fq} gives
\begin{equation}
\Phi = - \frac{3 \left(1 + w\right)}{5 + 3 w} {\cal R},
\end{equation}
so that in a matter-dominated universe,
\begin{equation}
\Phi = - \frac{3}{5} {\cal R}.
\end{equation}
In these expressions, ${\cal R}$ is the curvature perturbation measured on comoving hypersurfaces. To see qualitatively why comoving gauge corresponds correctly to the Newtonian limit in the current universe, consider the end of inflation. Since inflation ends at a particular field value $\phi = \phi_e$, comoving gauge corresponds to a foliation for which inflation ends at {\it constant time} at all points in space: all observers synchronize their clocks to $\tau = 0$ at the end of inflation. This means that the background, or unperturbed universe is exactly the homogeneous case diagrammed in Fig. \ref{fig:INFLdiagram}, and the comoving curvature perturbation ${\cal R}$ is the Newtonian potential measured relative to that background. 

To calculate ${\cal R}$, we start by calculating the four-velocity $u^\mu$ in terms of the perturbed metric.\footnote{This treatment closely follows that of Sasaki and Stewart \cite{Sasaki:1995aw}, except that we use the opposite sign convention for $N$.} 
If we specialize to comoving gauge, $u^i \equiv 0$, the norm of the four-velocity can be written
\begin{equation}
u^\mu u_\mu = a^2 \left(1 + 2 A\right) \left(u^0\right)^2 = 1,
\end{equation}
and the timelike component of the four-velocity is, to linear order,
\begin{eqnarray}
&&u^0 = \frac{1 }{ a} \left(1 - A\right)\cr
&&u_0 = a \left(1 + A\right).
\end{eqnarray}
The velocity divergence $\Theta$ is then
\begin{eqnarray}
\label{eq:thetacomoving}
\Theta &&= u^\mu{}_{\!;\mu} = u^0{}_{\!,0} + \Gamma^\alpha{}_{\!\alpha 0} u^0\cr
&&= 3 H \left[1 - A - \frac{1 }{ a H} \left(\frac{\partial {\cal R} }{ \partial \tau} + \frac{1 }{ 3} \partial_i \partial_i \frac{\partial H_T }{ \partial \tau}\right)\right],
\end{eqnarray}
where the unperturbed Hubble parameter is defined as
\begin{equation}
H \equiv \frac{1 }{ a^2} \frac{\partial a }{ \partial \tau}.
\end{equation}
Fourier expanding $H_T$,
\begin{equation}
\partial_i \partial_i H_T = k^2 H_T,
\end{equation}
we see that for long-wavelength modes $k \ll a H$, the last term in Eq. (\ref{eq:thetacomoving}) can be ignored, and the velocity divergence is
\begin{equation}
\label{eq:comovingtheta}
\Theta \simeq 3 H \left[1 - A - \frac{1 }{ a H} \frac{\partial {\cal R} }{ \partial \tau}\right].
\end{equation}
Remembering the definition of the number of e-folds in the unperturbed case,
\begin{equation}
N \equiv - \int{H dt}.
\end{equation}
we can define a generalized number of e-folds as the integral of the velocity divergence along comoving world lines:
\begin{equation}
{\cal N} \equiv - \frac{1 }{ 3} \int{\Theta d s} = - \frac{1 }{ 3} \int{\Theta \left[a \left(1 + A\right) d \tau\right]}.
\end{equation}
Using Eq. (\ref{eq:comovingtheta}) for $\Theta$ and evaluating to linear order in the metric perturbation results in
\begin{equation}
{\cal N} = {\cal R} - \int{H d t},
\end{equation}
and we have a simple expression for the curvature perturbation,
\begin{equation}
{\cal R} = {\cal N} - N.
\end{equation}

\begin{figure}
\begin{center}
%--%\psfig{file=deltaN.eps,width=4.5in}
\includegraphics[width=4.5in]{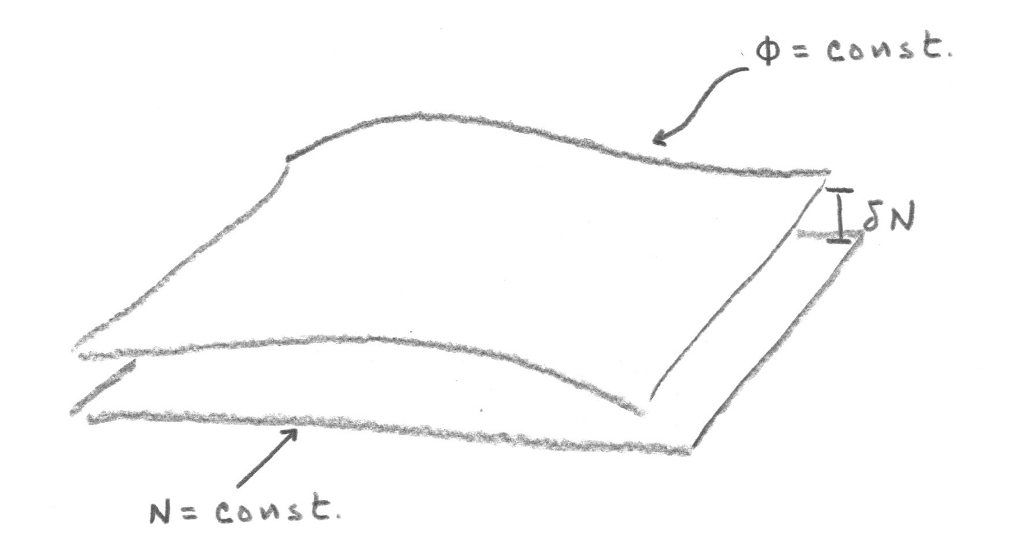}
\end{center}
\caption{Flat and comoving hypersurfaces.}
\label{fig:hypersurfaces}
\end{figure}

This requires a little physical interpretation:  we defined comoving hypersurfaces such that the field has no spatial variation,
\begin{equation}
\left(\nabla \phi\right)^\mu = 0\ \Rightarrow \phi = {\rm const.}
\end{equation}
Then $\cal N$ is the number of e-folds measured on comoving hypersurfaces. But we can equivalently foliate the spacetime such that spatial hypersurfaces are flat, and the field exhibits spatial fluctuations:
\begin{equation}
A = {\cal R} = 0\ \Rightarrow \phi \neq {\rm const.}
\end{equation}
On flat hypersurfaces, the field varies, but the curvature does not, so that the metric on these hypersurfaces is exactly of the FRW form (\ref{eq:generalFRWmetric}) with $k = 0$. We then see immediately that 
\begin{equation}
N = - \int{H d t} = {\rm const.}
\end{equation}
is the number of e-folds measured on flat hypersurfaces, and has no spatial variation. The curvature perturbation $\cal R$ is the difference in the number of e-folds between the two sets of hypersurfaces (Fig. \ref{fig:hypersurfaces}). This can be expressed to linear order in terms of the field variation $\delta\phi$ on flat hypersurfaces as 
\begin{equation}
{\cal R} = {\cal N} - N = - \frac{\delta N}{\delta \phi} \delta\phi
\end{equation}
where $\cal R$ is measured on comoving hypersurfaces, and $\delta N / \delta\phi$ and $\delta\phi$ are measured on flat hypersurfaces. 
%%%%%%%%%%%%%%%%%%%%%%%%%%%%%%%%%%%%%%%%%%%%%%%%%%%%%%%%%%%%%%%%%%%%%%%%%%%%%%%%%%%%%%%%%%%%%%%%%
%%%%%%%%%%%%%%%%%%%%%%%%%%%%%%%%%%%%%%%%%%%%%%%%%%%%%%%%%%%%%%%%%%%%%%%%%%%%%%%%%%%%%%%%%%%%%%%%%
We can express $N$ as a function of the field $\phi$:
\begin{equation}
\label{eq:defN}
N = - \int{H d t} = - \int{\frac{H }{ \dot\phi} d\phi}.
\end{equation}
For monotonic field evolution, we can express $\dot\phi$ as a function of $\phi$, so that
\begin{equation}
\frac{\delta N }{ \delta\phi} = - \frac{H }{ \dot\phi},
\end{equation}
and the curvature perturbation is given by
\begin{equation}
{\cal R} = {\cal N} - N = - \frac{\delta N }{ \delta\phi} \delta\phi = \frac{H }{ \dot\phi} \delta\phi.
\end{equation}
Note that this is an expression for the metric perturbation ${\cal R}$ on comoving hypersurfaces, calculated in terms of quantities defined on {\em flat} hypersurfaces. For $\delta\phi$ produced by quantum fluctuations in inflation, the two-point correlation function is
\begin{equation}
\sqrt{\left\langle \delta\phi^2\right\rangle} = \frac{H }{ 2 \pi},
\end{equation}
and the two-point correlation function for curvature perturbations is
\begin{equation}
\label{eq:curvaturepert}
\sqrt{\left\langle{\cal R}^2\right\rangle} = \frac{H^2 }{ 2 \pi \dot\phi} = \frac{H}{\mpl \sqrt{\pi \epsilon}},
\end{equation}
which is the needed result. 

\end{appendix}
\bibliographystyle{ws-procs9x6}

\end{document}